\documentstyle[preprint,aps]{revtex}
\tighten
\draft
\widetext
\input epsf
\preprint{CLNS 98/1549, HUTP-97/A044, NUB 3168}
\bigskip
\bigskip
\begin{document}
\title{Type IIB Orientifolds, F-theory, Type I Strings on Orbifolds\\ 
and Type I - Heterotic Duality}
\medskip
\author{Zurab Kakushadze$^{1,2}$\footnote{E-mail: 
zurab@string.harvard.edu},
Gary Shiu$^3$\footnote{E-mail: shiu@mail.lns.cornell.edu} and S.-H. Henry Tye$^3$\footnote{E-mail: tye@mail.lns.cornell.edu}}
\bigskip
\address{$^1$Lyman Laboratory of Physics, Harvard University, Cambridge, 
MA 02138\\
$^2$Department of Physics, Northeastern University, Boston, MA 02115\\
$^3$Newman Laboratory of Nuclear Studies, Cornell University,
Ithaca, NY 14853}
\date{\today}
\bigskip
\medskip
\maketitle

\begin{abstract}
{}We consider six and
four dimensional ${\cal N}=1$ supersymmetric orientifolds of Type IIB 
compactified on orbifolds. 
We give the conditions under which the perturbative world-sheet
orientifold approach is adequate, and list the four dimensional ${\cal N}=1$ 
orientifolds (which are rather constrained) that satisfy these conditions.
We argue that in most cases orientifolds contain 
non-perturbative sectors that are missing in the world-sheet approach.
These non-perturbative 
sectors can be thought of as arising from D-branes wrapping various collapsed 
2-cycles in the orbifold. 
Using these observations, we explain certain
``puzzles'' in the literature on four dimensional 
orientifolds. In particular, in some four dimensional orientifolds the ``naive''
tadpole cancellation conditions have no solution. However, 
these tadpole cancellation
conditions are derived using the world-sheet approach which we argue to
be inadequate in these cases due to appearance of additional non-perturbative
sectors. The main tools in our analyses are the map between 
F-theory and orientifold vacua and Type I-heterotic duality.
Utilizing the consistency conditions we have found in this paper, we discuss
consistent four dimensional chiral ${\cal N}=1$ Type I vacua which 
are non-perturbative
from the heterotic viewpoint.
 
\end{abstract}

\section{Introduction}\label{intro}

{}In ten dimensions there are five consistent string theories. The 
first four, Type IIA, Type IIB, $E_8\otimes E_8$ heterotic and 
${\mbox{Spin}}(32)/{\bf Z}_2$ heterotic, are theories of oriented 
closed strings. The last one, Type I, is a theory of both unoriented closed
and open strings. Perturbatively, these five theories are apparently 
different. In recent years, however, a unified picture has emerged, where
the five string theories appear as different regimes of an underlying theory
related via a web of conjectured dualities
in ten and lower dimensions. Most of these 
dualities are intrinsically non-perturbative, and often shed light on 
non-perturbative phenomena in one theory by mapping them to perturbative
phenomena in another theory.

{}As to the perturbative formulation, the four oriented closed string theories
are relatively well understood. Conformal field theory and modular invariance
serve as guiding principles for perturbative model building in closed string 
theories. Type I, however, still remains the least understood string theory
even perturbatively. This is in part due to lack of modular invariance, which 
is necessary for perturbative consistency of oriented closed string theories.

{}In the past years various unoriented closed plus open string vacua have been
constructed using orientifold techniques. Type IIB orientifolds are 
generalized orbifolds 
that involve world-sheet parity reversal along with geometric symmetries 
of the theory. The orientifold procedure results in an unoriented closed string theory.
Consistency then generically requires introducing open strings that can 
be viewed as starting and ending on D-branes \cite{Db}. 
In particular, Type I compactifications on toroidal orbifolds can be viewed as Type IIB
orientifolds with a certain choice of the orientifold projection. 
Global Chan-Paton charges
associated with D-branes manifest themselves as a gauge symmetry in 
space-time. D-branes (as well as orientifold planes) are coherent states \cite{nappi,PC}
built from a superposition of an infinite tower of closed string oscillators acting 
on the momentum and/or winding states.

{}To ensure that a given orientifold model gives rise to a consistent string theory
it is necessary to make sure that the underlying conformal field theory satisfies
certain self-consistency requirements. However, conformal field theories on 
world-sheets with boundaries (ultimately
present in an open sting theory) are still poorly understood. 
To circumvent these difficulties some techniques have been developed 
in the past (see, {\em e.g.}, \cite{nappi,PC,IO,Dunbar,dai}). The idea is to
implement factorization of loop amplitudes (to ensure, say, consistency
of closed-to-open string transitions), generalized GSO projections 
(to guarantee correct spin-statistics relation in space-time), and (at the
last step) tadpole cancellation (which is required for finiteness).
In this approach space-time anomaly 
cancellation is expected to be guaranteed 
by the world-sheet consistency of the theory, 
just as in oriented closed string theories.

{}These techniques have been (rather) successfully
applied to the construction of six dimensional ${\cal N}=1$ space-time supersymmetric 
orientifolds of Type IIB compactified on orbifold limits of K3 (that is, toroidal orbifolds
$T^4/{\bf Z}_N$, $N=2,3,4,6$). In particular, the ${\bf Z}_2$ orbifold case \cite{PS,GP}
has been studied in detail. This construction was subsequently generalized to
other orbifold limits of K3 (namely, ${\bf Z}_N$ with $N=3,4,6$) in Refs \cite{GJ,DP}. 
These orientifold models contain more than one tensor multiplet in their massless spectra,
and, therefore, describe six dimensional vacua which are non-perturbative from the 
heterotic viewpoint. 

{}It is natural to expect that these orientifold constructions should be generalizable
to the cases of four dimensional ${\cal N}=1$ space-time supersymmetric orientifolds
of Type IIB on orbifold limits of Calabi-Yau three-folds (that is, toroidal
orbifolds $T^4/G$ with $SU(3)$ holonomy). Understanding such compactifications is
extremely desirable as according to the conjectured Type I-heterotic duality \cite{PW}
certain non-perturbative heterotic phenomena are expected to
have perturbative Type I origins. In particular, non-perturbative 
dynamics of heterotic NS 5-branes
under this duality is mapped to (at least naively) perturbative dynamics of Type I 
D5-branes. 

{}The first example of a four dimensional ${\cal N}=1$ Type I vacuum was constructed in 
Ref \cite{BL} as an orientifold of Type IIB on a ${\bf Z}_2 \otimes {\bf Z}_2$ toroidal 
orbifold. This model has enhanced gauge symmetries from D5-branes which are
non-perturbative from the heterotic viewpoint. This vacuum is non-chiral, however. To 
obtain chiral vacua it is natural to try other orbifold groups. The first example of a chiral
${\cal N}=1$ Type I vacuum in four dimensions was constructed in Ref \cite{Sagnotti}
via an orientifold of Type IIB on the $Z$-orbifold. This vacuum contains no D5-branes,
and it was shown to be dual to a perturbative heterotic vacuum in Ref \cite{ZK}. (Other
examples of such Type I vacua have been constructed in Refs \cite{KS1,KS2} via
orientifolds of Type IIB on ${\bf Z}_7$ and ${\bf Z}_3\otimes {\bf Z}_3$ orbifolds.)

{}Subsequently, the first four dimensional chiral ${\cal N}=1$ Type I vacuum which is 
non-perturbative from the heterotic viewpoint was constructed in Ref \cite{KS2} via
an orientifold of Type IIB on a ${\bf Z}_6$ orbifold. This model has D5-branes giving rise
to enhanced gauge symmetries which are non-perturbative from the heterotic
viewpoint. 

{}In Ref \cite{Zw} an attempt was made to extend the work in Refs 
\cite{BL,Sagnotti,KS1,KS2} to the four dimensional ${\bf Z}_N\otimes {\bf Z}_M$ 
orbifold cases. However, a bothersome puzzle was encountered: in some of the models
the tadpole cancellation conditions (derived using the perturbative orientifold
approach, namely, via a straightforward generalization of the six dimensional 
tadpole cancellation conditions of Refs \cite{PS,GP,GJ,DP}) allowed for no solutions.
This, at least at the first sight, seems surprising as Type IIB compactifications on those
orbifolds are well defined, and so should be the corresponding orientifolds. This
clearly indicates that a better understanding of the orientifold construction is desirable.
This is precisely the subject to which this paper is devoted. 

{}We consider six and 
four dimensional ${\cal N}=1$ supersymmetric orientifolds of Type IIB 
compactified on orbifold
limits of K3 and Calabi-Yau three-folds, respectively.
We study conditions necessary for world-sheet
consistency of Type IIB orientifolds, that is, the conditions under which perturbative
orientifold approach is adequate. We argue that in most cases orientifolds contain sectors
which are non-perturbative ({\em i.e.}, these sectors have no world-sheet description). These
sectors can be thought of as arising from D-branes wrapping various collapsed 2-cycles
in the orbifold. In particular, we argue that such non-perturbative states are present
in the ``anomalous'' models of Ref \cite{Zw} (as well as in other examples of this type
recently discussed in Ref \cite{Iba}). This resolves the corresponding ``puzzles''.
Moreover, we point out certain world-sheet consistency conditions in four dimensional cases
(which are automatically satisfied in the six dimensional cases studied in Refs 
\cite{PS,GP,GJ,DP} so their relevance cannot be appreciated in those constructions)
which indicate that the only four dimensional orientifolds that have perturbative description
are those of Type IIB compactified on the ${\bf Z}_2\otimes {\bf Z}_2$ \cite{BL}, ${\bf Z}_3$
\cite{Sagnotti}, ${\bf Z}_7$ \cite{KS1}, ${\bf Z}_3\otimes {\bf Z}_3$ and ${\bf Z}_6$
\cite{KS2}, and ${\bf Z}_2\otimes{\bf Z}_2 \otimes {\bf Z}_3$ \cite{zk} orbifolds. In particular,
none of the other models considered in Refs \cite{Zw,Iba} have perturbative orientifold
description, and even in the models with all tadpoles cancelled the massless spectra given
in Refs \cite{Zw,Iba} miss certain non-perturbative states.   

{}The main tool in our analyses is the interplay between different string
theories via the web of dualities. The relations between Type IIB 
orientifolds, Type I, heterotic and F-theory are schematically 
depicted in Fig.1. Our goal in this paper is to understand
Type I compactifications and Type IIB orientifolds, and, in particular,
the relation between them (which is link ``b'' in Fig. 1).
In most cases, none of the above descriptions are
completely perturbative. Nonetheless, by combining 
various approaches together, we are able to get much of the qualitative 
as well as some quantitative properties of Type I compactifications and 
Type IIB orientifolds. On the other hand, by studying orientifolds in various
dimensions, one can obtain non-trivial information about F-theory and
non-perturbative heterotic string vacua. In the following we summarize
some of the important points in this approach.\\
$\bullet$ Type I-heterotic duality \cite{PW} (which is link
``c'' in Fig. 1) is crucial in checking the consistency of the models that do
have perturbative heterotic duals. (To be precise, these are orientifolds which
only contain D9-branes but no D5-branes. The ${\bf Z}_3$ \cite{Sagnotti}, 
${\bf Z}_7$ \cite{KS1} and ${\bf Z}_3 \otimes {\bf Z}_3$ \cite{KS2}
cases are examples of such orientifolds.) 
These checks are largely based on
the observations of Ref \cite{ZK} (as well as Refs \cite{KS1,KS2}).
Moreover, we are able to determine the
non-perturbative states that appear in the orientifold approach by
studying the perturbative spectrum of the heterotic dual.
This will be discussed in section \ref{het}.\\
$\bullet$ Having established the map between some orientifolds and 
their perturbative heterotic duals, one can use orientifold construction 
(with both D9- and D5-branes) as 
a tool to understand non-perturbative heterotic string vacua.
This will be discussed in section \ref{non-pert}.\\
$\bullet$ The map \cite{sen} between F-theory \cite{vafa}
and orientifolds (which is link ``a'' in Fig. 1) is an invaluable tool 
for understanding the qualitative features of the non-perturbative
states in Type IIB orientifolds (even in cases where perturbative
heterotic duals do not exist). In particular, one can identify
the non-perturbative states 
in the orientifold approach as arising from D-branes wrapping various 
collapsed two cycles in the orbifold. This will be discussed in
section \ref{F} and section \ref{FA}.\\
$\bullet$ By studying various orientifolds in six (and four)
dimensions,
one can obtain certain non-trivial information about Calabi-Yau three-fold 
(and four-fold) geometry along the lines of Refs \cite{gj,BG}. In 
section \ref{FA}, we will show that
the six-dimensional ${\bf Z}_N$ orientifolds ($N=2,3,4,6$) \cite{PS,GP,GJ,DP}
are equivalent to F-theory 
compactifications on certain elliptically fibered Calabi-Yau three-folds,
which can be regarded as extended Voisin-Borcea orbifolds 
\cite{Voisin,Borcea} (see Fig.2).
Similarly, the four-dimensional 
${\bf Z}_2 \otimes {\bf Z}_2$ orientifold \cite{BL}
is dual to F-theory compactification on a Borcea four-fold \cite{Borcea}.\\
$\bullet$ Finally, the duality between F-theory and heterotic
vacua (which is link ``d'' in Fig. 1) turns out to be useful in 
understanding certain 
aspects of Type I compactifications on K3. This will be discussed in
details in section \ref{FA}.

{}The remainder of this paper is organized as follows. 
In section \ref{prelim} we review some facts in conformal field theory of orbifolds and set
up our notations. In section \ref{IIB} we derive world-sheet consistency conditions 
for orientifolds of Type IIB on non-geometric conformal field theory orbifolds. In section \ref{class} we classify six and four dimensional orientifolds that satisfy this constraint. 
In section \ref{F} we give F-theory 
interpretation of the consistency condition derived in section \ref{IIB}.
In section \ref{IIBA} we extend these analyses to orientifolds of Type IIB on geometric 
conformal field theory orbifolds. In section \ref{other} we discuss six dimensional
orientifolds of Refs \cite{GJ,DP} and their possible generalizations to four dimensions.
In particular, we point out that there are two distinct choices for the orientifold
projection in six dimensions, whereas in four dimensions there is only one such choice.
This is basically the reason why there are subtleties in attempting to generalize
the tadpole cancellation conditions of Refs \cite{GJ,DP} to four dimensions.
In section \ref{FA} we give various F-theory checks for our arguments
in section \ref{other}. We also discuss F-theory duals of six and four dimensional 
orientifolds. In section \ref{het} we review the four dimensional Type I-heterotic
duality map studied in Ref \cite{ZK}. In section \ref{non-pert} we demonstrate how to
use this map to construct consistent four dimensional chiral ${\cal N}=1$ Type I vacua
which are non-perturbative from the heterotic viewpoint. In section \ref{anom}
we explain the ``puzzles'' encountered in the literature (in particular, in Refs \cite{Zw,Iba})
on four dimensional orientifolds and point out which of these have perturbative description. 
In section \ref{disc} we summarize the main conclusions of this paper. We also point out 
some directions for future research. Some of the details are relegated to appendices. As an
aside, in appendix \ref{CHL} we construct F-theory duals of six dimensional CHL
compactifications. Although various sections are interrelated, most of them 
are rather self-contained and can be read separately.

\section{Preliminaries}\label{prelim}

{}In this section we review some well-known facts in conformal field theory of 
orbifolds. This will serve the purpose of setting up our notations and 
conventions, as well as emphasizing certain points which will be important in 
the subsequent sections.

{}Consider a free closed string propagating in space-time. Its world-sheet is
a cylinder parametrized by a time-like coordinate $\sigma^0$ and a space-like 
coordinate $\sigma^1$. Let the circumference of the string be $2\pi$. Then we 
have the identification $\sigma^1=\sigma^1+2\pi$. Due to this identification one must 
specify periodicity conditions under $\sigma^1\rightarrow\sigma^1+2\pi$
for all the fields on the world-sheet. 

{}Instead of working with
$\sigma^0$ and $\sigma^1$, it is convenient to introduce the holomorphic and 
anti-holomorphic coordinates $z\equiv\exp(i(\sigma^0+\sigma^1))$ and
${\overline z}\equiv\exp(i(\sigma^0-\sigma^1))$, respectively. Then the left- and 
right-moving fields on the world-sheet depend only on $z$ and ${\overline z}$,
respectively.   

\subsection{Twist Fields}

{}Let $\phi_v(z)$ be a single free left-moving complex world-sheet 
boson with the monodromy
\begin{equation}
 \partial\phi_v(z{\mbox{e}}^{2\pi i})=\exp(-2\pi i v)\partial\phi_v(z)~,
\end{equation} 
where $0<v<1$. This monodromy implies 
that a twist field $\sigma_v (z)$ is located at the origin such that
\begin{eqnarray}
 && i\partial\phi_v(z)\sigma_v(0)\sim z^{-v} \tau_v(0)+\cdots~,\\
 && i\partial\phi_v^\dagger (z)\sigma_v(0)\sim z^{v-1} \tau_v^\prime(0)+\cdots~,
\end{eqnarray}
where $\phi_v^\dagger$ is the Hermitean conjugate of $\phi_v$, and 
$\tau_v,\tau_v^\prime$ are the excited twist fields. The basic twist fields $\sigma_v$
has conformal dimension $v(1-v)/2$.

{}Next, consider a single free right-moving complex world-sheet 
boson ${\overline \phi}_u({\overline z})$ with the monodromy
\begin{equation}
 {\overline \partial}{\overline \phi}_u({\overline z}
 {\mbox{e}}^{-2\pi i})=\exp(+2\pi iu){\overline \partial}{\overline \phi}_u
 ({\overline z})~,
\end{equation} 
where $0<u<1$. This monodromy implies 
that a twist field ${\overline \sigma}_u ({\overline z})$ is located at the 
origin such that
\begin{eqnarray}
 && i{\overline \partial}{\overline \phi}_u({\overline z})
 {\overline \sigma}_u(0)\sim {\overline z}^{-u} {\overline \tau}_u(0)+\cdots~,\\
 && i{\overline \partial}{\overline \phi}^\dagger_u({\overline z})
 {\overline \sigma}_u(0)\sim {\overline z}^{u-1} 
 {\overline \tau}^\prime_u(0)+\cdots~,
\end{eqnarray}
where ${\overline \phi}_u^\dagger$ is the Hermitean conjugate of ${\overline 
\phi}_u$,
and ${\overline \tau}_u,{\overline \tau}_u^\prime$ are the excited twist fields. 
The basic twist fields ${\overline\sigma}_u$ has conformal dimension $u(1-u)/2$.

{}The twist fields $\sigma_v$ and ${\overline \sigma}_v$ are identical. (By this 
we mean that $\sigma_v (x)={\overline \sigma}_v (x)$, where $x$ is an arbitrary
complex number.) The twist fields $\sigma_v$ and ${\overline \sigma}_{1-v}$,
on the other hand, are different except for $v=1/2$.

{}There are two inequivalent ways of combining the above left- and right-moving
fields into a world-sheet boson.\\
$\bullet$ ({\em i}) Let $\phi_v(\sigma^0,\sigma^1)=\phi_v(z)+
{\overline\phi}_v^\dagger ({\overline z})$.
This field has the following periodicity condition: $\phi_v(\sigma^0,\sigma^1+
2\pi)=\exp(-2\pi i v)\phi_v(\sigma^0,\sigma^1)$. The twisted ground state is given 
by $\sigma_v \vert 0\rangle_L\otimes {\overline \sigma}_v \vert 0\rangle_R$, 
where $\vert 0\rangle_L$ and $\vert 0\rangle_R$ are the left- and right-moving 
conformal ground states, respectively. Note that the twisted ground state in this 
case is left-right {\em symmetric}.\\
$\bullet$ ({\em ii}) Let ${\widetilde\phi}_v(\sigma^0,\sigma^1)=\phi_v(z)+
{\overline \phi}_{1-v} ({\overline z})$.
This field has the same periodicity condition as the field 
$\phi_v(\sigma^0,\sigma^1)$: $\phi_v(\sigma^0,\sigma^1+
2\pi)=\exp(-2\pi i v) \phi_v(\sigma^0,\sigma^1)$. However, the twisted ground 
state is now given 
by $\sigma_v \vert 0\rangle_L\otimes {\overline \sigma}_{1-v} \vert 0\rangle_R$. 
Note that the twisted ground state in this 
case is left-right {\em asymmetric} unless $v=1/2$.

{}Here we note that in case ({\em i}) the complexification for the
left- and right-movers is opposite. That is, $\phi_v (z)=\phi^1 (z) +i\phi^2 (z)$,
while ${\overline \phi}_v ({\overline z})={\overline \phi}^1 ({\overline z}) -
i{\overline \phi}^2 ({\overline z})$, where $\phi^1 (z),\phi^2 (z)$ are left-moving
real world-sheet bosons, and ${\overline \phi}^1 ({\overline z}),
{\overline \phi}^2 ({\overline z})$ are their right-moving counterparts. 
On the other hand, in case ({\em ii}) the complexification for the left- and 
right-movers is the same: $\phi_v (z)=\phi^1 (z) +i\phi^2 (z)$,
while ${\overline \phi}_{1-v} ({\overline z})={\overline \phi}^1 ({\overline z}) +i
{\overline \phi}^2 ({\overline z})$.

\subsection{``Symmetric'' {\em vs.} ``Asymmetric'' Orbifolds}

{}So far we have considered a single complex world-sheet boson. Now let us
discuss toroidal orbifolds which lead to Calabi-Yau $d$-folds ($d=2,3$). First 
consider the following orbifold: ${\cal M}_d=T^{2d}/G$, where $G=\{g_a\vert a=1,\dots,{\mbox{dim}}(G)\}$ is the orbifold group. Let the twisted ground states
in all of the $g_a$ twisted sectors be left-right symmetric as in case ({\em i}) above.
We will refer to such orbifolds as ``symmetric'' orbifolds. Next, let us consider the
following orbifold: ${\widetilde {\cal M}}_d=T^{2d}/{\widetilde G}$, where 
${\widetilde G}=\{{\widetilde g}_a\vert a=1,\dots,{\mbox{dim}}({\widetilde G})\}$ 
is the orbifold group. Let the twisted ground states
in all of the ${\widetilde g}_a$ twisted sectors be left-right asymmetric as 
in case ({\em ii}) above. (The ${\bf Z}_2$ twisted sectors, however, are 
automatically left-right symmetric.) We will refer to such orbifolds as 
``asymmetric'' orbifolds.

{}Throughout this paper we will assume that ${\cal M}_d$ 
(${\widetilde {\cal M}}_d$) are orbifold ``limits'' of Calabi-Yau $d$-folds
with $SU(d)$ holonomy. Let $z_s$, $s=1,\dots,d$, be complex coordinates 
parametrizing 
$T^{2d}$. The Calabi-Yau condition implies that $G$ (${\widetilde G}$)
must preserve the 
holomorphic $d$-form $dz_1\wedge\dots\wedge dz_d$ on ${\cal M}_d$ 
(${\widetilde {\cal M}}_d$), so that 
$g_a$ (${\widetilde g}_a$) must act as $d\times d$ matrices on $dz_s$ such that 
$\det(g_a)=1$ ($\det({\widetilde g}_a)=1$).

{}Here we note that the ``asymmetric'' orbifolds ${\widetilde {\cal M}}_d$ are the
``geometric'' orbifolds. That is, they correspond to conformal field theory realizations
of geometric quotients of the form $T^{2d}/{\widetilde G}$. On the other hand, the
``symmetric'' orbifolds ${\cal M}_d$ do not have an analogous geometric 
interpretation. They are conformal field theory constructions, and when referred to 
as ${\cal M}_d=T^{2d}/G$ orbifolds they should not be literally understood as 
geometric quotients. Rather, one has to bear in mind the action of the twists $g_a$
on left- and right-moving components of the conformal fields $z_s$. 

{}The relation between the ``symmetric'' ${\cal M}_d$ and ``asymmetric'' 
${\widetilde {\cal M}}_d$ orbifolds is that they are ``mirror pairs''. Thus, for
$d=2$ they give rise to K3 surfaces (for $G\approx {\widetilde G}\approx
{\bf Z}_N$, $N=2,3,4,6$) ${\cal M}_2$ and ${\widetilde {\cal M}}_2$ which 
are related by a mirror transform of K3. For $d=3$ they give rise to mirror 
Calabi-Yau three-folds with the Hodge numbers interchanged: 
$(h^{1,1},h^{2,1})=({\widetilde h}^{2,1},{\widetilde h}^{1,1})$. As an example 
consider the $Z$-orbifold generated by the following twist: $gz_s=\omega z_s$ 
($s=1,2,3$), where $\omega=\exp(2\pi i/3)$. The ``asymmetric'' $Z$-orbifold has 
the Hodge numbers $({\widetilde h}^{1,1},{\widetilde h}^{2,1})=(36,0)$, which 
are the same as for the familiar geometric $Z$-orbifold. The ``symmetric'' 
$Z$-orbifold has the Hodge numbers $(h^{1,1},h^{2,1})=(0,36)$, which are those
of the  manifold mirror to the geometric $Z$-orbifold.  

{}Here we should point out that the terminology ``symmetric'' and 
``asymmetric'' orbifolds, which we are using here, is non-standard. In particular, 
the standard orbifolds in Refs \cite{DHVW} are the geometric, that is, ``asymmetric'' 
orbifolds in our terminology. We will always use quotation marks when referring to
``symmetric'' and ``asymmetric'' orbifolds (as well as ``symmetric'' and ``asymmetric'' 
orientifolds - see below) as a reminder to avoid confusion.

\subsection{Torus}

{}For our purposes in the subsequent sections it will suffice to examine the 
untwisted sector 
contributions of the bosonic world-sheet degrees of freedom $z_s$ into the closed 
string one-loop vacuum amplitude. 

{}First, consider the one-loop vacuum amplitude for Type IIB compactified on 
${\cal M}_d$. The closed string world-sheet is a compact
Riemann surface of genus one, {\em i.e.}, a two-torus. The complex structure of this
two-torus is described by one complex parameter $\tau=\tau_1+i\tau_2$. (The 
one-loop vacuum amplitude is independent of the K{\"a}hler structure of this 
two-torus as a consequence of conformal invariance.) The untwisted sector 
contributions
of the bosonic world-sheet degrees of freedom $z_s$ into the torus amplitude
are given by (here we drop all the fermionic world-sheet degrees of freedom 
as well as the bosonic world-sheet degrees of freedom corresponding to 
non-compact coordinates, and the light-cone gauge is adapted throughout):
\begin{equation}\label{torus}
 {\cal T}={1\over{\mbox{dim}}(G)} \sum_{a=1}^{{\mbox{\small{dim}}}(G)} 
 {\cal T}_a= {1\over {\mbox{dim}}(G)}
 \sum_{a=1}^{{\mbox{\small{dim}}}(G)}
 {\mbox{Tr}}\left(g_a q^{L_0} {\overline q}^{{\overline L}_0}\right)~.
\end{equation}  
Here $q\equiv\exp(2\pi i\tau)$; $L_0$ and ${\overline L}_0$ are the left- and 
right-moving Hamiltonians, respectively;  
the trace is over the untwisted sector states 
corresponding to $z_s$ (oscillator excitations as well as 
momenta and windings).

{}Here we are considering left-right {\em symmetric}
orbifolds ${\cal M}_d$.
Then the operator $g_a$ (as it appears in Eq (\ref{torus}))
is given by:
\begin{equation}\label{ga}
 g_a=\prod_{s=1}^d
 \exp\left({2\pi i\phi_{as}} [M_{sL}-M_{sR}]\right)~.
\end{equation}
Here we are writing each $g_a$ in its own diagonal basis. The phases
$\exp({2\pi i\phi_{as}})$ are eigenvalues of $g_a$ (that is, in the diagonal basis
$g_a={\mbox{diag}}(\exp({2\pi i\phi_{a1}}),\dots,\exp({2\pi i\phi_{ad}}))$ with
$\prod_{s=1}^d \exp({2\pi i\phi_{as}})=1$, which follows from the condition 
$\det(g_a)=1$). The operators $M_{sL}$ and $M_{sR}$ are the left- and 
right-moving generators of
infinitesimal rotations in the $z_s$ plane. 
The important point here is the {\em Lorentzian} signature for the
left- and right-moving contributions ({\em i.e.}, the minus sign in front of 
$M_{sR}$ in Eq (\ref{ga})). This is required by modular invariance of the 
complete torus amplitude which also includes twisted sector states.
The fact that in Eq (\ref{ga}) we must have $M_{sL}-M_{sR}$ (and not 
$M_{sL}+M_{sR}$) can also be seen as follows: since the orbifold 
is left-right symmetric, all the left-right symmetric states must 
be invariant under the action of $g_a$, {\em i.e.}, the corresponding
operator $g_a=1$ on left-right symmetric states, hence Eq (\ref{ga}). Appendix
\ref{Orbifold} provides more detail concerning this point.  

{}The torus amplitude for Type IIB compactified on ${\widetilde {\cal M}}_d$ is 
given by Eq (\ref{torus}) with $g_a$ replaced by ${\widetilde g}_a$. In its 
diagonal basis the operator ${\widetilde g}_a$ is given by
\begin{equation}\label{ga1}
 {\widetilde g}_a=\prod_{s=1}^d
 \exp\left({2\pi i\phi_{as}} [M_{sL}+M_{sR}]\right)~.
\end{equation}  
Note the {\em Euclidean} signature for the left- and right-moving contributions.

\subsection{World-Sheet Parity}

{}Consider Type IIB compactification on ${\cal M}_d$ (${\widetilde 
{\cal M}}_d$).  
Let us confine our attention to Type IIB compactifications with zero NS-NS 
antisymmetric 
tensor $B_{ij}$ ($i=1,\dots,2d$)
backgrounds. The physical spectrum of Type IIB string theory 
compactified on ${\cal M}_d$ (${\widetilde 
{\cal M}}_d$) with $B_{ij}=0$ is left-right symmetric.
Thus, we can attempt to gauge the world-sheet 
parity symmetry generated by $\Omega$ that interchanges left- 
and right-movers.

{}Instead of gauging $\Omega$ we can consider a more general class of 
orientifolds corresponding to gauging $\Omega J I^{F_L}$, 
where: $J$ is a symmetry 
of ${\cal M}_d$ (${\widetilde 
{\cal M}}_d$) such that $J^2=1$; $J$ acts left-right symmetrically on 
${\cal M}_d$ (${\widetilde 
{\cal M}}_d$); $I\equiv\det(J)$ (see below); $F_L$ is the operator
that flips the sign of the left-moving Ramond (R) sector states
but leaves the right-moving Ramond sector states and all the Neveu-Schwarz 
(NS) sector states unaffected. Then for $\Omega J I^{F_L}$ orientifolds of
Type IIB on ${\cal M}_d$ we have the following orientifold group: 
${\cal O}=\{g_a,\Omega J_a I^{F_L}\vert a=1,\dots, {\mbox{dim}}(G)\}$,
where $J_a\equiv Jg_a$. The orientifold group for $\Omega J I^{F_L}$ 
orientifolds of Type IIB on ${\widetilde {\cal M}}_d$ is defined similarly. 

{}It is important to understand what are the allowed choices of $J$. 
Type IIB compactification on ${\cal M}_d$ (${\widetilde 
{\cal M}}_d$) results in a $10-2d$ dimensional 
theory with 
${\cal N}=2$ space-time supersymmetry. After orientifolding we should have 
${\cal N}=1$ space-time supersymmetry. This implies that $J$ must preserve 
complex structure on ${\cal M}_d$ (${\widetilde 
{\cal M}}_d$), so that $J$ must act as a $d\times d$ 
matrix on $dz_s$. That is, $J$ must act on $dz_s$ as an $SU(d)\otimes{\bf Z}_2$
matrix (such that $J^2=1$).

{}Before we end this section let us make two comments.\\
$\bullet$ For ``symmetric'' orbifolds ${\cal M}_d$ the  
twisted ground states are left-right 
symmetric. Thus, the world-sheet parity operator $\Omega$ in this case 
is defined to interchange left- and right-moving oscillators and momenta.
However, it does not affect the twisted ground states.\\ 
$\bullet$ On the other hand, for ``asymmetric'' orbifolds ${\widetilde 
{\cal M}}_d$ the twisted ground states are left-right asymmetric (except for
${\bf Z}_2$ twisted sectors). Thus, 
the world-sheet parity operator $\Omega$ (defined as in the case of 
``symmetric'' orbifolds ${\cal M}_d$) is not a symmetry of the theory in
this case. At least naively, therefore, it must always be accompanied by an 
operator that interchanges the left- and right-moving ground states. We will discuss 
this issue in detail in sections \ref{IIBA}, \ref{other} and \ref{FA}.

\section{``Symmetric'' Type IIB Orientifolds}\label{IIB}

{}In this section we consider ``symmetric'' Type IIB orientifolds, {\em i.e.}, 
orientifolds of Type IIB compactified on ``symmetric'' orbifolds ${\cal M}_d$.
Here we derive a condition necessary for consistent 
world-sheet description of ``symmetric'' Type IIB orientifolds to exist. 

\subsection{Klein Bottle} 

{}Next, consider the one-loop vacuum amplitude for the $\Omega J I^{F_L}$ 
orientifold of Type IIB (compactified on ${\cal M}_d$). We are still interested
only in the closed untwisted sector contributions
of the bosonic world-sheet degrees of freedom $z_s$. 
For the sake of simplicity we will 
assume that $J$ and $g_a$ act homogeneously on $z_s$, {\em i.e.}, 
without shifts. It is not difficult to see that the following argument can be 
repeated even if $J$ and $g_a$ act inhomogeneously on $z_s$. 
The conclusions, however, do not depend on whether $J$ and $g_a$ 
include shifts (since the argument intrinsically depends only on how 
$J$ and $g_a$ act on $dz_s$). 
Since we are not
looking at world-sheet fermions, the $I^{F_L}$ factor in $\Omega J I^{F_L}$
will be irrelevant in the following. (Also, $\Omega \vert\Psi_L,\Psi_R\rangle=\pm\vert\Psi_R,\Psi_L\rangle$
with the positive and negative signs corresponding to 
the NS-NS and R-R sectors, respectively. These signs will be of no 
relevance in the following discussion either.)
The corresponding one-loop vacuum 
amplitude for the orientifold theory reads ${\cal T}/2+{\cal K}$, where 
${\cal K}$ is the Klein bottle contribution:   
\begin{equation}\label{Klein}
 {\cal K}={1\over 2{\mbox{dim}}(G)} \sum_{a=1}^{{\mbox{\small{dim}}}(G)} 
 {\cal K}_a= {1\over 2{\mbox{dim}}(G)}
 \sum_{a=1}^{{\mbox{\small{dim}}}(G)}
 {\mbox{Tr}}\left(\Omega J_a q^{L_0} {\overline q}^{{\overline L}_0}\right)~.
\end{equation}

{}Let us first consider the oscillator contributions. (Note that oscillator 
contributions and momentum plus winding contributions factorize.)
The presence of the $\Omega$ projection in the Klein bottle amplitude
implies that only left-right symmetric states contribute. The discussion in
section \ref{prelim} (see Eq (\ref{ga})) implies that left-right symmetric oscillator
excitations
do not contribute any non-trivial phase into $J_a$. That is, (in the diagonal 
basis for $J_a$) $J_a\vert\Psi_L,\Psi_R\rangle=\vert\Psi_L,\Psi_R\rangle$ 
for a left-right symmetric state with $\Psi_L=\Psi_R$. Thus, the oscillator 
contributions to ${\cal K}_a$ are given by $1/\eta^{2d}(q{\overline q})$, and are 
independent of $a$.

{}Next, consider the momentum and winding contributions. For Type IIB 
on $T^{2d}$ the left- and right-moving momenta $(p_L,p_R)$ corresponding to 
$z_s$ span an even self-dual Lorentzian lattice $\Gamma^{2d,2d}$.
Here we are considering Type IIB compactifications with zero NS-NS 
antisymmetric tensor $B_{ij}$ backgrounds. 
We can therefore write the left- and right-moving momenta in 
$\Gamma^{2d,2d}$ as
\begin{eqnarray}
 p_{L,R}={1\over 2} m_i{\tilde e}^i \pm n^i e_i\equiv p\pm w~.
\end{eqnarray}
Here $m_i$ and $n^i$ are integers; $e_i$ are constant vielbeins;
$e_i\cdot e_j=G_{ij}$ is the constant 
background metric on $T^{2d}$; $e_i \cdot {\tilde e}^j={\delta_i}^j$. 
Note that
the windings $w\in\Lambda$, and the momenta 
$p\in{\widetilde\Lambda}/2$, where $\Lambda$ is the lattice spanned by the
vectors $e_i n^i$ ($n^i\in {\bf Z}$), and ${\widetilde\Lambda}$ is the 
lattice dual 
to $\Lambda$. Instead of describing the momentum states as 
$\vert p_L,p_R\rangle$, we can use the $\vert p,w\rangle$ basis. 
This is convenient as the action of $g_a\in G$ 
on $\vert p,w\rangle$ is simply given by $g_a\vert p,w\rangle=
\vert g_ap,g_aw\rangle$. The action of $\Omega$ on $p_L$ and $p_R$ 
reads: $\Omega p_L=p_R$, $\Omega p_R=p_L$. 
This implies $\Omega p=p$, $\Omega w=-w$.

{}The momentum and winding contributions to ${\cal K}_a$ are given by
(our normalization convention is  
$\langle p,w\vert p^\prime,w^\prime\rangle=\delta_{pp^\prime} 
\delta_{ww^\prime}$):
\begin{eqnarray}
 &&\sum_{p\in{1\over 2}{\widetilde \Lambda},w\in\Lambda}
 q^{{1\over 2}(p+w)^2}
 {\overline q}^{{1\over 2}(p-w)^2}\langle p,w\vert\Omega J_a\vert
 p,w\rangle=\nonumber\\
 \label{MW}
 &&\sum_{p\in{1\over 2}{\widetilde \Lambda}(J_a)}
 (q{\overline q})^{{1\over 2}p^2}
 \sum_{w\in\Lambda(RJ_a)}
 (q{\overline q})^{{1\over 2}w^2}~.
\end{eqnarray}    
Here ${\widetilde \Lambda}(J_a)\subset{\widetilde \Lambda}$ is the lattice dual to $\Lambda(J_a)\subset\Lambda$, where $\Lambda(J_a)$ is the sublattice 
of $\Lambda$ invariant under the action of $J_a$.  
Similarly, $\Lambda(RJ_a)\subset\Lambda$ is the sublattice of $\Lambda$ 
invariant under the action of $RJ_a$. (Its dual lattice will be denoted 
by ${\widetilde \Lambda}(RJ_a)\subset{\widetilde \Lambda}$.) The 
appearance of $R$, 
which acts as $Rz_s=-z_s$, in $\Lambda(RJ_a)$ is due to the
non-trivial action of $\Omega$ on windings.  

{}Combining the oscillator contributions with those of momenta and windings, 
we have the following expression for ${\cal K}_a$:
\begin{equation}
 {\cal K}_a={1\over\eta^{2d}({\mbox{e}}^{-2\pi t})}
 \sum_{p\in{1\over 2}{\widetilde \Lambda}(J_a)}
 \exp(-\pi t p^2)
 \sum_{w\in\Lambda(RJ_a)}
 \exp(-\pi t w^2)~.
\end{equation} 
Here we have introduced $t\equiv2\tau_2$.

\subsection{Cylinder with Two Cross-Caps}

{}Under the modular transformation $t\rightarrow1/t$ the Klein bottle turns into a 
cylinder with two cross-caps as its boundaries. The Klein bottle amplitude is a one-loop
unoriented closed string amplitude. The cylinder with two cross-caps corresponds 
to a tree-level amplitude for closed strings propagating between the boundary states
describing the cross-caps. These boundary states cannot be arbitrary but must 
correspond to coherent closed string states (built from a superposition of an infinite 
tower of closed string oscillator and momentum plus winding states) \cite{nappi,PC} 
(also see, {\em e.g.}, Refs \cite{IO,Dunbar}). The consistency therefore requires
the Klein bottle ({\em i.e.}, loop-channel amplitude) upon $t\rightarrow1/t$ transformation
agree with the cylinder with 
two cross-caps ({\em i.e.}, tree-channel amplitude). This  constraint is often 
referred to as (loop-tree) factorization condition.

{}The cross-cap boundary states describe the familiar orientifold planes. (Similarly,
other boundary states describe D-branes). The orientifold planes arise due to the
action of the orientifold group elements $\Omega J_a$. We will refer to
the corresponding cross-cap boundary states as $\vert C_a\rangle$. 

{}The most general expression for the tree-channel amplitude corresponding to
the cylinder with two cross-caps, call it ${\widetilde {\cal K}}$, has the following 
form:
\begin{equation}\label{tree}
 {\widetilde {\cal K}}=
 \sum_{a,b}\sum_{\vert s\rangle} D_{ab} \langle C_a\vert
 \exp\left(-\pi t(L_0+{\overline L}_0)\right)\vert s \rangle\langle s\vert
 C_b\rangle~.
\end{equation}
Here the sum runs over all the (untwisted) closed string states 
$\vert s\rangle$. The matrix
$D_{ab}$ must be Hermitean for ${\widetilde {\cal K}}$ must be real. 
Moreover, neither 
$D_{ab}$ nor $\vert C_a\rangle$ can depend upon the ``proper time'' $t$. 

{}To see what are the cross-cap boundary states $\vert C_a\rangle$ 
we must perform the modular 
transformation $t\rightarrow1/t$ on the Klein bottle amplitude ${\cal K}$. Let 
${\widetilde {\cal K}}=(1/2{\mbox{dim}}(G))\sum_a {\widetilde {\cal K}}_a$ be 
the resulting 
tree-channel amplitude. The contributions ${\widetilde {\cal K}}_a$ are obtained
from ${\cal K}_a$ via $t\rightarrow1/t$:  
\begin{equation}\label{cross}
 {\widetilde {\cal K}}_a={(\sqrt{t})^{-2d}\over\eta^{2d}({\mbox{e}}^{-2\pi t})}
 \left((2\sqrt{t})^{d(J_a)} V(J_a)\right)
 \sum_{{\widetilde p}\in 2\Lambda(J_a)}
 \exp(-\pi t {\widetilde p}^2)
 \left({(\sqrt{t})^{d(RJ_a)}\over V(RJ_a)}\right)
 \sum_{{\widetilde w}\in{\widetilde\Lambda}(RJ_a)}
 \exp(-\pi t {\widetilde w}^2)~.
\end{equation}
Here $d(J_a)$ and $d(RJ_a)$ are the numbers of dimensions
of the lattices $\Lambda(J_a)$ and $\Lambda(RJ_a)$, whereas $V(J_a)$
and $V(RJ_a)$ are the volumes of their unit cells.

{}The important point about Eq (\ref{cross}) is presence of extra
factors of $\sqrt{t}$. They cancel if and only if $d(J_a)+d(RJ_a)=2d$ for all $a$. 
Suppose this 
condition is not satisfied. Then it is impossible to rewrite Eq (\ref{cross}) in the
form of Eq (\ref{tree}). We therefore conclude that the orientifold consistency
requires the following constraint be satisfied:
\begin{equation}\label{const0}
 \forall a~d(J_a)+d(RJ_a)=2d~. 
\end{equation}
Subject to this condition, Eq (\ref{cross}) can be rewritten in the
form of Eq (\ref{tree}) with $D_{ab}=(1/2{\mbox{dim}}(G))\delta_{ab}$, and
\begin{equation}
 \vert C_a\rangle=
 \left[{2^{d(J_a)}V(J_a)\over{V(RJ_a)}}\right]^{1\over 2}
 \sum_{{\widetilde p}\in 2\Lambda(J_a)}
 \sum_{{\widetilde w}\in{\widetilde\Lambda}(RJ_a)}
 \sum_{\bf n} \zeta_{\bf n} V_L({\bf n}) V_R({\bf n}) \vert 
 {\widetilde w},{\widetilde p}\rangle~. 
\end{equation}
Here $V_L({\bf n})$ and $V_R({\bf n})$ are strings of the left- and right-moving
(untwisted sector) oscillator creation operators. These strings are labeled
by the occupation number vector ${\bf n}$ (which is infinite dimensional). 
Note that both $V_L({\bf n})$ and 
$V_R({\bf n})$ are labeled by the same occupation number vector ${\bf n}$,
so that the corresponding oscillator states are left-right symmetric.
$\vert {\widetilde w},{\widetilde p}\rangle$ denotes a state of momentum 
${\widetilde w}$ and winding ${\widetilde p}$. The coefficients $\zeta_{\bf n}$
are pure phases ($\vert \zeta_{\bf n}\vert=1$) whose precise values
are not relevant here.

\subsection{World-Sheet Consistency Condition}

{}Next, we would like to rewrite the condition (\ref{const0}) in a more 
convenient form. Consider any given $J_a$ in its diagonal basis:
$J_a={\mbox{diag}}(\lambda_{a1},\dots,\lambda_{ad})$, where 
$\lambda_{as}$ are the eigenvalues of the matrix $J_a$ when acting on
$dz_s$ complex coordinates. Let $n_{\pm} (J_a)$ be the numbers of 
eigenvalues $\lambda_{as}=\pm1$, respectively. Then the dimension 
$d(J_a)$ of the lattice $\Lambda(J_a)$ is given by $d(J_a)=2n_+ (J_a)$,
which follows from the definition of $\Lambda(J_a)$ being the sublattice
of $\Lambda$ invariant under $J_a$. Similarly, the dimension 
$d(RJ_a)$ of the lattice $\Lambda(RJ_a)$ is given by $d(RJ_a)=2n_- (J_a)$.
This can be seen by noting that in the diagonal basis $RJ_a=
{\mbox{diag}}(-\lambda_{a1},\dots,-\lambda_{ad})$. Thus, $d(J_a)+d(RJ_a)=2d$ if 
and only if $n_+ (J_a)+n_- (J_a)=d$, {\em i.e.}, all the eigenvalues of $J_a$
are either $+1$ or $-1$. This implies that $J_a^2=1$. We can therefore
rewrite the condition (\ref{const0}) as follows:
\begin{equation}\label{wsc}
 \forall a~J^2_a=1~,~{\mbox{or, equivalently}},~Jg_a=g_a^{-1}J~.
\end{equation}  
This constraint is necessary for world-sheet consistency of the 
orientifold. In the next
section we will classify six and four dimensional orientifolds that satisfy Eq 
(\ref{wsc}).

\section{6D and 4D ``Symmetric'' Type IIB Orientifolds}\label{class}

{}In this section we classify six and four dimensional ``symmetric'' Type IIB
orientifolds that satisfy the world-sheet consistency constraint (\ref{wsc}) 
derived in section II.

\subsection{6D Orientifolds}

{}Consider Type IIB compactifications on orbifold limits of K3: 
${\cal M}_2=T^4/{\bf Z}_N$ ($N=2,3,4,6$). Let $z_1$ and 
$z_2$ be complex coordinates on ${\cal M}_2$. Then we 
can write the action of the orbifold group 
$G=\{g^k \vert k=0,1,\dots,N-1 \} \approx {\bf Z}_N$ 
as follows:
\begin{equation}\label{gact}
 gz_1=\omega z_1~, ~~~gz_2=\omega^{-1} z_2~,
\end{equation}     
where $\omega=\exp(2\pi i/N)$.   

{}The world-sheet consistency condition (\ref{wsc}) implies that
\begin{equation}\label{con1}
 Jg=g^{-1}J~.
\end{equation} 

{}Let us first consider the case $G\approx {\bf Z}_2$. Eq (\ref{con1}) then 
implies that $J$ and $g$ commute (since $g^2=1$ in this case). 
Consider the action of $J$ on $dz_1$ 
and $dz_2$. We will represent it as a $2\times 2$ matrix. Note that in these
notations $g=-{\bf 1}$, where ${\bf 1}$ is a $2\times 2$ identity matrix. 
There are only three inequivalent choices for $J$ that satisfy $J^2=1$ 
condition:\\
$\bullet$ $J={\bf 1}$;\\
$\bullet$ $J=-{\bf 1}$;\\
$\bullet$ $J={\vec \alpha}\cdot {\vec{\sigma}}$.\\
Here ${\vec{\sigma}}=({\sigma}_1,{\sigma}_2,{\sigma}_3)$; 
${\sigma}_1,{\sigma}_2,{\sigma}_3$ are the Pauli matrices; 
${\vec \alpha}$ is a unit 3-vector: ${\vec \alpha}^2=1$.
(When acting on $z_1$ and $z_2$ (instead of $dz_1$ and $dz_2$), $J$
can also include shifts. We will not list them here for brevity since
they are not difficult to classify.) The first two choices of $J$ given above 
lead to various 
orientifolds of Type IIB on the ${\bf Z}_2$ orbifold limit of K3 
\cite{PS,GP,dp,Pol}. We will discuss the models corresponding to the third 
choice\footnote{Here we note that ${\vec\alpha}$ must be such that the resulting $J$ 
is a symmetry of $T^4$.} in section \ref{FA}.

{}Next, let us consider the cases $G\approx{\bf Z}_N$, $N=3,4,6$. 
In these cases we must have non-trivial $J$ to satisfy Eq ({\ref{con1}). 
For simplicity we can assume that $T^4=T^2\otimes T^2$, and $z_1,z_2$ are
complex coordinates parametrizing these 2-tori.
Then it is not difficult to show that (in the basis where $g$ is defined as in 
Eq (\ref{gact})) the most general $J$ that satisfies Eq (\ref{con1}) 
is given by:
\begin{equation}\label{J}
 Jz_1=\eta z_2+b~,~~~Jz_2=\eta^{-1} z_1-\eta^{-1} b~.
\end{equation} 
Here $\eta=\pm\omega^m$, $m=0,1,\dots,N-1$. Note that $J$ interchanges the two 
$T^2$'s which therefore must be identical. The shift $b$ is fixed under the 
action of $g$ on $T^2$, {\em i.e.}, $(1-\omega)b\sim 0$ (where the 
identification is modulo a lattice shift on $T^2$). For a given $N$ all choices of 
$\eta$ and $b$
lead to the same orientifold (see section \ref{FA}), so we can take 
$\eta=1$ and $b=0$.
Then $Jz_1=z_2$, $Jz_2=z_1$. (If we write $J$ as a $2\times 2$ matrix, then
$J={\sigma}_1$.) We give the spectra of the resulting models in section 
\ref{FA}.  

\subsection{4D Orientifolds}

{}The discussion of the previous subsection can be readily generalized to 
orientifolds of Type IIB  compactifications on orbifold limits of Calabi-Yau
three-folds: ${\cal M}_3=T^6/G$, where $G=\{g_a\vert a=1,\dots,
{\mbox{dim}}(G)\}$
is the orbifold group. Here we assume that ${\cal M}_3$ has $SU(3)$ holonomy.
We can classify all possible orbifold groups $G$ compatible with this
requirement. For the following discussion it is going to be irrelevant
whether $J$ and $g_a$ act with or without shifts on $z_1,z_2,z_3$, so we will
confine our attention to the actions of $J$ and $g_a$ on $dz_1,dz_2,dz_3$.  
We will mainly concentrate on Abelian orbifolds and briefly consider some 
non-Abelian orbifolds at the end of this section.

{}For Abelian orbifolds the possible choices of $G$ can be divided in two 
categories: ({\em i}) 
$G\approx
{\bf Z}_N$; ({\em ii}) $G\approx{\bf Z}_N\otimes {\bf Z}_M
(\not\approx{\bf Z}_{NM})$.   

{}Next, we list all possible choices in each of these categories that are 
compatible with the $SU(3)$ holonomy condition. (For the orbifold 
${\cal M}_3=T^6/G$ to be consistent, the action of $G$ must be a symmetry of 
$T^6$. In particular, this requirement guarantees that the number of fixed 
points (or two-tori) $n_a$ in the 
$g_a$ twisted sector is a positive integer.)\\
$\bullet$ ({\em i}) $G\approx {\bf Z}_N$. Let $g$ be the generator of 
this ${\bf Z}_N$. Then 
we have the following choices for $g$ (where we write $g$ as a diagonal
$3\times 3$ matrix acting on $dz_1,dz_2,dz_3$):\\
${\bf Z}_3$: $g={\mbox{diag}}(\omega,\omega,\omega)$, 
$\omega=\exp(2\pi i/3)$;\\
${\bf Z}_7$: $g={\mbox{diag}}(\omega,\omega^2,\omega^4)$, 
$\omega=\exp(2\pi i/7)$;\\
${\bf Z}_4$: $g={\mbox{diag}}(\omega,\omega,\omega^2)$, 
$\omega=\exp(2\pi i/4)$;\\
${\bf Z}_6$: $g={\mbox{diag}}(\omega,\omega,\omega^4)$, 
$\omega=\exp(2\pi i/6)$;\\
${\bf Z}_6^\prime$: $g={\mbox{diag}}(\omega,\omega^2,\omega^3)$, 
$\omega=\exp(2\pi i/6)$;\\
${\bf Z}_8$: $g={\mbox{diag}}(\omega,\omega^2,\omega^5)$, 
$\omega=\exp(2\pi i/8)$;\\
${\bf Z}_8^\prime$: $g={\mbox{diag}}(\omega,\omega^3,\omega^4)$, 
$\omega=\exp(2\pi i/8)$;\\
${\bf Z}_{12}$: $g={\mbox{diag}}(\omega,\omega^4,\omega^7)$, 
$\omega=\exp(2\pi i/12)$;\\
${\bf Z}_{12}^\prime$: $g={\mbox{diag}}(\omega,\omega^5,\omega^6)$, 
$\omega=\exp(2\pi i/12)$.\\
$\bullet$ ({\em ii}) $G\approx {\widetilde {\bf Z}}_N\otimes {\widetilde {\bf Z}}_M
(\not\approx{\bf Z}_{NM})$. (For later convenience we put tilde sign on the 
${\widetilde {\bf Z}_N}$ and ${\widetilde {\bf Z}}_M$ subgroups to distinguish 
them from the $G\approx{\bf Z}_N$ cases considered in category ({\em i}).)
Let $g$ and $h$ be the generators of 
the ${\widetilde {\bf Z}_N}$ and ${\widetilde {\bf Z}}_M$ subgroups, respectively. 
Let us write $g$ and $h$ as a diagonal
$3\times 3$ matrices acting on $dz_1,dz_2,dz_3$:
\begin{equation}\label{gh}
 g={\mbox{diag}}(\omega,\omega^{-1},1)~,~~~
 h={\mbox{diag}}(1,\eta,\eta^{-1})~.
\end{equation}
Where $\omega=\exp(2\pi i/N)$ and $\eta=\exp(2\pi i/M)$.
Then we have the following choices for $N$ and $M$:\\
${\widetilde {\bf Z}}_2\otimes {\widetilde {\bf Z}}_2$;\\
${\widetilde {\bf Z}}_2\otimes {\widetilde {\bf Z}}_4 (\supset {\bf Z}_4)$;\\
${\widetilde {\bf Z}}_2\otimes {\widetilde {\bf Z}}_6 (\supset {\bf Z}_6^\prime)$;\\
${\widetilde {\bf Z}}_3\otimes {\widetilde {\bf Z}}_3 (\supset {\bf Z}_3)$;\\
${\widetilde {\bf Z}}_3\otimes {\widetilde {\bf Z}}_6 (\supset {\bf Z}_3, {\bf Z}_6,
{\bf Z}_6^\prime)$;\\
${\widetilde {\bf Z}}_6\otimes {\widetilde {\bf Z}}_6 (\supset {\bf Z}_3, {\bf Z}_6,
{\bf Z}_6^\prime)$;\\
${\widetilde {\bf Z}}_4\otimes {\widetilde {\bf Z}}_4 (\supset {\bf Z}_4)$;\\
${\widetilde {\bf Z}}_2\otimes {\widetilde {\bf Z}}^\prime_6 (\supset {\bf Z}_3, 
{\bf Z}_6)$.\\
In brackets we have indicated the subgroups of 
${\widetilde {\bf Z}}_N\otimes {\widetilde {\bf Z}}_M$ that have already appeared 
in category ({\em i}). In the last case of ${\widetilde {\bf Z}}_2\otimes 
{\widetilde {\bf Z}}^\prime_6$
the generators $g$ and $h$ of the ${\widetilde {\bf Z}}_2$ and 
${\widetilde {\bf Z}}^\prime_6$
subgroups cannot be written as in Eq (\ref{gh}). These generators are given by:
\begin{equation}
 g={\mbox{diag}}(-1,-1,1)~,~~~
 h={\mbox{diag}}(\omega,-\omega,-\omega)~,
\end{equation}
where $\omega=\exp(2\pi i/3)$. Note that ${\widetilde {\bf Z}}^\prime_6=
{{\bf Z}}_6$, where ${{\bf Z}}_6$
has already appeared in category ({\em i}).

{}Next, let us solve the constraint (\ref{wsc}) for $J$ for each of the above
two categories.\\
$\bullet$ ({\em i}) For $G\approx{\bf Z}_N$ this condition reads:
\begin{equation}\label{con2}
 g^{-1}=JgJ~.
\end{equation}
Here we note that ${\mbox{Tr}}(g^{-1})={\mbox{Tr}}(JgJ)={\mbox{Tr}}(gJ^2)=
{\mbox{Tr}}(g)$. Thus, Eq (\ref{con2}) implies that such $J$
exists only if ${\mbox{Tr}}(g)$ is a real number. None of the $G\approx{\bf Z}_N$
cases in category ({\em i}) satisfy this requirement, so the consistency condition
(\ref{wsc}) for orientifolds of Type IIB on these $G\approx{\bf Z}_N$
orbifolds is not satisfied.\\
$\bullet$ ({\em ii}) In all the cases $G\approx {\widetilde {\bf Z}}_N\otimes 
{\widetilde {\bf Z}}_M
(\not\approx{\bf Z}_{NM})$, with the only 
exception of $G\approx {\widetilde {\bf Z}}_2\otimes 
{\widetilde {\bf Z}}_2$, $G$ contains a subgroup
which appears in category ({\em i}). This implies that the only case for which 
the constraint (\ref{wsc}) can be satisfied is $G\approx {\widetilde {\bf Z}}_2\otimes 
{\widetilde {\bf Z}}_2$. The two generators $g$ and $h$ in this case must commute
with $J$. This gives four inequivalent solutions (where we are writing $J$ as a 
$3\times 3$ matrix acting on $dz_1,dz_2,dz_3$):\\  
$\bullet$ $J={\mbox{diag}}(1,1,1)$;\\
$\bullet$ $J={\mbox{diag}}(1,1,-1)$;\\
$\bullet$ $J={\mbox{diag}}(1,-1,-1)$;\\
$\bullet$ $J={\mbox{diag}}(-1,-1,-1)$.\\
(When acting on $z_1,z_2,z_3$ (instead of $dz_1,dz_2,dz_3$), $J$
can also include shifts. As in the six dimensional case, these shifts are not 
difficult to classify, so 
we will not list them here for brevity.) The above choices of $J$ lead to 
orientifolds discussed in Ref \cite{BL}.

{}For brevity we will not consider all possible non-Abelian orbifolds $T^6/G$ 
with $SU(3)$ holonomy, but confine our discussion to the following two examples:
$G\approx D_N$ (non-Abelian
dihedral group), and $G\approx T$ (non-Abelian tetrahedral group).\\ 
$\bullet$ $G\approx D_N$ ($N=3,4,6$). The dihedral group
$D_N$ has two generators $g$ and $r$. Here 
$g$ is the generator of the ${\bf Z}_N\subset D_N$, and $r$ is the generator of 
${\bf Z}_2\subset D_N$ (where ${\bf Z}_2\not\subset {\bf Z}_N$). Note that $g$ and 
$r$ do not commute: $rg=g^{-1}r$.
Up to equivalent representations we have:
\begin{eqnarray}
 &&gdz_1=dz_1~,~~~gdz_2=\omega dz_2~,~~~gdz_3=\omega^{-1} dz_3~,\\
 &&rdz_1=-dz_1~,~~~rdz_2=dz_3~,~~~rdz_3=dz_2~.
\end{eqnarray} 
Here $\omega=\exp(2\pi i/N)$. We have the following allowed choices:
$N=3,4,6$.\\
$\bullet$ $G\approx T$. The tetrahedral group
$T$ has two generators $g$ and $r$. Here 
$g$ is the generator of the ${\bf Z}_3\subset T$, and $r$ is the generator of 
${\bf Z}_2\subset T$. Note that $g$ and 
$r$ do not commute: $rg=gr^\prime$, $r^\prime g=grr^\prime$. Here 
$r^\prime$ is the generator of ${\bf Z}_2^\prime\subset T$. The two generators
$r$ and $r^\prime$ commute, and together generate the subgroup
${\bf Z}_2\otimes{\bf Z}_2^\prime\subset T$.  
Up to equivalent representations we have:
\begin{eqnarray}
 &&gdz_1=dz_2~,~~~gdz_2=dz_3~,~~~gdz_3=dz_1~,\\
 &&rdz_1=-dz_1~,~~~rdz_2=-dz_2~,~~~rdz_3=dz_3~.
\end{eqnarray} 

{}Let us see whether the constraint (\ref{wsc}) is satisfied for these orbifolds.\\ 
$\bullet$ For $G\approx D_N$ the constraint (\ref{wsc}) can be
summarized by the following three equations:
\begin{equation}\label{con3}
 g^{-1}=JgJ~,~~~Jr=rJ~,~~~(Jgr)^2=1~.
\end{equation}
Let us assume that the first two of these equations are satisfied and check 
if the third one is compatible with this assumption. We have: $(Jgr)^2=
JgrJgr=JgJrgr=g^{-2}\not=1$ (where we have used $rgr=g^{-1}$). We 
therefore conclude that the consistency condition
(\ref{wsc}) for orientifolds of Type IIB on these $G\approx D_N$
orbifolds is not satisfied.\\
$\bullet$ For $G\approx T$ the constraint (\ref{wsc}) can be
summarized by the same equations  (\ref{con3}) as in the $D_N$ case.
As in the $D_N$ case, let us assume that the first two of these equations 
are satisfied and check 
if the third one is compatible with this assumption. We have: $(Jgr)^2=
JgrJgr=JgJrgr=r^\prime r\not=1$ (where we have used $rg=gr^\prime$).
We therefore conclude that the consistency condition
(\ref{wsc}) for orientifolds of Type IIB on this $G\approx T$
orbifold is not satisfied.\\
It is not difficult to show that the constraint (\ref{wsc}) is not satisfied for any 
other non-Abelian orbifold (such as $G\approx D_N\otimes {\bf Z}_N$, 
$N=3,4,6$, and $G\approx T\otimes {\bf Z}_3$) with $SU(3)$ holonomy. 

{}Thus, the only choice of ${\cal M}_3=T^6/G$ that satisfies the constraint
(\ref{wsc}) is $G\approx {\widetilde {\bf Z}}_2
\otimes {\widetilde {\bf Z}}_2$. The corresponding solutions for $J$ have been given
above.

{}Here we note that in the cases $G\approx{\widetilde {\bf Z}}_N\otimes 
{\widetilde {\bf Z}}_M (\not\approx{\bf Z}_{NM})$ we can have discrete
torsion. This only affects the twisted sectors of the orbifold, but not the 
untwisted sector. Since the world-sheet consistency condition (\ref{wsc})
was derived by examining only the untwisted sector contributions, 
the conclusions of this section are independent of whether we have
discrete torsion. 

\section{F-theory Interpretation}\label{F}

{}The results of section \ref{class} may at first appear 
surprising as the number of orientifolds that satisfy the world-sheet 
consistency condition (\ref{wsc}) is very limited. This may raise the question
of whether the consistency condition (\ref{wsc}) is indeed necessary. In this 
section we give F-theory \cite{vafa} interpretation of this condition. We will first consider 
six dimensional orientifolds, and then generalize our discussion to four 
dimensional cases.

\subsection{6D Orientifolds}

{}Consider an $\Omega J (-1)^{F_L}$ orientifold of Type IIB on 
${\cal M}_2=T^4/{\bf Z}_N$
($N=2,3,4,6$), where in the diagonal basis $J={\mbox{diag}}(-1,+1)$,
so that only D7-branes can be present in the open string sector. (Here we 
are writing $J$ as a $2\times 2$ matrix acting on $dz_1$ and $dz_2$.) 
The only assumption we will make about $J$ is that $J$ and $g$ (where $g$
is the generator of ${\bf Z}_N$) form a group. This is necessary for the set 
${\cal O}=\{g^k,\Omega J_k (-1)^{F_L}\vert k=0,\dots,N-1\}$ to form a group. 
(Here $J_k\equiv Jg^k$.) In particular, we will not assume that $J$ satisfies
Eq (\ref{wsc}). 

{}Note that $J$ reverses the sign of the holomorphic 2-form $dz_1\wedge dz_2$
on ${\cal M}_2$. 
Following Refs \cite{sen}, we can map this orientifold to (a limit of) F-theory on a 
Calabi-Yau three-fold ${\cal X}_3$ defined as
\begin{equation}
 {\cal X}_3=(T^2\otimes {\cal M}_2)/X~,
\end{equation}     
where $X=\{1,S\}\approx{\bf Z}_2$, and $S$ acts as $Sz_0=-z_0$ on $T^2$ 
($z_0$ is a complex coordinate on $T^2$), and as $J$ on ${\cal M}_2$. 
Note that ${\cal X}_3$ is an elliptically fibered Calabi-Yau three-fold with 
the base ${\cal B}_2={\cal M}_2/B$, where $B\equiv\{1,J\}\approx{\bf Z}_2$.

{}From this viewpoint, six dimensional $\Omega J (-1)^{F_L}$
orientifolds are described as F-theory 
compactifications, which were studied in detail in Refs \cite{MV}, 
on Calabi-Yau three-folds ${\cal X}_3$.
Such Calabi-Yau three-folds 
are known as the Voisin-Borcea orbifolds \cite{Voisin,Borcea}. (The action of
$J$ on ${\cal M}_2$ is known as a Nikulin involution \cite{Nik} of K3.) 
Since ${\cal M}_2=T^4/{\bf Z}_N$, the resulting Voisin-Borcea orbifold ${\cal X}_3=
(T^2\otimes T^4)/{\cal G}$, where ${\cal G}=\{g^k,S_k\vert k=0,\dots,N-1\}$, and
$S_k\equiv Sg^k$. 
The generators $S$ and $g$ of the ${\bf Z}_2\subset{\cal G}$ and ${\bf Z}_N\subset
{\cal G}$ subgroups have the following action on $dz_0,dz_1,dz_2$:
\begin{eqnarray}
 &&Sdz_0=-dz_0~,~~~Sdz_1=Jdz_1~,~~~Sdz_2=Jdz_2~,\\
 &&gdz_0=dz_0~,~~~gdz_1=\omega dz_1~,~~~gdz_2=\omega^{-1} dz_2~,
\end{eqnarray}   
where $\omega=\exp(2\pi i/N)$.

{}The map between the orientifold and F-theory descriptions is as follows.
The untwisted sector in F-theory corresponds to the untwisted closed string
sector of the orientifold. The $g^k$, $k=1,\dots,N-1$, twisted sectors in F-theory
correspond to the twisted closed string sectors of the orientifold.
The $S_k$ twisted sectors of F-theory (are supposed to) correspond to the open 
string sectors of the orientifold.

{}Let us examine these $S_k$ twisted sectors in more detail. In the diagonal 
basis $S_k={\mbox{diag}}(-1,-\rho_k,\rho^{-1}_k)$, where $\vert\rho_k\vert=1$. 

{}First consider the case $\rho_k=1$ (which for our purposes here
is equivalent to the case 
$\rho_k=-1$).  
Then $(S_k)^2=1$ which implies that $J_k^2=1$. In this case
$S_k={\mbox{diag}}(-1,-1,1)$, and the set of points ${\cal F}_k$
fixed under the action of 
$S_k$ is a one complex dimensional submanifold of the base ${\cal B}_2$.
This implies that the orientifold description of the states in the $S_k$ twisted 
sectors in F-theory is given by open strings stretched between the
corresponding D7-branes whose transverse directions lie in ${\cal F}_k$
$\subset{\cal B}_2$ \cite{sen}. 

{}Next, let us focus on the cases $\rho_k\not=\pm1$, which implies
that $(S_k)^2\not=1$, {\em i.e.}, $J_k^2\not=1$. The fixed point set
is now  discrete. The corresponding states in F-theory are no longer
described (in the orientifold language)
in terms of open strings stretched between D7-branes. Instead,
they are more appropriately viewed as F-theory seven-branes wrapping
the collapsed two-cycles (corresponding to the fixed points in the base).
Locally this corresponds to having D7-branes with ${\bf C}/{\bf Z}_N$ 
($N=3,4,6$) singularities in their world-volumes.
These states are non-perturbative from the orientifold viewpoint, and cannot be
described in conformal field theory.

{}Let us try to understand in more detail 
why such states do not have (perturbative) orientifold 
description. Open strings required to describe these states 
would have to have boundary conditions which are neither Neumann (N) nor
Dirichlet (D) but mixed. Such boundary conditions can be written as follows:
\begin{eqnarray}\label{mixed1}
 &&\left.\left(\cos(\pi v_s) \partial_\sigma z_s -\sin(\pi v_s) \partial_\tau z_s
 \right)\right|_{\sigma=0}=0~,\\
 \label{mixed2}
 &&\left.\left(\cos(\pi u_s) \partial_\sigma z_s -\sin(\pi u_s) \partial_\tau z_s
 \right)\right|_{\sigma=\pi}=0~,  
\end{eqnarray} 
where $\sigma$ and $\tau$ are the space-like and time-like world-sheet 
coordinates, and $v_s=m_s/N$, $u_s=n_s/N$ ($m_s,n_s\in{\bf Z}$). 
It is not difficult to see that the $z_s$ oscillators for the above boundary 
conditions are moded as $\pm(v_s-u_s)~(\mbox{mod}~1)$ (and, therefore,
can be fractional yet different from $1/2$).
Note that for $v_s=u_s=0$ we have NN boundary conditions in the $z_s$ 
direction. The corresponding open strings have momenta in this direction
but no windings, and the oscillators are integer moded. 
For $v_s=u_s=1/2$ we have DD boundary conditions in 
the $z_s$ direction. 
The corresponding open strings have windings in this 
direction but no momenta, and the oscillators are also integer moded. 
For $v_s=0$, $u_s=1/2$ and $v_s=1/2$, $u_s=0$ we have ND and DN
boundary conditions, respectively.  The corresponding open strings have
no momenta or windings, and half odd
integer moded oscillators. In all the other cases, however, we have
mixed boundary conditions. In particular, in the cases
$v_s=u_s\not=0,1/2$ we have open strings with no momenta or windings, and
{\em integer} moded oscillators. Such open string sectors pose no problem at
the tree level, but at the one-loop level we run into a difficulty. The  contribution
of such states into the annulus partition function would be proportional to 
$1/\eta^4({\mbox{e}}^{-2\pi t})$ not accompanied by a momentum or winding sum. 
After the transformation $t\rightarrow1/t$ we will therefore have uncompensated
factor of $1/(\sqrt{t})^4$ in complete analogy with the discussion of section \ref{IIB}.
This poses a problem since upon the transformation $t\rightarrow1/t$ the annulus
(that is, open string loop) amplitude turns into a tree-channel amplitude that 
describes closed strings propagating between two boundary states. Normally, 
these would be D-branes. Here, however, we see that we cannot construct the 
boundary states due to the extra factor of $1/(\sqrt{t})^4$ in the tree-channel
amplitude. The reason is that the open strings with mixed boundary conditions
simply do not end on D-branes: it is not difficult to see (by solving equations
(\ref{mixed1}) and (\ref{mixed2}) as discussed in appendix \ref{MIX}) 
that an open string endpoint 
(for a mixed boundary condition) is not stuck on a rigid manifold but rather
it harmonically oscillates around a fixed point. This is not 
necessarily inconsistent as far as physics is concerned. In fact, F-theory 
provides a non-perturbative framework for describing such 
``breathing'' boundary states. On the other hand, there is no consistent 
world-sheet, {\em i.e.}, perturbative description of these phenomena
within the orientifold approach. The sectors with mixed boundary
conditions were also recognized (from
a somewhat different viewpoint) in Ref \cite{blum} where they were referred
to as ``twisted (open) strings''.      

{}The above discussion has the implication that unless $S_k^2=1$, or, 
equivalently, unless $J_k^2=1$, the world-sheet, {\em i.e.}, the orientifold
description does not capture all the sectors of the theory. In particular, the 
D-brane picture is no longer applicable unless the condition $J^2_k=1$ is 
satisfied. This is the same condition as derived in section \ref{IIB} from a 
world-sheet approach, namely, Eq (\ref{wsc}). There, however, we looked 
at the untwisted
contributions into the Klein bottle amplitude and found that the cross-cap
states could not be constructed. In the F-theory description we looked at the
$S_k$ twisted sectors that turn out to correspond to open strings stretched 
between D7-branes if and only if $J^2_k=1$. Alternatively, we can examine
the action of the twists $S_k$ in
the untwisted sector in F-theory. The set of points ${\cal F}_k$ 
fixed under $S_k$
then would have to correspond to the space transverse to the
orientifold 7-planes \cite{sen}. This, 
however, cannot be the case unless $S_k^2=1$ (which follows from the
previous discussion). Thus, F-theory 
provides a geometric setting for understanding the world-sheet
consistency condition derived in section \ref{IIB}.

{}The above discussion has the following implications. Perturbative
world-sheet description requires the condition (\ref{wsc}) be satisfied.
On the other hand, other six-dimensional orientifolds of 
Type IIB on (symmetric) orbifolds ${\cal M}_2=T^4/{\bf Z}_N$ are not 
necessarily inconsistent. In particular, the $\Omega J(-1)^{F_L}$ 
orientifolds (where $J$ reverses the sign of the holomorphic two-form 
on ${\cal M}_2$) have a non-perturbative description via F-theory
regardless of whether (\ref{wsc}) is satisfied.

\subsection{4D Orientifolds}

{}Consider an $\Omega J (-1)^{F_L}$ orientifold of Type IIB on 
${\cal M}_3=T^6/G$, where in the diagonal basis $J={\mbox{diag}}(-1,+1,+1)$.
(Here we 
are writing $J$ as a $3\times 3$ matrix acting on $dz_1,dz_2,dz_3$.)
As before, the orbifold group $G=\{g_a\vert a=1,\dots,{\mbox{dim}}(G)\}$.
The orientifold group is given by 
${\cal O}=\{g_a,\Omega J_a (-1)^{F_L}\vert a=1,\dots,{\mbox{dim}}(G)\}$,
where $J_a=Jg_a$.

{}Note that $J$ reverses the sign of the holomorphic 3-form 
$dz_1\wedge dz_2\wedge dz_3$ on ${\cal M}_3$. Following Refs \cite{sen},
we can map this orientifold to (a limit of) F-theory on a 
Calabi-Yau four-fold ${\cal X}_3$ defined as
\begin{equation}\label{CY_4}
 {\cal X}_4=(T^2\otimes {\cal M}_3)/X~,
\end{equation}     
where $X=\{1,S\}\approx{\bf Z}_2$, and $S$ acts as $Sz_0=-z_0$ on $T^2$ 
($z_0$ is a complex coordinate on $T^2$), and as $J$ on ${\cal M}_3$. 
Note that ${\cal X}_4$ is an elliptically fibered Calabi-Yau four-fold with 
the base ${\cal B}_3={\cal M}_2/B$, where $B\equiv\{1,J\}\approx{\bf Z}_2$.

{}So far the story for the 4D orientifolds has been the same as for the 6D orientifolds.
In four dimensions, however, there is a new ingredient: three-branes. On general
grounds it is known \cite{SVW} that to cancel space-time anomaly in F-theory
on a Calabi-Yau four-fold ${\cal X}_4$ one needs $\chi/24$ three-branes, where
$\chi$ is the Euler characteristic of ${\cal X}_4$.

{}Let us try to understand the map between the $\Omega J (-1)^{F_L}$
orientifold and F-theory in more detail. Note that we can write
${\cal X}_4=
(T^2\otimes T^6)/{\cal G}$, where ${\cal G}=\{g_a,S_a\vert a=1,
\dots,{\mbox{dim}}(G)\}$, and
$S_a\equiv Sg_a$. As in six dimensions, the untwisted sector in F theory
corresponds to the untwisted closed string sector of the orientifold. Also,
the $g_a$ ($g_a\not=1$) twisted sectors in F-theory correspond to the twisted
closed string sectors of the orientifold. What we need to understand is what
corresponds to the $S_a$ twisted sectors in F-theory on the orientifold side.

{}In the diagonal basis $S_a=(-1,-\rho_a,\rho^\prime_a,
(\rho_a\rho^\prime_a)^{-1})$,
where $\vert \rho_a\vert=\vert \rho^\prime_a\vert=1$. Here we have the 
following possibilities.\\
$\bullet$ $\rho_a=\rho^\prime_a=1$. Then we have D7-branes without 
any singularities.\\ 
$\bullet$ $\rho_a=1$, $\rho^\prime_a=-1$. Then we have D7-branes with
${\bf C}^2/{\bf Z}_2$, {\em i.e.}, $A_1$ singularities in their world-volumes.
These are equivalent to perturbative (from the orientifold viewpoint) 
D3-branes.\\
$\bullet$ $\rho_a=1$, $\rho^\prime_a\not=\pm1$. Then we have D7-branes with
${\bf C}^2/{\bf Z}_N$ ($N=3,4,6$), {\em i.e.}, $A_{N-1}$
singularities in their world-volumes. These states are non-perturbative from
the orientifold viewpoint.\\
$\bullet$ $\rho_a\not=\pm1$, $\rho^\prime_a=1$. Then we have D7-branes with
${\bf C}/{\bf Z}_N$ ($N=3,4,6$) 
singularities in their world-volumes. These states have already
appeared in six dimensional cases, and are non-perturbative from
the orientifold viewpoint.\\
$\bullet$ $\rho_a,\rho^\prime_a\not=\pm1$. Then we have D7-branes with
${\bf C}^2/\Gamma$ ($\Gamma\subset G$, $\Gamma\not\approx{\bf Z}_2$) 
singularities in their world-volumes. These states are non-perturbative from
the orientifold viewpoint. They should also have a description as F-theory
three-branes, at least for certain choices of $\Gamma$.\\
All the other cases are equivalent (for our purposes here) to the previous
possibilities. 

{}The above analysis implies that unless $S_a^2=1$, which is equivalent to
$J^2_a=1$, the states in the $S_a$ twisted sectors do not have perturbative
orientifold description. Thus, as in six dimensions, we have recovered the 
world-sheet consistency condition (\ref{wsc}) in four dimensional cases from
F-theory viewpoint. Here we have only analyzed the $S_a$ twisted sectors of 
F-theory. The analysis of the untwisted sector parallels that in six 
dimensions, and the same constraint can be obtained by requiring that we
have perturbatively well defined orientifold planes.  

{}It would be important to understand F-theory compactifications on Calabi-Yau
four-folds defined in Eq (\ref{CY_4}). Orbifold compactifications 
(some examples of F-theory compactifications on orbifold
Calabi-Yau four-folds were studied in Ref \cite{GM}) might be under
greater control than more generic four dimensional compactifications of F-theory
which are rather involved \cite{SVW,FMW} (also see, {\em e.g.}, Refs \cite{F4}). 
One might expect, at least naively, that 
all of the models with ${\cal X}_4$ as in Eq  (\ref{CY_4}) have gauge groups 
which are products 
of $SO(8)$'s: the singularity in the fibre is always of $D_4$ type. If this were the 
case all of these models would be
non-chiral. However, here we need
to take into account that unlike in six dimensions (where specifying the 
Calabi-Yau three-fold is enough to determine the massless spectrum) F-theory
compactifications on Calabi-Yau four-folds require specification of additional 
data \cite{FMW}, and the question of whether a given model is chiral requires 
a more careful
examination. We will encounter an example of necessity for specifying such 
additional data in section \ref{FA}.

\subsection{Comments}

{}The analyses of the previous subsection indicate that the 
perturbative orientifold description is inadequate unless the world-sheet
consistency condition (\ref{wsc}) is satisfied for otherwise it misses the
corresponding sectors which are non-perturbative. This may at first appear
surprising as the underlying conformal field theory is well defined, and one 
does not expect non-perturbative effects to arise unless the
conformal field theory goes bad. We believe this point deserves 
further clarification to which we now turn.

{}By now it has been well appreciated that the geometric and conformal field
theory orbifolds are not the same. Geometric orbifolds are singular spaces
which should, at least classically, lead to enhanced gauge symmetries and,
perhaps, some other non-perturbative effects in string theory. On the other
hand, the description of string theory on conformal field theory orbifolds is
non-singular, and no enhanced gauge symmetries are expected. The 
resolution of this discrepancy is the following \cite{aspinwall}. Quantum
geometry can modify the classical picture and move the theory away from
the singular point in the moduli space. This is realized via non-zero value
of twisted sector $B$-fields corresponding to the blow-up modes of the orbifold.
Thus, for zero values of these twisted sector modes the conformal field theory
description would be inadequate due to the singularity.

{}In F-theory all the $B$-fields (including those coming from the twisted sectors)
must be zero \cite{vafa}. The orbifold there then corresponds to the true
geometric orbifold with real singularities. This is why it is not surprising that
we see effects in F-theory that have no perturbative description in 
orientifolds. On the other hand, for the perturbative orientifold description to
be adequate we must work with the conformal field theory orbifolds with 
non-zero twisted $B$-fields turned on. Let us see what the implications of this
fact are for the orientifold consistency\footnote{We would like to
thank A. Sen and C. Vafa for valuable discussions on this point.}. 
First consider the case of $T^4/{\bf Z}_2$.
In Ref \cite{aspinwall} it was shown that the twisted $B$-field must be $1/2$ in this
case (where the normalization convention is such that the $B$-field is defined up
to an integer). Taking into account the discrete symmetry of the ${\bf Z}_N$ orbifold it 
is reasonable to believe that in $g^k$ twisted sectors the $B$-field takes values
$k/N$ (where $g$ is the generator of ${\bf Z}_N$) \cite{DM}. 
Note that the $B$-field is odd 
under the action of the orientifold reversal $\Omega$. Under its action, therefore,
the $B$-field in the $g^k$ twisted sector changes the sign: $\Omega B=-B=-k/N=
(N-k)/N~({\mbox{mod}}~1)$. For the left-right symmetric orbifolds considered in 
section \ref{IIB},
$\Omega$ maps $g^k$ twisted sector to itself, yet it changes the $B$-field from
its $g^k$ twisted value to the $g^{N-k}$ twisted value. Thus, for consistency 
$\Omega$ should be accompanied by $J$ such that $J^2=1$, and
 $J$ maps $g^k$ twisted
sector to $g^{N-k}$ twisted sector (but leaves the $B$-field unchanged). But this 
implies that
\begin{equation}
 Jg^kJ^{-1}=g^{N-k}~.
\end{equation}       
This is precisely the world-sheet consistency constraint (\ref{wsc}) derived in
section \ref{IIB}. Here we looked at the ${\bf Z}_N$ cases, but the generalization 
to an arbitrary (Abelian) group $G$ should be clear (as one can consider  
${\bf Z}_N$ subgroups of $G$).

{}The above discussion implies that to have non-singular 
conformal field theory description to start with it is necessary to turn
on the twisted $B$-fields, but then the constraint (\ref{wsc}) must be
satisfied or else orientifolding is not a symmetry of the theory. On the other
hand, if we turn off the twisted $B$-fields then the constraint (\ref{wsc})
need not be satisfied, but the perturbative description is no longer adequate
and one needs to appeal to F-theory. Thus, (``symmetric'' Type IIB) 
orientifolds do not seem to provide 
us with a ``free lunch''. This calls for caution when dealing with orientifolds.

{}Finally, we would like to make the following remark. We derived the 
world-sheet consistency condition (\ref{wsc}) in section \ref{IIB}
without any reference to F-theory or the argument of 
this subsection based on the twisted $B$-field. On the other hand, the 
connection between orientifolds and F-theory in the light of the classical 
{\em vs.} quantum geometry argument of Ref \cite{aspinwall}
indicates that we may view the results of section \ref{IIB} and
this section as (albeit, perhaps, indirect) evidence for extending the
conclusions of Ref \cite{aspinwall} about the presence of non-zero 
twisted $B$-fields in conformal field theory orbifolds to more 
general cases ({\em e.g.}, ${\bf Z}_N$).

\section{``Asymmetric'' Type IIB Orientifolds}\label{IIBA}

{}In the previous section we saw that the perturbative orientifold description 
captured only the sectors of the theory that correspond to the following elements
of the orientifold group ${\cal O}=\{g_a,\Omega J_a I^{F_L} \vert a=1,\dots,
{\mbox{dim}}(G)\}$ ($J_a=Jg_a$): ({\em i}) the $g_a$ twisted sectors corresponding 
to closed string sectors (including the untwisted sector); ({\em ii}) 
the $\Omega J_a I^{F_L}$ twisted sectors with $J_a^2=1$ corresponding to open 
string sectors where open strings are stretched between D-branes. However, the
perturbative orientifold description is inadequate for the $\Omega J_a I^{F_L}$ 
twisted sectors with $J_a^2\not=1$. 

{}At least naively, we expect similar conclusions to hold in the case of ``asymmetric''
Type IIB orientifolds, that is, orientifolds of Type IIB compactified on ``asymmetric'' 
orbifolds ${\widetilde {\cal M}}_d$. However, there are additional subtleties arising 
in ``asymmetric'' Type IIB orientifolds, and this section is devoted to understanding
precisely these new issues.

\subsection{Klein Bottle}

{}Consider the one-loop vacuum amplitude for the $\Omega J I^{F_L}$ 
orientifold of Type IIB compactified on ${\widetilde {\cal M}}_d$. (We will denote 
the orientifold group as ${\widetilde {\cal O}}=\{{\widetilde g}_a,\Omega 
{\widetilde J}_a I^{F_L} \vert a=1,\dots,{\mbox{dim}}({\widetilde G})\}$, where
${\widetilde J}_a\equiv J{\widetilde g}_a$.)
For now let us concentrate on the closed untwisted sector contributions
of the bosonic world-sheet degrees of freedom $z_s$. 
As in section \ref{IIB},  for the sake of simplicity we will 
assume that $J$ and ${\widetilde g}_a$ act homogeneously on $z_s$, {\em i.e.}, 
without shifts. The Klein bottle contribution is given by:  
\begin{equation}\label{KleinA}
 {\cal K}={1\over 2{\mbox{dim}}({\widetilde G})} \sum_{a=1}^{{\mbox{\small{dim}}}
({\widetilde G})} 
 {\cal K}_a= {1\over 2{\mbox{dim}}({\widetilde G})}
 \sum_{a=1}^{{\mbox{\small{dim}}}({\widetilde G})}
 {\mbox{Tr}}\left(\Omega {\widetilde J}_a q^{L_0} 
 {\overline q}^{{\overline L}_0}\right)~.
\end{equation}

{}Let us first consider the oscillator contributions. (Note that oscillator 
contributions and momentum plus winding contributions factorize.)
The presence of the $\Omega$ projection in the Klein bottle amplitude
implies that only left-right symmetric states contribute. 
The discussion in
section \ref{prelim} (see Eq (\ref{ga1})) implies
that the oscillator contribution into ${\cal K}_a$ is given by $\prod_{s=1}^d
X^0_{2\varphi_{as}}(q{\overline q})$. Here the phases
$\exp(2\pi i\varphi_{as})$ are eigenvalues of ${\widetilde J}_a$ 
(that is, in the diagonal basis ${\widetilde J}_a=
{\mbox{diag}}(\exp(2\pi i\varphi_{a1}),\dots,\exp(2\pi i\varphi_{ad}))$).
The characters $X^0_u$, $u\not=0$, are 
defined in appendix \ref{Orbifold}. The character $X^0_0$ is defined as
$X^0_0\equiv\eta^{-2}$. (Note that $X^0_u=X^0_{u+1}$.)
Next, consider the momentum and winding 
contributions. 
It is the same as in the case of ``symmetric'' orientifolds, and is given by Eq 
(\ref{MW}). Combining the oscillator contributions with those of momenta and 
windings, we have the following expression for ${\cal K}_a$:
\begin{equation}\label{Ka}
 {\cal K}_a=\prod_{s=1}^d X^0_{2\varphi_{as}} ({\mbox{e}}^{-2\pi t})
 \sum_{p\in{1\over 2}{\widetilde \Lambda}({\widetilde J}_a)}
 \exp(-\pi t p^2)
 \sum_{w\in\Lambda(R{\widetilde J}_a)}
 \exp(-\pi t w^2)~.
\end{equation} 
Here (and in the following subsection) 
we are using some of the same notations as in section \ref{IIB}.

\subsection{Cylinder with Two Cross-Caps}

{}Under the modular transformation $t\rightarrow1/t$ the Klein bottle turns into a 
cylinder with two cross-caps as its boundaries. Let 
${\widetilde {\cal K}}=(1/2{\mbox{dim}}({\widetilde G}))\sum_a {\widetilde {\cal K}}_a$ be 
the resulting 
tree-channel amplitude. The contributions ${\widetilde {\cal K}}_a$ are obtained
from ${\cal K}_a$ via $t\rightarrow1/t$:  
\begin{eqnarray}
 {\widetilde {\cal K}}_a=&&
 \prod_{s=1}^d X^{2\varphi_{as}}_0 ({\mbox{e}}^{-2\pi t})\xi(\varphi_{as})
 \nonumber\\
 &&\label{crossA}
 \left((2\sqrt{t})^{d({\widetilde J}_a)} V({\widetilde J}_a)\right)
 \sum_{{\widetilde p}\in 2\Lambda({\widetilde J}_a)}
 \exp(-\pi t {\widetilde p}^2)
 \left({(\sqrt{t})^{d(R{\widetilde J}_a)}\over V(R{\widetilde J}_a)}\right)
 \sum_{{\widetilde w}\in{\widetilde\Lambda}(R{\widetilde J}_a)}
 \exp(-\pi t {\widetilde w}^2)~.
\end{eqnarray}
Here $\xi(\varphi_{as})=(\sqrt{t})^{-2}$ if $2\varphi_{as}\in{\bf Z}$, and
$\xi(\varphi_{as})=2\vert \sin(2\pi\varphi_{as}) \vert$ otherwise.

{}As in the case of ``symmetric'' orientifolds we must make sure that there are
no overall factors of $\sqrt{t}$ in ${\widetilde {\cal K}}_a$ or else we will not 
have perturbatively well defined cross-cap boundary states. 
We therefore conclude that the orientifold consistency
requires the following constraint be satisfied:
\begin{equation}\label{const0A}
 \forall a~d({\widetilde J}_a)+d(R{\widetilde J}_a)=2n_{+-}({\widetilde J}_a)~. 
\end{equation}
Here $n_{+-}({\widetilde J}_a)=n_{+}({\widetilde J}_a)+
n_{-}({\widetilde J}_a)$, where $n_{\pm} ({\widetilde J}_a)$ are 
the numbers of ${\widetilde J}_a$ eigenvalues equal $\pm1$,
respectively. On the other hand, the dimension 
$d({\widetilde J}_a)$ of the lattice $\Lambda({\widetilde J}_a)$ is given by 
$d({\widetilde J}_a)=2n_+ ({\widetilde J}_a)$,
which follows from the definition of $\Lambda({\widetilde J}_a)$ being 
the sublattice
of $\Lambda$ invariant under ${\widetilde J}_a$. Similarly, the dimension 
$d(R{\widetilde J}_a)$ of the lattice $\Lambda(R{\widetilde J}_a)$ is given by 
$d(R{\widetilde J}_a)=2n_- ({\widetilde J}_a)$.
Thus, the world-sheet consistency condition (\ref{const0A}) is always satisfied 
for ``asymmetric'' Type IIB orientifolds.

\subsection{Twisted Sectors}

{}The analysis in the previous subsections indicates that the untwisted closed 
string sector does not pose any (obvious) problems for ``asymmetric'' orientifold
consistency from the world-sheet viewpoint. Next, let us examine whether twisted 
sectors require any additional constraints. Suppose there are ${\bf Z}_2$ twisted 
closed string sectors. These are left-right symmetric, and it is not difficult to see 
that their contributions to the Klein bottle amplitude (at least at the
level of the present analysis) do not pose any problem for constructing 
perturbatively consistent cross-cap boundary states. The story with twisted 
sectors other than the ${\bf Z}_2$ twisted sectors, however, is quite different.

{}Let ${\widetilde g}_a\in{\widetilde G}$ such that ${\widetilde g}^2_a\not=1$.
The ground state in the ${\widetilde g}_a$ twisted sector is left-right asymmetric,
and the world-sheet parity operator $\Omega$ by itself is not a symmetry of the
theory. To flip the ground state $\sigma_{{\widetilde g}_a}\vert 0\rangle_L\otimes
{\overline \sigma}_{{\widetilde g}_a^{-1}}\vert 0\rangle_R$ to 
$\sigma_{{\widetilde g}_a^{-1}}\vert 0\rangle_L\otimes
{\overline \sigma}_{{\widetilde g}_a}\vert 0\rangle_R$, $\Omega$ {\em must} be 
accompanied by an operator $J$ (more precisely, we must also include $I^{F_L}$,
where, as before, $I\equiv\det(J)$), where: $J$ is a symmetry of ${\widetilde 
{\cal M}}_d$ such that $J^2=1$; $J$ acts left-right symmetrically on ${\widetilde 
{\cal M}}_d$; $J$ maps the ${\widetilde g}_a$ twisted sector into its inverse 
${\widetilde g}_a^{-1}$ twisted sector. (The need for such $J$ was recognized 
in Ref \cite{Pol}.) The latter statement implies that 
\begin{equation}\label{wscA}
 \forall a~J{\widetilde g}_a J^{-1}={\widetilde g}_a^{-1}~,~
 {\mbox{or, equivalently}},~{\widetilde J}_a^2=1~.
\end{equation} 
Note that this constraint is the same as the world-sheet consistency constraint
(\ref{wsc}) we derived for ``symmetric'' orientifolds in section \ref{IIB}. In section
\ref{class} we saw that solution to this constraint for six dimensional 
orientifolds exist only in two cases: ({\em i}) ${\widetilde G}\approx{\bf Z}_2$
(and the corresponding solutions for $J$ were given in subsection A of section
\ref{class}); ({\em ii}) ${\widetilde G}\approx{\bf Z}_N$ ($N=3,4,6$), and the most
general solution for $J$ was given in Eq (\ref{J}). Here we note that in case 
({\em ii}) the ``asymmetric'' orientifold models (for all choices of $J$) are the same as 
the corresponding ``symmetric'' orientifold models ({\em i.e.}, the $\Omega 
J (-1)^{F_L}$ orientifold of Type IIB on ${\cal M}_2=T^4/{\bf Z}_N$ with $J$ given
by Eq (\ref{J}))\footnote{Here we note that if we relax $J^2=1$ condition then there is
an additional possibility. Namely, consider Type IIB  on ${\cal M}_2=T^4/{\bf Z}_N$ or 
${\widetilde {\cal M}}_2=T^4/{\bf Z}_N$. Next, consider the $\Omega J$ orientifold
of this theory where the action of $J$ on the complex coordinates $z_1,z_2$ is given by
$Jz_1=z_2$, $Jz_2=-z_1$. Note that $J^2=-1$. These orientifolds satisfy the world-sheet
consistency conditions (\ref{wsc}) and (\ref{wscA}), respectively. In these models, however, 
there are no D-branes as the unoriented closed string sector does not give rise to
any tadpoles. All of these orientifolds have the same massless spectrum which
arises solely from the closed string sector (as there are no open strings in these models),
which consists of $H=12$ hypermultiplets and $T=9$ tensor multiplets. This corresponds
to F-theory compactification on a Voisin-Borcea orbifold with $(r,a,\delta)=(10,10,0)$
with Hodge numbers $(h^{1,1},h^{2,1})=(11,11)$ (see section \ref{FA} for notations).
An alternative orientifold realization of this vacuum was discussed in Ref \cite{GJ}.}. 
We discuss these models in section \ref{FA}. 
As to the four dimensional orientifolds, in section \ref{class} we found that 
solutions to the constraint (\ref{wscA}) exist only for ${\widetilde G}\approx
{\widetilde {\bf Z}}_2\otimes {\widetilde {\bf Z}}_2$ (and the corresponding solutions 
for $J$ were given in subsection B of section \ref{class}).  
  
\section{Other Orientifold Constructions}\label{other}

{}``Asymmetric'' orientifolds of Type IIB on orbifolds ${\widetilde {\cal M}}_d=
T^{2d}/{\widetilde G}$ have been 
extensively studied in the literature.

{}In six dimensions we have the following known examples
with ${\cal N}=1$ space-time supersymmetry.\\ 
$\bullet$ Orientifolds of Type IIB on the ${\bf Z}_2$ orbifold limit of K3, {\em i.e.}, 
${\widetilde {\cal M}}_2=
{\cal M}_2=T^{4}/{\bf Z}_2$ \cite{PS,GP} (also see Refs \cite{dp,Pol}). (The models
of Refs \cite{PS,GP} have been studied in the context of 
Type I-heterotic duality \cite{PW} in Ref \cite{berkooz}. The F-theory
realizations of these models have been discussed in Refs \cite{sen1}
(also see Refs \cite{BZ,dp1}).)\\
$\bullet$ ``Asymmetric'' orientifolds of Type IIB on the ${\bf Z}_N$ ($N=3,4,6$) 
orbifold limits of K3, {\em i.e.}, 
${\widetilde {\cal M}}_2=T^{4}/{\bf Z}_N$ \cite{GJ,DP}. (These models have
been discussed in the context of Type I-heterotic duality in Ref \cite{gj},
and attempts have been made to construct their F-theory \cite{gj,dp1,blum}
and M-theory \cite{gj,blum} realizations.)

{}In four dimensions the following ${\cal N}=1$ space-time supersymmetric 
examples have been constructed. (Here we are using the notations of 
subsection B of section
\ref{class}.)\\
$\bullet$ Orientifolds of Type IIB on a ${\widetilde {\bf Z}}_2\otimes 
{\widetilde {\bf Z}}_2$ 
orbifold \cite{BL}. (The ``symmetric'' and ``asymmetric'' orientifolds
in this case coincide.) These models have been obtained by generalizing the
tadpole cancellation conditions of Refs \cite{PS,GP} for six dimensional
${\bf Z}_2$ orientifolds.\\
$\bullet$ ``Asymmetric'' orientifolds of Type IIB on the ${\bf Z}_3$ orbifold
\cite{Sagnotti}. (This model has been discussed in the context of Type I-heterotic
duality in Ref \cite{ZK}.)\\
$\bullet$ ``Asymmetric'' orientifolds of Type IIB on ${\bf Z}_7$, ${\widetilde 
{\bf Z}}_3\otimes
{\widetilde {\bf Z}}_3$
and ${\bf Z}_6$ orbifolds \cite{KS1,KS2}.
(The ${\bf Z}_7$ and ${\widetilde {\bf Z}}_3\otimes
{\widetilde {\bf Z}}_3$
models have been discussed in the context of Type I-heterotic
duality in Refs \cite{KS1} and \cite{KS2} (also see Ref \cite{Alda}), respectively.)\\
$\bullet$ ``Asymmetric'' orientifolds of Type IIB on ${\bf Z}_6^\prime$,
${\widetilde {\bf Z}}_2\otimes{\widetilde {\bf Z}}_6$, ${\widetilde {\bf Z}}_3
\otimes{\widetilde {\bf Z}}_6$, 
${\widetilde {\bf Z}}_6\otimes{\widetilde {\bf Z}}_6$, 
${\widetilde {\bf Z}}_2\otimes{\widetilde {\bf Z}}_4$ and 
${\widetilde {\bf Z}}_4\otimes{\widetilde {\bf Z}}_4$ orbifolds \cite{Zw}. 
(In Ref \cite{Zw} 
it was found impossible to cancel all tadpoles 
in the ${\widetilde {\bf Z}}_2\otimes{\widetilde {\bf Z}}_4$ and 
${\widetilde {\bf Z}}_4\otimes{\widetilde {\bf Z}}_4$ cases, 
which would render these orientifolds inconsistent.  
We will discuss these cases in more detail in 
section \ref{anom}.)\\
$\bullet$ ``Asymmetric'' orientifolds of Type IIB on ${\bf Z}_4$, ${\bf Z}_{8}$,
${\bf Z}_8^\prime$, ${\bf Z}_{12}$ and ${\bf Z}_{12}^\prime$ orbifolds \cite{Iba}. 
(In Ref \cite{Iba} 
it was found impossible to cancel all tadpoles 
in the ${\bf Z}_4$, ${\bf Z}_{8}$,
${\bf Z}_8^\prime$, and ${\bf Z}_{12}^\prime$ cases, 
which would render these orientifolds inconsistent.  
We will discuss these cases in more detail in 
section \ref{anom}.)\\
The models of Refs \cite{Sagnotti,KS1,KS2,Zw,Iba} (also see Ref \cite{O'D}) 
have been obtained by 
generalizing the tadpole cancellation conditions of Refs \cite{GJ,DP} 
for six dimensional ``asymmetric'' orientifolds of Type IIB on
${\bf Z}_N$ ($N=3,4,6$) orbifolds.

{}The results of section \ref{IIBA} raise certain issues concerning some of the above 
examples, namely those of Refs \cite{GJ,DP,Sagnotti,KS1,KS2,Zw,Iba}. In the remainder of
this section we elaborate on these issues. We will first focus on
orientifolds of Type IIB on ${\bf Z}_N$ ($N=3,4,6$) limits of K3:
${\widetilde {\cal M}}_2=T^4/{\bf Z}_N$. These cases have been discussed 
in Refs 
\cite{GJ,DP}. We will then discuss
four dimensional cases studied in Refs 
\cite{Sagnotti,KS1,KS2,Zw,Iba}.

\subsection{6D Orientifolds}

{}Let us consider ``asymmetric'' orientifolds of Type IIB on ${\widetilde
{\cal M}}_2=T^4/{\bf Z}_N$, $N=3,4,6$. In Refs \cite{GJ,DP} the orientifold
action was assumed to be $\Omega J^\prime$ (here we use prime 
to avoid confusion with $J$ discussed
throughout this paper) where $J^\prime$ acts as follows \cite{Pol}:
({\em i}) in the untwisted sector it acts as identity; ({\em ii})
in ${\widetilde g}^k$ twisted sectors ($k\not=0$) it acts only on ground
states $\sigma_k\vert 0\rangle_L\otimes {\overline \sigma}_{N-k}\vert 0
\rangle_R$, and takes the $\sigma_k\vert 0\rangle_L\otimes 
{\overline \sigma}_{N-k}
\vert 0 \rangle_R$ ground state to the $\sigma_{N-k}\vert 0\rangle_L
\otimes {\overline \sigma}_{k}
\vert 0 \rangle_R$ ground state in the ${\widetilde g}^{N-k}$ twisted sector. 
(Here ${\widetilde g}$
is the generator of the orbifold group ${\widetilde G}\approx {\bf Z}_N$.)
Such $J^\prime$ would solve the problem pointed out in subsection C of 
section \ref{IIBA}. However, such $J^\prime$ is not a symmetry 
of the operator
product expansions (OPEs) in the ${\bf Z}_N$ ($N\not=2$) orbifold 
conformal field theory (which was pointed out in Ref \cite{Pol}).
(This can be seen by considering the action of $J^\prime$ on 
an OPE $V_k V_{N-k}\sim V_0$, where $V_k$, $V_{N-k}$, $V_0$ are
vertex operators of states in the ${\widetilde g}^k$ twisted sector, 
${\widetilde g}^{N-k}$ twisted 
sector, and untwisted sector, respectively.) That is, $J^\prime$ 
is not a symmetry of the ${\bf Z}_N$ orbifold conformal field theory. 
(Attempts to understand $J^\prime$ have also been made 
in Refs \cite{gj,blum}.)

{}Note that the models of Refs \cite{GJ,DP} are free of gravitational
and gauge anomalies. On the other hand, the fact that $J^\prime$ is not 
a symmetry of the underlying orbifold conformal
field theory raises the question about consistency of such a construction.  
In the following we will argue that a consistent description 
does exist provided that we are away from the orbifold conformal field theory 
points. 

{}For illustrative purposes we will first consider a specific example:
``asymmetric'' orientifold of Type IIB on ${\widetilde {\cal M}}_2=
T^4/{\bf Z}_3$, and then we will generalize our discussion to other 
cases. The quotient ${\widetilde {\cal M}}_2=T^4/{\bf Z}_3$ 
corresponds to an orbifold limit of K3 whose Hodge number 
$h^{1,1}=20$. The untwisted sector contributes 2 into $h^{1,1}$. 
The ${\widetilde g}$ twisted sector and its inverse ${\widetilde g}^{-1}$
twisted sector therefore
contribute 18 into $h^{1,1}$. On the other hand, there are 9 fixed points
under the action of the ${\bf Z}_3$ twist. This implies that each fixed point
contributes 2 into $h^{1,1}$. Let us now consider blowing up the orbifold
singularities. The blow-ups correspond to inserting two-spheres at fixed 
points. Each ${\bf P}^1$ has Hodge number $h^{1,1}=1$. We therefore
conclude that blowing up requires inserting 2 ${\bf P}^1$'s per fixed point.

{}Now consider orientifolding Type IIB on such a blown-up orbifold K3 (which is no 
longer singular but smooth). Let $\Omega$ be the world-sheet parity operator
corresponding to orientifolding Type IIB on a generic smooth K3. (This $\Omega$ 
is the same as that in the case of ``symmetric'' Type IIB orientifolds.) Then we have 
(at least) two inequivalent choices for orientifolding Type IIB on blown-up 
${\widetilde {\cal M}}_2=T^4/{\bf Z}_3$: ({\em i}) we can simply orientifold by
$\Omega$; ({\em ii}) we can orientifold by $\Omega J^\prime$, where $J^\prime$ 
permutes the 2 ${\bf P}^1$'s at each of the nine fixed points. In case ({\em ii}) the
action of $J^\prime$ has the same effect as that of $J^\prime$ (which mapped
the ${\widetilde g}$ twisted sector to the ${\widetilde g}^{-1}$ twisted sector) 
discussed above except that the latter was not a symmetry of the orbifold
conformal field theory, whereas the former is a symmetry of the blown-up 
(that is, smooth) K3. Note that the action of $J^\prime$ on the blown-up K3
only affects the ${\bf P}^1$'s at fixed points in the ``twisted'' sectors but has no 
effect on the ``untwisted'' sector. (The action of $J^\prime$ on blown-up
orbifold K3 was recognized in Ref \cite{Pol} from a slightly different, although, 
we believe, equivalent viewpoint.)

{}Let us consider case ({\em ii}) in more detail. (We will discuss case ({\em i}) in
the next subsection.) This would correspond to
orientifolds discussed in Refs \cite{GJ,DP} for blown-up K3. Since at the
orbifold conformal field theory point $J^\prime$ is not a symmetry of the theory, we can
view the orientifolds of Refs \cite{GJ,DP} in the context
of blown-up K3 as discussed above. This construction, however, does not 
correspond to free-field conformal field theory approach. Any analyses along the
lines of sections \ref{IIB} and \ref{IIBA}, therefore, become exceedingly difficult. 
On the other hand, the action
of $\Omega J^\prime$ maps states in the ${\widetilde g}$ ``twisted'' sector to the 
${\widetilde g}^{-1}$ ``twisted'' sector, so these ``twisted'' sector states should not
contribute into the Klein bottle amplitude. Also, here we do not expect 
any additional states
coming from the $\Omega J^\prime {\widetilde g}^k$
($k=1,2$) twisted sectors as the latter
are not well defined. That is, we expect that such sectors are simply absent in
such an orientifold construction. (We will give more evidence
supporting this conclusion from the F-theory viewpoint in section \ref{FA}. 
Absence of $\Omega J^\prime {\widetilde g}^k$ 
``twisted'' sectors in orientifolds of Refs 
\cite{GJ,DP} was also recognized in Ref \cite{blum} from a somewhat different
viewpoint.) Thus, we expect the ``naive'' tadpole cancellation conditions derived
in Refs \cite{GJ,DP} to produce models free of gravitational and gauge anomalies
without adding any extra states. On the other hand, the spectra of the models of Refs \cite{GJ,DP} were worked out at the orbifold conformal field theory points. These 
spectra have certain enhanced gauge symmetries.  
Since the above construction involves
blowing up the orbifold singularities, we, at least naively, might expect that these gauge
symmetries might be reduced after blow-ups are performed.
There are, however, certain quantitative features that must be robust:
first, the number of tensor multiplets $T$ and the number of hypermultiplets $H_c$ 
(the latter are neutral) in 
the closed string sectors must be the same everywhere in the moduli space. Also,
we always have $T+H_c=21$. On the other hand, in the open string sectors the
number of vector multiplets ${\widetilde V}$ and the number of hypermultiplets 
${\widetilde H}_o$ must 
obey the rule that ${\widetilde H}_o-{\widetilde V}$ is the same everywhere 
in the moduli space. This follows
from the fact that in six dimensions there is no superpotential, and Higgsing cannot 
affect ${\widetilde H}_o-{\widetilde V}$. 

{}The above considerations lead us to the conclusion that the ``asymmetric'' 
$\Omega J^\prime$ orientifold of Type IIB on {\em blown-up} ${\widetilde 
{\cal M}}_2=T^4/{\bf Z}_3$ as described above has $T=10$, $H_c=11$, and
${\widetilde H}_o-{\widetilde V}=-28$ as can be deduced from the ``naive'' 
spectrum presented in Refs \cite{GJ,DP}. We can push this a bit further if we 
compactify this model on $T^2$ to four dimensions. Then after Higgsing 
we can deduce the number of $U(1)$ vector multiplets $V$ and the number of
neutral hypermultiplets $H_o$ descending from six dimensions ({\em i.e.}, not
taking into account the extra 2 vector multiplets coming from $T^2$). From the
spectrum given in Refs \cite{GJ,DP}, we obtain the following data: $V=8$ and
$H_o=4$. Now assuming that there is a heterotic dual of this model (which 
would ultimately have to be non-perturbative due to the fact that $T\not=1$), we can
further use Type IIA-heterotic duality to deduce the Calabi-Yau three-fold 
on which Type IIA would produce this spectrum. The Hodge numbers of this
Calabi-Yau three-fold would have to be given by $h^{1,1}=T+V+2=20$,
$h^{2,1}=H-1=14$, where $H=H_c+H_o$.
Such a Calabi-Yau three-fold does exist: it is one of the 
Voisin-Borcea orbifolds discussed in section \ref{F}. (This Voisin-Borcea orbifold has
$(r,a,\delta)=(11,9,0)$. See section \ref{FA} for notation.) Since it is an elliptically
fibered Calabi-Yau three-fold, we expect that Type IIA on this three-fold is dual to
F-theory on the same three-fold further compactified on $T^2$. This in turn implies that
there must exist F-theory dual of the above orientifold model directly in six dimensions
(that is, F-theory compactified on the Calabi-Yau three-fold with Hodge numbers
$(h^{1,1},h^{2,1})=(20,14)$ must be dual to the above orientifold model). In section 
\ref{FA} and appendix \ref{VoBo} we will give an explicit map of this orientifold 
model to F-theory.  

{}We can generalize the above discussion to the ``asymmetric'' orientifolds
of Type IIB on $T^4/{\bf Z}_4$ and $T^4/{\bf Z}_6$ presented in Refs 
\cite{GJ,DP}. In the ${\bf Z}_4$ case we have the ${\widetilde g}$ and 
${\widetilde g}^3$ twisted sectors with 4 fixed points in each, plus the
${\widetilde g}^2$ twisted sector with 10 fixed points\footnote{Here we use
the terminology ``fixed point'' loosely. For instance, 10 fixed points in the 
${\widetilde g}^2$ twisted sector are ``linear combinations'' of the original 
16 fixed points in the ${\bf Z}_2$ twisted sector that are invariant under the
${\bf Z}_4$ twist.}. After blowing-up, 
the ${\widetilde g}$ and ${\widetilde g}^3$ ``twisted'' sectors together 
contain four fixed points with 2 ${\bf P}^1$'s per fixed point. The 
${\widetilde g}^2$ ``twisted'' sector (which is left-right symmetric for it is a 
${\bf Z}_2$ twisted sector) contains 10 fixed points with only 1 ${\bf P}^1$
per fixed point. Now consider $\Omega J^\prime$ orientifold of Type IIB on 
this blown-up orbifold with the following action of $J^\prime$: it acts as identity
in the ``untwisted'' and ${\bf Z}_2$ ``twisted'' sectors; it permutes 2 ${\bf P}^1$'s
at each fixed point in the ${\bf Z}_4$ ``twisted'' sectors. Then we have $T=5$, 
$H_c=16$. 
(Here we are closely following the discussion of Ref \cite{gj}.) From the 
corresponding spectrum given in Refs \cite{GJ,DP} we deduce that 
${\widetilde H}_o-{\widetilde V}=112$. In fact, one can Higgs the gauge group 
completely in this model. Thus, we have $V=0$, $H_o=112$, and $H=128$. 
Upon further
compactification on $T^2$ the corresponding Type IIA dual would have to be given
by a compactification on the Calabi-Yau three-fold with Hodge numbers
$(h^{1,1},h^{2,1})=(7,127)$. Here we note that this is not a Voisin-Borcea orbifold.

{}One can consider the ${\bf Z}_6$ case similarly. After blowing up the ${\bf Z}_6$
``twisted'' sectors together contain 1 fixed point with 2 ${\bf P}^1$'s per fixed point,
the ${\bf Z}_3$ ``twisted'' sectors together contain 5 fixed point with 2 ${\bf P}^1$'s
per fixed point, and the ${\bf Z}_2$ ``twisted'' sector contains 6 fixed points with
only 1 ${\bf P}^1$ per fixed point. Now consider $\Omega J^\prime$ orientifold of 
Type IIB on 
this blown-up orbifold with the following action of $J^\prime$: it acts as identity
in the ``untwisted'' and ${\bf Z}_2$ ``twisted'' sectors; it permutes 2 ${\bf P}^1$'s
at each fixed point in the ${\bf Z}_6$ and ${\bf Z}_3$ ``twisted'' sectors. Then we
have $T=7$, $H_c=14$. From the 
corresponding spectrum given in Refs \cite{GJ,DP} we deduce that 
${\widetilde H}_o-{\widetilde V}=56$. In fact, one can Higgs the gauge group 
completely in this model. Thus, we have $V=0$, $H_o=56$, and $H=70$.
Upon further
compactification on $T^2$ the corresponding Type IIA dual would have to be given
by a compactification on the Calabi-Yau three-fold with Hodge numbers
$(h^{1,1},h^{2,1})=(9,69)$. Here we note that, just as in the ${\bf Z}_4$ case,
this is not a Voisin-Borcea orbifold.

{}At first it might appear surprising that the Type IIA duals in the ${\bf Z}_4$ and 
${\bf Z}_6$
cases would have to correspond to compactifications on Calabi-Yau three-folds 
that are not among the Voisin-Borcea orbifolds since from the map of Refs \cite{sen}
between the orientifold and F-theory descriptions (which we discussed in section 
\ref{F}) one expects the F-theory duals of Type IIB orientifolds to be elliptically fibered
Calabi-Yau three-folds of the Voisin-Borcea type. This is, however, correct only if
the corresponding three-fold on the F-theory side is non-singular (or can be blown 
up to a smooth Calabi-Yau three-fold). We will explain this point in detail in section
\ref{FA}. Here for completeness we note that the Type IIA dual of the ${\bf Z}_2$ 
model of Refs \cite{PS,GP} is given by a compactification on the Calabi-Yau 
three-fold with Hodge numbers $(h^{1,1},h^{2,1})=(3,243)$. This is not among 
Voisin-Borcea orbifolds either. We will put off the discussion of this issue until
section \ref{FA} and turn to Type I compactifications on K3 in the next subsection.

\subsection{Type I on K3}

{}In the previous subsection we pointed out two possibilities for orientifolding
Type IIB on (blown-up) ${\widetilde {\cal M}}_2=T^4/{\bf Z}_N$ ($N=3,4,6$). 
There we discussed $\Omega J^\prime$ orientifolds in detail. In this subsection 
we will consider $\Omega$ orientifolds. These always contain only one tensor 
multiplet, and are equivalent to Type I compactifications on K3 (which in this 
case is blown-up ${\widetilde {\cal M}}_2$). Just as in ``symmetric'' orientifolds
of Type IIB on ${\cal M}_2=T^4/{\bf Z}_N$ ($N=3,4,6$) (in which case 
orientifolding amounts to gauging $\Omega$), here we expect extra sectors,
namely, $\Omega {\widetilde g}^k$ sectors ($k=1,\dots,N-1$) to contribute into 
the massless spectrum. Unless ${\widetilde g}^{2k}=1$, these sectors are 
non-perturbative from the orientifold viewpoint in complete parallel with our 
discussion in section \ref{F}. One way to see that these sectors are important
is as follows. 

{}Type I on K3 is expected to be dual to heterotic on K3. 
For example, consider a perturbative heterotic compactification 
on the ${\bf Z}_3$ orbifold limit of K3. The twisted sectors in such a model
contribute states charged under the unbroken gauge group 
(which is a subgroup of $SO(32)$). On the other hand, in the corresponding 
Type I model all the matter charged under the gauge group (which is the same as 
on the heterotic side) comes from the 99 open string sector, that is, from the sector
corresponding to open strings stretched between D9-branes. (The tadpole 
cancellation conditions in the Type I model imply that there are no D5-branes
in the compactification of Type I on the ${\bf Z}_3$ orbifold limit of K3 \cite{GJ,DP}). 
The 99 
open string sector gives rise to the same gauge group and the matter content
as the untwisted sector on the heterotic side (provided that the gauge bundle, 
that is, the action of the ${\bf Z}_3$ twist on the Chan-Paton charges on the Type I
side and on the ${\mbox{Spin}}(32)/{\bf Z}_2$ lattice on the heterotic side is the 
same). Thus, perturbative orientifold approach to Type I misses the charged matter
fields that arise in the twisted sectors of the heterotic dual. These states are 
necessary for cancellation of (gravitational and gauge) anomalies in six dimensions. 
We, therefore, conclude that the orientifold approach is inadequate
in this case. In section 
\ref{FA} we will discuss the F-theory description of Type I on K3 which will enable us
to understand the non-perturbative (from the orientifold viewpoint) origin of these 
extra states.   
 
\subsection{4D Orientifolds}

{}In this section we will consider four dimensional ``asymmetric'' orientifolds
of Type IIB compactified on ${\widetilde {\cal M}}_3=T^6/{\widetilde G}$ (which 
we will assume to have $SU(3)$ holonomy). In Type IIB compactifications
on ${\widetilde {\cal M}}_3$ orbifolds we have two (possible) 
types of twisted sectors:\\
$\bullet$ ({\em i}) ${\widetilde g}_a$ twisted sectors where in the diagonal basis 
${\widetilde g}_a={\mbox{diag}}(\rho_a,\rho_a^{-1},1)$;\\ 
$\bullet$ ({\em ii}) 
${\widetilde g}_a$ twisted sectors where in the diagonal basis 
${\widetilde g}_a={\mbox{diag}}(\rho_a,\rho_a^\prime,
(\rho_a\rho_a^\prime)^{-1})$ with $\rho_a,\rho_a^\prime,
(\rho_a\rho_a^\prime)\not=1$.\\
The first type of sectors may or may not be present in a given 
${\widetilde {\cal M}}_3$ orbifold. The second type of sectors is always 
present in {\em Abelian} ${\widetilde {\cal M}}_3$ orbifolds with $SU(3)$ holonomy
(except for the ${\widetilde {\bf Z}}_2\otimes {\widetilde {\bf Z}}_2$ case).

{}Let us first consider
{\em non-Abelian} orbifolds. In some non-Abelian orbifolds (such as 
$D_N$ orbifolds, $N=3,4,6$) there are no twisted sectors of type ({\em ii}).
So naively one might hope that the situation in such cases will be similar 
to the six dimensional orientifolds considered in subsection A: we could {\em a priori}
attempt to include $J^\prime$ in the orientifold projection. However, as recently
pointed out in Ref \cite{zura}, additional complications arise in non-Abelian cases. 
Here we will review the discussion in Ref \cite{zura}. 

{}Instead of being most general, we will focus
on the case of $D_N$ orbifolds ($N=3,4,6$).
(The generalization to other non-Abelian cases should be clear.) Thus, consider Type IIB on ${\widetilde {\cal M}}_3=(T^2\otimes T^2\otimes T^2)/{\widetilde G}$ 
where ${\widetilde G}\approx D_N$ (non-Abelian
dihedral group), and the action of ${\widetilde G}$ on the complex coordinates $z_i$ ($i=1,2,3$)
on ${\widetilde {\cal M}}_3$ is given by ($\omega=\exp(2\pi i/N)$):
\begin{eqnarray}
 &&{\widetilde g}z_1=z_1~,~~~{\widetilde g}z_2=\omega z_2~,~~~
 {\widetilde g}z_3=\omega^{-1} z_3~,\\
 &&rz_1=-z_1~,~~~rz_2=z_3~,~~~rz_3=z_2~,
\end{eqnarray}
where ${\widetilde g},r$ are the generators of $D_N$. Note that ${\widetilde g}$
and $r$ do not commute:
$r{\widetilde g}={\widetilde g}^{-1}r$.

{}Now consider the $\Omega J$ orientifold of this theory where $Jz_i=-z_i$. 
The orientifold group is ${\cal O}=\{{\widetilde g}^k,r{\widetilde g}^k,
\Omega J{\widetilde g}^k,\Omega Jr{\widetilde g}^k\vert
k=0,\dots,N-1\}$. Note that $(Jr{\widetilde g}^k)^2=1$, and
the set of points in ${\widetilde {\cal M}}_3$ fixed under the action of $Jr{\widetilde g}^k$
has real dimension 
two. This implies that there are $N$ kinds of orientifold 7-planes corresponding to the
elements $\Omega Jr{\widetilde g}^k$. Note, however, that due to non-commutativity between 
${\widetilde g}$ and $r$ (and, therefore, between different $Jr{\widetilde g}^k$), 
these orientifold 7-planes
(as well as the corresponding D7-branes) are mutually non-local. This implies that
this orientifold does not have a world-sheet description. In this case we appear to
have no choice but to invoke the F-theory description via the map of Refs \cite{sen}.
Note that appearance of mutually non-local D-branes is a generic feature of orientifolds
of Type IIB on non-Abelian toroidal orbifolds.

{}Next, let us consider {\em Abelian} ${\widetilde {\cal M}}_3$ orbifolds 
(with $SU(3)$ holonomy). As we 
already mentioned above, twisted sectors of type ({\em ii}) are always present
in such cases. After blowing up we have
one ${\bf P}^1$ per fixed point in such sectors. This implies that in these 
sectors the action of $J^\prime$ (that acts as identity in untwisted sectors, 
and maps the ${\widetilde g}_a$ ``twisted'' sector to the ${\widetilde g}_a^{-1}$ 
``twisted'' sector) would not be well defined. That is, in type ({\em ii}) sectors 
we can orientifold by $\Omega$ but not by $\Omega J^\prime$ 
(after blow-ups). We will give evidence for correctness of this statement from the
F-theory viewpoint in section \ref{FA}. 
We therefore (at least {\em a priori}) expect additional 
non-perturbative (from the orientifold viewpoint)
contributions coming from the $\Omega {\widetilde g}_a$ sectors. Since such 
${\widetilde g}_a$ twisted sectors are always present 
in Abelian ${\widetilde {\cal M}}_3$ orbifolds with $SU(3)$ holonomy (except for the
${\widetilde {\bf Z}}_2\otimes {\widetilde {\bf Z}}_2$ case), we conclude that
orientifolds of Type IIB on Abelian ${\widetilde {\cal M}}_3$ orbifolds (other than
the ${\widetilde {\bf Z}}_2\otimes {\widetilde {\bf Z}}_2$ orbifold),
at least naively, are always
expected to receive non-perturbative contributions from the corresponding
$\Omega {\widetilde g}_a$ sectors.

{}Let us now consider twisted sectors of type ({\em i}). These have 
the structure given by twisted sectors of ${\widetilde {\cal M}}_2\otimes T^2$ 
(where ${\widetilde {\cal M}}_2$ is an orbifold limit of K3) projected to 
${\widetilde G}$ invariant states. Thus, in the ``twisted'' sectors (other than the
${\bf Z}_2$ ``twisted'' sectors) descending from those in the 
${\widetilde {\cal M}}_2$ orbifold after the appropriate blow-ups
we have (at least) two different choices for the orientifold projection: $\Omega$
and $\Omega J^\prime$. Here $J^\prime$ acts in the same way as in six 
dimensional orientifolds discussed in subsection B. In the first case we 
{\em a priori} expect additional non-perturbative (from the orientifold viewpoint)
contributions coming from the $\Omega {\widetilde g}_a$ sectors. In the second
case such contributions would be absent just as in the six dimensional models
discussed in subsection A. However, it is not difficult to see that the $\Omega J^\prime$
orientifold projection in the twisted sectors of type ({\em i}) is not consistent with
the choice of the $\Omega$ projection in the twisted sectors of type ({\em ii}). To
see this define the operator $J^\prime$ as follows:
\begin{equation}
 J^\prime\vert {\widetilde g}_a\rangle=\vert {\widetilde g}^{\epsilon_a}_a\rangle~,
\end{equation}
where $\epsilon_a=\pm1$. (Note that in ${\bf Z}_2$ twisted sectors
both choices $\epsilon_a=\pm1$ are equivalent.)
We must require that $\epsilon_a=+1$ in the 
${\widetilde g}_a$ twisted sectors of type ({\em ii}). Let 
${\widetilde g}_c={\widetilde g}_a {\widetilde g}_b$ where $a\not=b\not=c\not=a$. 
To have a consistent action
of $J^\prime$, we must assume that
\begin{equation}
 J^\prime\vert {\widetilde g}_a {\widetilde g}_b\rangle=
 \vert {\widetilde g}^{\epsilon_a}_a {\widetilde g}_b^{\epsilon_b}\rangle~.
\end{equation}
This, in particular, implies that $\epsilon_c=\epsilon_a=\epsilon_b$. Note that $a,b,c$
are arbitrary here, so we conclude that all $\epsilon_a=+1$ if $J^\prime$ acts trivially
in the twisted sectors of type ({\em ii})\footnote{Here we have used the fact that the orbifold
group is {\em Abelian}.}. In section \ref{anom} we will present additional evidence that 
the above constraint is indeed necessary.

{}Thus, in four dimensions
``asymmetric'' Type IIB orientifolds do not seem to provide us
with a ``free lunch'' either. In section \ref{het}, however, using Type I-heterotic 
duality as a guiding principle we will be able to circumvent 
difficulties with these additional states in ``asymmetric'' orientifolds of Type IIB
on certain Abelian ${\widetilde {\cal M}}_3$ orbifolds, which in turn will lead us to
the construction of chiral ${\cal N}=1$ vacua in four dimensions that are non-perturbative
from the heterotic viewpoint. In other cases we can map the
corresponding orientifold models to F-theory compactifications on Calabi-Yau
four-folds which provide additional (albeit, sometimes limited) insight into the 
structure of four dimensional orientifolds. We will discuss this map in section 
\ref{FA}. We will see that in most cases one has to be careful as 
non-perturbative contributions are crucial. Examples of such models
will be discussed in section \ref{anom}.  

{}Let us summarize the above discussion. In four dimensional orientifolds of
Type IIB on Abelian ${\widetilde {\cal M}}_3$ orbifolds the orientifold projection
must be $\Omega J$ (where $J$ is a geometric symmetry of  ${\widetilde {\cal M}}_3$)
in {\em all} twisted sectors. This, in particular, implies that all orbifold singularities, except
for the ${\bf Z}_2$ singularities, must be blown up (or else $\Omega J$ is not a symmetry
of the theory). If $J=1$, then the orientifold corresponds to a Type I compactification
on blown up ${\widetilde {\cal M}}_3$ (${\bf Z}_2$ singularities need not be blown up).  
We, therefore,
expect non-perturbative
states to appear in the sectors of the form $\Omega  {\widetilde g}_a$ (where
${\widetilde g}^2_a\not=1$). The ``naive'' tadpole calculation (which is 
a generalization of the corresponding calculation in six dimensional cases
of orientifolds of Type IIB on $T^4/{\bf Z}_N$) is performed (at the orbifold
conformal field theory point) as though the orientifold projection is accompanied
by $J^\prime$ (which is {\em not} a symmetry of the underlying orbifold
conformal field theory). This, in particular, implies that in the ``naive'' tadpole
calculation there are no contributions coming from the ${\widetilde g}_a$ twisted
sectors with ${\widetilde g}^2_a\not=1$. Note
that in the case of the $\Omega$ orientifold of Type IIB on 
(blown up) ${\widetilde {\cal M}}_3$, the massless closed string sector
states are given by $h^{1,1}+h^{2,1}$ {\em chiral} neutral 
supermultiplets\footnote{In the twisted sectors of type ({\em ii}) 
the ``naive'' orientifold approach would give one chiral multiplet 
(for each point fixed under ${\widetilde g}_a$) 
which is a linear combination of the corresponding chiral multiplets coming from the
${\widetilde g}_a$ and ${\widetilde g}_a^{-1}$ twisted sectors. However, this identification
of states is not completely precise. The correct projection in this case would be the
$\Omega$ projection in ${\widetilde g}_a$ plus ${\widetilde g}_a^{-1}$ twisted sectors
after {\em blowing up} the orbifold singularities. For each fixed point we then get a 2-sphere 
${\bf P}^1$. The orientifold projection here is the same as for a smooth Calabi-Yau three-fold,
{\em i.e.}, that of the Type I compactification on blown-up ${\widetilde {\cal M}}_3$. Each
${\bf P}^1$ gives rise to a chiral multiplet. So the counting of states in this picture is the same 
as in the ``naive'' orientifold approach albeit the vertex operators may not be the same.
As to the twisted
sectors of type ({\em i}), the ``naive'' orientifold approach would give
$h^{1,1}_a+{1\over 2} h^{2,1}_a$ chiral multiplets and ${1\over 2} h^{2,1}_a$
vector multiplets, where $(h^{1,1}_a, h^{2,1}_a)$ is a combined contribution of the
${\widetilde g}_a$ and ${\widetilde g}^{-1}_a$ twisted sectors (assuming
${\widetilde g}^2_a\not=1$) into the Hodge numbers of ${\widetilde {\cal M}}_3$.
(Note that both $h^{1,1}_a$ and $h^{2,1}_a$ are even for such twisted sectors.)
This is clearly different from the correct answer which is
$h^{1,1}_a+h^{2,1}_a$ chiral multiplets and no vector multiplets. The discrepancy
is due to the incorrect $\Omega J^\prime$ projection in the
``naive'' orientifold approach.}, where  
$(h^{1,1},h^{2,1})$ are the Hodge numbers of ${\widetilde {\cal M}}_3$.     

{}Before we conclude this section, the following comments are in order.
Both ``symmetric'' ${\cal M}_2$ and ``asymmetric'' ${\widetilde {\cal M}}_2$ 
orbifolds after the appropriate blow-ups give rise to smooth K3 surfaces.
This implies that these two cases can be treated in the same way for both
$\Omega$ and $\Omega J^\prime$ orientifold projections. On the other hand,
``symmetric'' ${\cal M}_3$ and ``asymmetric'' ${\widetilde {\cal M}}_3$ 
orbifolds are mirror pairs, so their orientifolds (generically) are not the same.  
  
\section{Map to F-Theory}\label{FA}

{}In this section we discuss ``asymmetric'' Type IIB orientifolds from the F-theory
viewpoint. We will first consider six dimensional orientifolds, and then generalize 
our discussion to four dimensional cases.

\subsection{Voisin-Borcea Orbifolds}

{}From the discussion in section \ref{F} it is clear that F-theory realizations of orientifold
vacua in six dimensions are related to F-theory compactifications on Voisin-Borcea
orbifolds. We will therefore review some facts about these Calabi-Yau three-folds which
will prove useful later. Let ${\cal W}_2$ be a K3 surface (which is not necessarily an 
orbifold) which admits an involution $J$ such that it reverses the sign of the holomorphic
two-form $dz_1\wedge dz_2$ on ${\cal W}_2$. Consider the following quotient:
\begin{equation}
 {\cal Y}_3= (T^2\otimes {\cal W}_2)/Y~,
\end{equation}
where $Y=\{1,S\}\approx {\bf Z}_2$, and $S$ acts as $Sz_0=-z_0$ on $T^2$ 
($z_0$ being a complex coordinate on $T^2$), and as $J$ on ${\cal W}_2$. This
quotient is a Calabi-Yau three-fold with $SU(3)$ holonomy which is elliptically 
fibered over the base ${\cal B}_2={\cal W}_2/B$, where 
$B=\{1,J\}\approx{\bf Z}_2$. 

{}Nikulin gave a classification \cite{Nik} 
of possible involutions of K3 surfaces in terms of three invariants $(r,a,\delta)$
(for a physicist's discussion, see, {\em e.g.}, \cite{paul,MV}).
The result of this classification is plotted in Fig.2 according to the values of $r$ 
and $a$. The open and closed circles correspond to the cases with $\delta=0$
and $\delta=1$, respectively. (The cases denoted by ``$\otimes$'' are outside
of Nikulin's classification, and we will discuss them shortly.) In the case 
$(r,a,\delta)=(10,10,0)$ the base ${\cal B}_2$ is an Enriques surface, and the 
corresponding ${\cal Y}_3$ has Hodge numbers $(h^{1,1},h^{2,1})=(11,11)$.
In all the other cases the Hodge numbers are given by:
\begin{eqnarray}\label{hodge1}
 &&h^{1,1}=5+3r-2a~,\\
 \label{hodge2}
 &&h^{2,1}=65-3r-2a~.
\end{eqnarray} 

{}For $(r,a,\delta)=(10,10,0)$ the ${\bf Z}_2$ twist $S$ is freely acting 
(that is, it has no fixed points). For $(r,a,\delta)=(10,8,0)$ the fixed point set
of $S$ consists of two curves of genus 1. The base ${\cal B}_2$ in this case 
is ${\bf P}^2$ blown up at 9 points. In all the other cases the fixed point set
of $S$ consists of one curve of genus $g$ plus $k$ rational curves where
\begin{eqnarray}
 &&g={1\over 2}(22-r-a)~,\\
 &&k={1\over 2}(r-a)~.
\end{eqnarray}

{}Note that except for the cases with $a=22-r$, $r=11,\dots,20$, the mirror pair
of ${\cal Y}_3$ is given by the Voisin-Borcea orbifold 
${\widetilde {\cal Y}}_3$ with ${\widetilde r}=20-r$, ${\widetilde a}=a$.
Under the mirror transform we have: ${\widetilde g}=f$, ${\widetilde f}=g$, where
$f=k+1$. 

{}In the cases $a=22-r$, $r=11,\dots,20$, the mirror would have to have
${\widetilde r}=20-r$ and ${\widetilde a}=a={\widetilde r}+2$, where 
${\widetilde r}=0,\dots,9$. We have depicted these cases in Fig.1 using 
``$\otimes$'' symbol. In 
particular, we have plotted cases with $a=r+2$, $r=0,\dots,10$. (The reason for 
including $r=10$ will become clear in a moment.) The Hodge numbers for these 
cases are still given by Eqs (\ref{hodge1}) and (\ref{hodge2}) (which follows from
their definition as mirror pairs of the cases with $a=22-r$, $r=11,\dots,20$). (This is
true for $a=r+2$, $r=0,\dots,9$. Extrapolation to $r=10$ is motivated by 
the fact that
in this case we get $(h^{1,1},h^{2,1})=(11,11)$ which is the same as for 
$(r,a,\delta)=(10,10,0)$.) The question that
arises in the above extrapolation of mirror symmetry for Voisin-Borcea orbifolds
is whether the corresponding Calabi-Yau three-folds (denoted by ``$\otimes$''
symbol in Fig.1) indeed exist. To answer this question we will first consider
compactifications of F-theory on known Voisin-Borcea orbifolds. 

{}F-theory compactification on ${\cal Y}_3$ with $(r,a,\delta)\not=
(10,10,0)$ or $(10,8,0)$ gives rise to the following massless spectrum in six 
dimensions. The number of tensor multiplets is $T=r-1$. The number of neutral 
hypermultiplets is $H=22-r$. The gauge group is $SO(8)\otimes SO(8)^k$.
There are $g$ adjoint hypermultiplets of the first $SO(8)$. There are no 
hypermultiplets charged under the other $k$ $SO(8)$'s. Under mirror symmetry
$g$ and $f=k+1$ are interchanged. Thus, the vector multiplets in the adjoint
of $SO(8)^k$ are traded for $g-1$ hypermultiplets in the adjoint of the first 
$SO(8)$. That is, gauge symmetry turns into global symmetry and {\em vice-versa}.
We can push this a bit further to understand what F-theory compactifications on
Calabi-Yau three-folds with $a=r+2$, $r=1,\dots,10$, would give\footnote{Here we 
must exclude the case with $r=0$, $a=2$ for the F-theory prediction would be $T=-1$
tensor multiplets. This Calabi-Yau three-fold, as we will argue in a moment, does exist, 
but it is singular and F-theory compactification on such a space does not appear to
have a local Lagrangian description. However, 
an extremal transition \cite{MV}
between this Calabi-Yau 
three-fold and another Voisin-Borcea orbifold could lead to a phase transition into
a well defined vacuum.}. The number of tensor multiplets is $T=r-1$. There are 
$H=22-r$ neutral hypermultiplets. In addition there are $g=10-r$ hypermultiplets
transforming as adjoints under a global $SO(8)$ symmetry. There are no gauge bosons,
however. It is not difficult to verify that this massless spectrum is free of gravitational
anomalies in six dimensions.

{}Let us try to understand these examples better. For $a=r+2$, $r=0,\dots,10$, the
Hodge numbers are given by $(h^{1,1},h^{2,1})=(r+1,61-5r)$. Let us use the above 
spectrum to see what the four dimensional 
Type IIA duals of F-theory compactifications on these three-folds
would be upon further compactification on $T^2$. It is not difficult to check
that the Type IIA duals would have to
correspond to compactifications on Calabi-Yau three-folds with Hodge numbers
$({\hat h}^{1,1},{\hat h}^{2,1})=(r+1,301-29r)$, $r=1,\dots,10$. These two sets of
Hodge numbers coincide only for $r=10$ (in which case we have a smooth 
Calabi-Yau three-fold). For all the other values of $r$ they differ, however. 
At first this might appear surprising as F-theory compactified on a Calabi-Yau
three-fold times $T^2$ is expected to be dual to Type IIA compactified on the
same Calabi-Yau three-fold. This is correct if the three-fold on the F-theory side
is non-singular (or can be blown up to a smooth Calabi-Yau three-fold). If, however,
the three-fold on the F-theory side is singular (and cannot be blown up to a smooth one)
this need not be the case. From these considerations we get a hint that the three-folds
with Hodge numbers $(h^{1,1},h^{2,1})=(r+1,61-5r)$ ($r=1,\dots,9$), if they exist, 
should be singular. On the other hand, existence of these Calabi-Yau three-folds
would prompt us to assume that there must exist (smooth) Calabi-Yau three-folds
with Hodge numbers $({\hat h}^{1,1},{\hat h}^{2,1})=(r+1,301-29r)$ ($r=1,\dots,9$). 
Moreover, we would be led to the following statement:
\begin{eqnarray}
 &&{\mbox{F-theory on ${\cal Y}_3$ with $(h^{1,1},h^{2,1})=(r+1,61-5r)$ is equivalent
 to}}\nonumber\\
 &&{\mbox{F-theory on ${\widehat{\cal Y}}_3$ with $({\hat h}^{1,1},{\hat h}^{2,1})
 =(r+1,301-29r)$  ($r=1,\dots,9$)}}~.
\end{eqnarray}
In the following we present evidence for correctness of these assumptions. Note that 
for $r=2$ we get $({\hat h}^{1,1},{\hat h}^{2,1})=(3,243)$, which is known to exist. For
$r=6$ we get $({\hat h}^{1,1},{\hat h}^{2,1})=(7,127)$. This Calabi-Yau three-fold has 
been recently constructed in Ref \cite{BG}. Also, in Ref \cite{KST} 
it was shown that orientifolds of Type IIB on $T^4/{\bf Z}_4$ and $T^4/{\bf Z}_6$
are on the same moduli as orientifolds of Type IIB on $T^4/{\bf Z}_2$ with non-zero
NS-NS antisymmetric tensor backgrounds. The latter orientifolds do not involve 
$J^\prime$ (see section \ref{other} for details) 
in the orientifold projection. Thus, they can be explicitly constructed at the
orbifold conformal field theory points. As pointed out in Ref \cite{KST} (just as
in the case of the original orientifolds of Type IIB on $T^4/{\bf Z}_4$ and $T^4/{\bf Z}_6$
\cite{gj}), their F-theory duals must correspond to compactifications on elliptic Calabi-Yau
three-folds with Hodge numbers $({\hat h}^{1,1},{\hat h}^{2,1})=(7,127)$ and
$({\hat h}^{1,1},{\hat h}^{2,1})=(9,69)$, respectively.
  
{}First, let us consider the case $r=0$, $a=2$. The Hodge numbers are
$(h^{1,1},h^{2,1})=(1,61)$. It is not difficult to check that the ``symmetric'' 
$T^6/G$ orbifold with 
$G\approx{\widetilde {\bf Z}}_2\otimes {\widetilde {\bf Z}}_4$
(see section \ref{class} for details) and no discrete torsion 
has these Hodge numbers. This is not a geometric orbifold, but it can be
constructed as a conformal field theory orbifold, and the corresponding 
Calabi-Yau three-fold should exist. (Here we note that this is a mirror manifold
of the ``asymmetric'' 
$T^6/{\widetilde G}$ orbifold with 
${\widetilde G}\approx{\widetilde {\bf Z}}_2\otimes {\widetilde {\bf Z}}_4$
and no discrete torsion which corresponds to $r=20$, $a=2$, and has Hodge 
numbers $(h^{1,1},h^{2,1})=(61,1)$.) This three-fold, however, would be singular
as the K{\"a}hler moduli required for blow-ups are missing\footnote{Here we note that
in this particular case {\em a priori} we {\em cannot} argue for existence of the 
corresponding three-fold ${\widehat{\cal Y}}_3$ with the Hodge numbers
$({\hat h}^{1,1},{\hat h}^{2,1})=(1,301)$.}.

{}Next, consider the case $r=2$, $a=4$. We have $(h^{1,1},h^{2,1})=(3,51)$
and $({\hat h}^{1,1},{\hat h}^{2,1})=(3,243)$. The first of these Calabi-Yau
three-folds is nothing but the orbifold $T^6/({\widetilde {\bf Z}}_2\otimes
{\widetilde {\bf Z}}_2)$ with discrete torsion. (The 
$T^6/({\widetilde {\bf Z}}_2\otimes
{\widetilde {\bf Z}}_2)$ orbifold without discrete torsion has Hodge numbers
$(h^{1,1},h^{2,1})=(51,3)$.) This Calabi-Yau is indeed singular \cite{dis}.
On the other hand, using the map of Refs \cite{sen} between F-theory and
orientifolds (discussed in section \ref{F}) it is not difficult to see that F-theory
on $T^6/({\widetilde {\bf Z}}_2\otimes
{\widetilde {\bf Z}}_2)$ with discrete torsion should be dual to an orientifold
which is T-dual of the ${\bf Z}_2$ model of Refs \cite{PS,GP} 
(see the next subsection 
for details). 
On the other hand, upon further compactification on $T^2$ the latter model is
dual to Type IIA on the Calabi-Yau three-fold with Hodge numbers
$({\hat h}^{1,1},{\hat h}^{2,1})=(3,243)$ \cite{berkooz}. This supports our 
assumption that F-theory on a (singular) Calabi-Yau threefold with Hodge 
numbers $(h^{1,1},h^{2,1})=(r+1,61-5r)$ ($r=1,\dots,9$) is the same as 
F-theory on a (smooth) Calabi-Yau threefold with Hodge 
numbers $({\hat h}^{1,1},{\hat h}^{2,1})=(r+1,301-29r)$ ($r=1,\dots,9$). 

{}Note that for $r=6$ and $r=8$ we have $({\hat h}^{1,1},{\hat h}^{2,1})=(7,127)$
and $({\hat h}^{1,1},{\hat h}^{2,1})=(9,69)$, respectively. These are the Hodge 
numbers of Calabi-Yau three-folds compactification on which would be dual
to the ${\bf Z}_4$ and ${\bf Z}_6$ orientifold models of Refs \cite{GJ,DP} 
further compactified on $T^2$ \cite{gj}. Then we should be able to map the
these ${\bf Z}_4$ and ${\bf Z}_6$ orientifold models to F-theory on
Calabi-Yau three-folds with Hodge numbers $(h^{1,1},h^{2,1})=(7,31)$ and
$(h^{1,1},h^{2,1})=(9,21)$, respectively. We present the details of this map in
appendix \ref{VoBo}. There we also give the map between the ${\bf Z}_3$ 
model of Refs \cite{GJ,DP} and F-theory. 
The explicit construction of $r=6$, $a=8$
and $r=8$, $a=10$ cases in appendix \ref{VoBo} gives more evidence in favor
of the existence of $(h^{1,1},h^{2,1})=(r+1,61-5r)$ and $({\hat h}^{1,1},{\hat h}^{2,1})=(r+1,301-29r)$ Calabi-Yau manifolds. 

\subsection{Explicit 6D Examples}

{}In this subsection we discuss explicit six dimensional examples of 
``asymmetric'' Type IIB orientifolds that do not suffer from presence of 
additional non-perturbative states discussed above. Here we present
such examples from the F-theory viewpoint. Some of the details of 
explicitly mapping the corresponding orientifold models to their F-theory
duals are relegated to appendix \ref{VoBo}.

$\bullet$ Let ${\widetilde {\cal M}}_2=(T^2\otimes T^2)/{\bf Z}_2$, where 
the generator ${\widetilde g}$
of ${\bf Z}_2$ acts on $z_s$ (that is, complex coordinates parametrizing the
2-tori) as ${\widetilde g}z_s=-z_s$, $s=1,2$. Consider the $\Omega J (-1)^{F_L}$ 
orientifold
of Type IIB on this ${\widetilde {\cal M}}_2$, where $Jz_1=-z_1$, $Jz_2=z_2$. 
The F-theory 
dual of this orientifold is given by F-theory on the Calabi-Yau three-fold $(T^2\otimes
T^2\otimes T^2)/({\bf Z}_2\otimes {\bf Z}_2)$, where the generators $S$ and 
${\widetilde g}$ of the
two ${\bf Z}_2$ subgroups act as follows ($z_0$ parametrizes the first $T^2$):
\begin{eqnarray}
 &&{\widetilde g}z_0=z_0~,~~~{\widetilde g}z_1=-z_1~,~~~
 {\widetilde g}z_2=-z_2~,\nonumber\\
 &&Sz_0=-z_0~,~~~Sz_1=-z_1~,~~~Sz_2=z_2~.
\end{eqnarray} 
First consider the case with no discrete torsion between ${\widetilde g}$ 
and $S$. The
corresponding Hodge numbers are $(h^{1,1},h^{2,1})=(51,3)$. In this case we
have (in the notations of the previous subsection 
for Nikulin's classification) $r=18,a=4$.
This model
has $T=17$ tensor multiplets and $H_c=4$ hypermultiplets in the closed
string sector, whereas the open string sector gives rise to gauge group
$SO(8)^8$ with no charged matter. This model is T-dual of the model
obtained via orientifolding Type IIB on ${\widetilde {\cal M}}_2$ by $\Omega {\hat J}$ where
${\hat J}=1$ in the untwisted sector, while ${\hat J}=-1$ in the twisted sector 
\cite{BZ,dp1}. Such an action of ${\hat J}$ is equivalent to nothing but introducing
discrete torsion between $\Omega$ and ${\widetilde g}$.\\
Next, consider the case with discrete torsion between 
${\widetilde g}$ and $S$. The
corresponding Hodge numbers are $(h^{1,1},h^{2,1})=(3,51)$. 
In this case we
have  (this is one of the cases depicted as $\otimes$ in
Fig.1) $r=2,a=4$. This model
has $T=1$ tensor multiplets and $H_c=20$ hypermultiplets in the closed
string sector, whereas the open string sector gives rise to 8 hypermultiplets
transforming as adjoints under a global $SO(8)$ symmetry. 
This model is T-dual of the model
obtained via orientifolding Type IIB on ${\widetilde {\cal M}}_2$
by $\Omega$ ({\em i.e.}, there
is no discrete torsion between $\Omega$ and 
${\widetilde g}$), which is the ${\bf Z}_2$ orientifold
model of Refs \cite{PS,GP}.
 
$\bullet$ Let ${\widetilde {\cal M}}_2$ be the same as in the above example. 
Consider the $\Omega J (-1)^{F_L}$ orientifold
of Type IIB on this ${\widetilde {\cal M}}_2$, where $Jz_1=z_2$, $Jz_2=z_1$. 
The F-theory dual is given by a compactification on a Calabi-Yau three-fold
with Hodge numbers $(h^{1,1},h^{2,1})=(21,9)$ if there is no discrete 
torsion between
$J$ and ${\widetilde g}$, and $(h^{1,1},h^{2,1})=(9,21)$, otherwise.
Note that the cases with and 
without discrete torsion are related by mirror symmetry. In the case 
$(h^{1,1},h^{2,1})=(21,9)$ we have $r=12,a=10$. In the closed string sector we have 
$T=11$
tensor multiplets and $H_c=10$ hypermultiplets. In the open string sector we have 
$SO(8)^2$ gauge group with no charged matter. In the case 
$(h^{1,1},h^{2,1})=(9,21)$ we have $r=8,a=10$. In the closed string sector we have 
$T=7$ tensor multiplets and 
$H_c=14$ hypermultiplets. In the open string sector we have 
2 hypermultiplets transforming as adjoints under a global $SO(8)$ symmetry.

$\bullet$ Let ${\widetilde {\cal M}}_2$ be the same as in the above example 
but with the restriction that
each of the 2-tori factorize as products of two identical circles: 
$T^2=S^1\otimes S^1$. Let
$J$ act as follows: $J$ permutes the two circles that make up the first $T^2$; it acts as a 
reflection on one of the two circles that make up the second $T^2$, while leaving the 
other circle untouched. The corresponding Hodge numbers are 
$(h^{1,1},h^{2,1})=(31,7)$ if there is no discrete torsion between
$J$ and ${\widetilde g}$, and $(h^{1,1},h^{2,1})=(7,31)$, otherwise. 
Note that the cases with and 
without discrete torsion are related by mirror symmetry. In the case 
$(h^{1,1},h^{2,1})=(31,7)$ we have $r=14,a=8$. In the closed string sector we have 
$T=13$
tensor multiplets and $H_c=8$ hypermultiplets. In the open string sector we have 
$SO(8)^4$ gauge group with no charged matter. In the case 
$(h^{1,1},h^{2,1})=(7,31)$ we have $r=6,a=8$. In the closed string sector we have 
$T=5$ tensor multiplets and 
$H_c=16$ hypermultiplets. In the open string sector we have 
4 hypermultiplets transforming as adjoints under a global $SO(8)$ symmetry.

$\bullet$ Let ${\widetilde {\cal M}}_2=(T^2\otimes T^2)/{\bf Z}_3$, 
where the generator ${\widetilde g}$
of ${\bf Z}_3$ acts on $z_s$ as ${\widetilde g}z_1=\omega z_1$,
${\widetilde g}z_2=\omega^{-1} z_2$ ($\omega=\exp(2\pi i/3)$).
Consider the $\Omega J (-1)^{F_L}$ orientifold
of Type IIB on this ${\widetilde {\cal M}}_2$, where 
$J$ is given by Eq (\ref{J}). It is not difficult to
show that the corresponding Hodge numbers are the same (that is, the model is the 
same) for all choices of $\eta,b$, so for simplicity we can take $\eta=1$, $b=0$.
We have $(h^{1,1},h^{2,1})=(15,15)$, which corresponds to the Voisin-Borcea 
orbifold with $(r,a,\delta)=(10,10,1)$.
At generic points this model contains $T=9$ tensor multiplets, $H=16$ 
hypermultiplets, and $V=4$ $U(1)$ vector multiplets \cite{MV}. 
At orbifold points 
we get gauge symmetry enhancement to $SO(8)$. 
The charged matter consists of one adjoint hypermultiplet of 
$SO(8)$ (hence ${\cal N}=2$ global supersymmetry in the gauge, 
that is, open string sector). The uncharged matter (in the closed string 
sector) is $T=9$ tensor multiplets and  $H=12$ hypermultiplets. 

$\bullet$ Let ${\widetilde {\cal M}}_2=(T^2\otimes T^2)/{\bf Z}_N$, $N=4,6$, 
where the 
generator ${\widetilde g}$
of ${\bf Z}_N$ acts on $z_s$ as ${\widetilde g}z_1=\omega z_1$,
${\widetilde g}z_2=\omega^{-1} z_2$ ($\omega=\exp(2\pi i/N)$).
Consider the $\Omega J (-1)^{F_L}$ orientifold
of Type IIB on this ${\widetilde {\cal M}}_2$, 
where $J$ is given by Eq (\ref{J}). It is not difficult to
show that the corresponding Hodge numbers are the same
for all choices of $\eta,b$, so we can take $\eta=1$, $b=0$.
We have $(h^{1,1},h^{2,1})=(21,9)$ if there is no discrete
torsion between $J$ and $g^{N/2}$, and $(h^{1,1},h^{2,1})=(9,21)$, otherwise.
We have discussed this model above.

$\bullet$ Let ${\widetilde {\cal M}}_2=(T^2\otimes T^2)/{\bf Z}_3$, 
where the generator ${\widetilde g}$
of ${\bf Z}_3$ acts on $z_s$ as ${\widetilde g}z_1=\omega z_1$,
${\widetilde g}z_2=\omega^{-1} z_2$ ($\omega=\exp(2\pi i/3)$).
Consider the $\Omega JJ^\prime (-1)^{F_L}$ orientifold
of Type IIB on this ${\widetilde {\cal M}}_2$, where $J$ acts as 
$Jz_1=-z_1$, $Jz_2=z_2$,
and the action of $J^\prime$ was discussed in subsection A of section \ref{other}. 
We have $(h^{1,1},h^{2,1})=(20,14)$
(see appendix \ref{VoBo} for details), which corresponds to the Voisin-Borcea 
orbifold with $r=11,a=9$.
At orbifold points we have the following massless spectrum. 
There are $T=10$ tensor multiplets and $H_c=11$ hypermultiplets
in the closed string sector.
The open string sector gives rise to $SO(8)\otimes SO(8)$ with one hypermultiplet
transforming in the adjoint of the first $SO(8)$, and no matter charged under the second 
$SO(8)$. This model is on the same moduli as the ${\bf Z}_3$ orientifold model
of Refs \cite{GJ,DP}. In particular, it is ``T-dual'' of the latter\footnote{We put 
``T-dual'' in quotes as T-duality in this and the following two cases is subtle. We will
discuss these subtleties in the next subsection.}. 

$\bullet$ Let ${\widetilde {\cal M}}_2=(T^2\otimes T^2)/{\bf Z}_4$, 
where the generator ${\widetilde g}$
of ${\bf Z}_3$ acts on $z_s$ as ${\widetilde g}z_1=\omega z_1$,
${\widetilde g}z_2=\omega^{-1} z_2$ ($\omega=\exp(2\pi i/4)$).
Consider the $\Omega JJ^\prime (-1)^{F_L}$ orientifold
of Type IIB on this ${\widetilde {\cal M}}_2$, where $J,J^\prime$ act as in the previous 
example. 
We have $(h^{1,1},h^{2,1})=(41,5)$ if there is no discrete torsion between $J$ and
${\widetilde g}^2$, and $(h^{1,1},h^{2,1})=(7,31)$, otherwise
(see appendix \ref{VoBo} for details). The case with  $(h^{1,1},h^{2,1})=(41,5)$
corresponds to $r=16,a=6$. In this model  
there are $T=15$ tensor multiplets and $H_c=6$ hypermultiplets in the closed string 
sector.
The open string sector gives rise to $SO(8)^6$ gauge group
with no charged matter. The model with $(h^{1,1},h^{2,1})=(7,31)$ has
been discussed above.
This model is ``T-dual'' to the ${\bf Z}_6$ orientifold model of Refs \cite{GJ,DP}.

$\bullet$ Let ${\widetilde {\cal M}}_2=(T^2\otimes T^2)/{\bf Z}_6$, 
where the generator ${\widetilde g}$
of ${\bf Z}_6$ acts on $z_s$ as ${\widetilde g}z_1=\omega z_1$,
${\widetilde g}z_2=\omega^{-1} z_2$ ($\omega=\exp(2\pi i/6)$).
Consider the $\Omega JJ^\prime (-1)^{F_L}$ orientifold
of Type IIB on this ${\cal M}_2$, where $J,J^\prime$ act as in the previous 
example. 
We have $(h^{1,1},h^{2,1})=(31,7)$ if there is no discrete torsion between $J$ 
and
${\widetilde g}^3$, and $(h^{1,1},h^{2,1})=(9,21)$, otherwise
(see appendix \ref{VoBo} for details).
We have discussed these models above. Note that the $(h^{1,1},h^{2,1})=(9,21)$
model is ``T-dual'' to the ${\bf Z}_6$ orientifold model of Refs \cite{GJ,DP}.

{}As an aside, in appendix \ref{CHL} we present two (singular) Calabi-Yau 
three-folds with $SU(2)$ holonomy. F-theory compactifications on these manifolds
are dual to CHL heterotic strings (with ${\cal N}=2$ supersymmetry) in six dimensions.

\subsection{Type I on K3}

{}As we already discussed in the previous section, in ``asymmetric'' $\Omega$ 
orientifolds of Type IIB on ${\widetilde {\cal M}}_2=
T^4/{\bf Z}_N$, $N=2,3,4,6$, which correspond 
to Type I compactifications on K3, we expect additional non-perturbative (from 
the orientifold viewpoint) contributions from the $\Omega {\widetilde g}^k$ 
sectors, $k=1,\dots,N-1$,
$2k\not=N$. To understand the structure of these sectors we can attempt to map these 
orientifolds to F-theory. In doing so some care is required. 
Thus, consider K3 as a $T^2$ fibration over ${\bf P}^1$.
Naively, T-duality will map
the $\Omega$ orientifold to the $\Omega J (-1)^{F_L}$ orientifold, where $J$ reverses the
sign of the complex coordinate $z_1$ on $T^2$ while leaving the complex coordinate
$z_2$ of the base ${\bf P}^1$ unaffected. However, this is only correct if the singularities
in the fibre are invariant under the action of $J$. This is the case for 
${\widetilde {\cal M}}_2=
(T^2\otimes T^2)/{\bf Z}_2$ and ${\widetilde {\cal M}}_2=(T^2\otimes T^2)/{\bf Z}_4$,
but does not hold for ${\widetilde {\cal M}}_2=(T^2\otimes T^2)/{\bf Z}_3$ and 
${\widetilde {\cal M}}_2=(T^2\otimes T^2)/{\bf Z}_6$. In the last two cases the fibration is
modified by the action of T-duality, and one ends up with K3 surfaces which are not 
orbifold K3's. For this reason we will use T-duality in the fibre only for the ${\bf Z}_2$ 
and ${\bf Z}_4$ cases, and then use a different approach to analyze the other two cases.   

{}In the case of the ${\bf Z}_2$ orbifold limit of K3 we already know  the answer: if there
is no discrete torsion between $J$ and ${\widetilde g}$ (the generator of the ${\bf Z}_2$
twist on K3), then this corresponds to F-theory on the Calabi-Yau three-fold with Hodge 
numbers $(51,3)$. (This model has $T=17$ tensor multiplets.) If there is discrete torsion 
between $J$ and ${\widetilde g}$, then the Hodge numbers are $(3,51)$. This model 
corresponds to Type I compactification on K3, and the number of tensor multiplets is
$T=1$.

{}In the ${\bf Z}_4$ case we also consider two cases. Suppose there is no discrete 
torsion between $J$ and ${\widetilde g}^2$ (where ${\widetilde g}$ is 
the generator of the ${\bf Z}_2$ twist on K3). Then the Hodge numbers are $(61,1)$. 
This model has $T=19$ tensor multiplets. On the other hand, if there is discrete 
torsion between $J$ and ${\widetilde g}^2$, then the Hodge numbers can be computed
to be $(3,51)$, just as in the ${\bf Z}_2$ case.

{}Let us try to understand the other two cases, namely, Type I compactifications
on the ${\bf Z}_3$ and ${\bf Z}_6$ orbifold limits of K3. To do this let us 
consider\footnote{Some dynamical aspects of Type I-heterotic duality
for compactifications on 
K3$\otimes T^2$ were studied in \cite{Taylor}.}
Type I on K3$\otimes T^2$. (Let the complex coordinate parametrizing this new $T^2$ 
be $z_3$.) Then to map to F-theory we can T-dualize this extra $T^2$.
The resulting compactification of F-theory is that on K3$\otimes$K3, where the first
K3 is obtained by orbifolding $T^2\otimes T^2$ by ${\bf Z}_2$ whose generator $S$ acts
on the corresponding complex coordinates $z_0$ and $z_3$ as $Sz_{0,3}=-z_{0,3}$. The second K3 is the original K3 we compactified Type I on to begin with. This K3 is given
by $T^2\otimes T^2$ orbifolded by ${\bf Z}_N$ whose generator ${\widetilde g}$ acts accordingly on the corresponding complex coordinates $z_1,z_2$.

{}The Euler characteristic of K3$\otimes$K3 is $\chi=24^2$. Thus, we need 24 
three-branes to cancel the space-time anomaly. However, we have a choice of where
to place the three-branes: ({\em i}) we can keep them in the bulk; from the heterotic
viewpoint these correspond to small instantons, while from the Type I viewpoint 
these correspond to dynamical five-branes (made of some number of D5-branes);
({\em ii}) alternatively, we can ``dissolve'' them into the seven-branes; from the heterotic
(Type I) viewpoint this corresponds to embedding a certain gauge bundle into 
${\mbox{Spin}}(32)/{\bf Z}_2$ ($SO(32)$). The corresponding instantons are no longer 
point-like (at generic points). Thus, we see that we need to specify additional data in 
F-theory. The total number of instantons must be 24 to cancel the anomaly. If we embed 
all of them in the gauge bundle, then we get a perturbative heterotic vacuum. On the other
hand, perturbative Type I vacuum (from the orientifold viewpoint) does not correspond
to such an embedding. Thus, in the ${\bf Z}_2$ model of Refs \cite{PS,GP} it is not difficult
to see that only 16 instantons are embedded into $SO(32)$. The other 8 are dynamical
five-branes (corresponding to NS 5-branes on the heterotic side). Each of these is made
of 4 D5-branes. Here two pairings take place: one due to the $\Omega$ projection, and the
other one due to the ${\bf Z}_2$ orbifold projection.

{}Let us consider the ${\bf Z}_3$ example for illustrative purposes. Let us choose the
gauge bundle in the following fashion. The action of the orbifold group on the 
Chan-Paton factors can be described in terms of $16\times16$ matrices $\gamma_k$,
$k=0,\dots,N-1$. (We have chosen to work with $16\times16$ matrices for we are not 
counting the orientifold images of the D9-branes.) Let us choose 
\begin{equation}
 \gamma_1=
 {\mbox{diag}}(\omega~(4~{\mbox{times}}),~\omega^2~(4~{\mbox{times}}),~1~
 (8~{\mbox{times}}))~,
\end{equation}
where $\omega=\exp(2\pi i/3)$. This choice of the gauge bundle
corresponds to embedding 24 instantons in $SO(32)$ (that is, it would lead to a 
perturbative heterotic model). Thus, we do not have any five-branes on the Type I side.
In fact, the tadpole cancellation conditions derived in the orientifold approach tell us
that there are no D5-branes in this model, and, moreover, all the untwisted and
twisted tadpoles cancel with this choice of the gauge bundle \cite{GJ,DP}. (See 
subsection A of section \ref{other} for a related discussion.) The ``naive'' orientifold 
approach, however, would give us an inadequate answer for the massless spectrum.
In six dimensional terms, the closed string sector gives rise to $T=1$ tensor multiplet,
and $H_c=20$ hypermultiplets. The open string sector (99 sector in the Type I 
language) gives rise to gauge bosons in the $U(8)\times SO(16)$ subgroup of $SO(32)$,
plus 1 hypermultiplet in $({\bf 28},{\bf 1})$ and $({\bf 8},{\bf 16})$ irreps of the unbroken
gauge group. This matches (as far as the charges under the non-Abelian subgroup of the unbroken gauge group goes) the heterotic massless spectrum except for the twisted sector
massless states that the latter possesses: there are 9 hypermultiplets in the
$({\bf 28},{\bf 1})$ irrep on the heterotic side. (The multiplicity 9 comes from the number 
of fixed points in the twisted sectors.) These states are non-perturbative from the 
orientifold viewpoint as they cannot be viewed as 99 open string states. Let us use the
F-theory picture to see the non-perturbative origin of these states from the orientifold 
viewpoint. After T-dualizing we have seven-branes only (as the three-branes have been 
``dissolved'' into the gauge bundle). The $S$ twisted sector in F-theory gives rise to
the T-duals of 99 sector in the Type I description. However, in F-theory we also see the
states that arise in the $S{\widetilde g}$ and $S{\widetilde g}^2$
twisted sectors. These correspond to D7-branes
with ${\bf C}^2/{\bf Z}_3$ (that is, $A_2$) singularities in their world-volumes. These
states are clearly non-perturbative from the orientifold viewpoint, and are precisely
the 9 hypermultiplets in $({\bf 28},{\bf 1})$ of $U(8)\times SO(16)$. We cannot ignore 
these states in Type I compactification on $T^4/{\bf Z}_3$ as the gauge and gravitational anomalies do not cancel unless they are taken into account.
  
{}We end this subsection with the following remark. Suppose we start from Type I on K3
(with only one tensor multiplet). Let K3 be a $T^2$ fibration over ${\bf P}^1$. Then we can
attempt to T-dualize the fibre $T^2$.
The net result should be an $\Omega J (-1)^{F_L}$ orientifold
of Type IIB on a mirror K3$^\prime$ where $J$ is Nikulin's involution that reverses the sign 
of the holomorphic 2-form on K3$^\prime$ \cite{SS}. From 
the F-theory viewpoint this corresponds to
a compactification on a Voisin-Borcea orbifold. Note that the integer $r$ for such
Voisin-Borcea
orbifolds must ultimately be equal 2 since the number of tensor multiplets is given by $T=r-1$,
and we have $T=1$. We thus conclude that these Voisin-Borcea orbifolds must be within the following set:\\
$\bullet$ $r=2$, $a=4$, $(h^{1,1},h^{2,1})=(3,51)$. This is a ${\bf Z}_2\otimes {\bf Z}_2$ 
orbifold with discrete torsion.\\
$\bullet$ $(r,a,\delta)=(2,0,0)$. This is a $T^2$ fibration over ${\bf F}_4$.\\
$\bullet$ $(r,a,\delta)=(2,2,0)$. This is a $T^2$ fibration over ${\bf F}_0$.\\
$\bullet$ $(r,a,\delta)=(2,2,1)$. This is a $T^2$ fibration over ${\bf F}_1$.\\  
(Here ${\bf F}_n$ are Hirzebruch surfaces.)

{}Note that only the first of the above cases corresponds to a toroidal orbifold. Thus,
as we already mentioned in the beginning of this subsection, starting from an orbifold K3
(say, $(T^2\otimes T^2)/{\bf Z}_3$) we may end up with a mirror K3$^\prime$ which is
not a (geometric) toroidal orbifold. 

\subsection{4D Orientifolds}

{}We start our discussion by considering the F-theory dual of the $\Omega$ orientifold of 
Type IIB on ${\widetilde {\cal M}}_3=
T^6/({\widetilde {\bf Z}}_2\otimes {\widetilde{\bf Z}}_2)$ constructed 
in Ref \cite{BL}. For simplicity we can take $T^6=T^2\otimes T^2\otimes T^2$. Let $z_i$ ($i=1,2,3$) be
the complex coordinates parametrizing these three 2-tori. Then the action of the 
orbifold group
${\widetilde G}=\{1,R_1,R_2,R_3\}\approx {\widetilde {\bf Z}}_2\otimes {\widetilde {\bf Z}}_2$ is given 
by: $R_iz_j=-(-1)^{\delta_{ij}} z_j$. (Note that $R_3=R_1R_2$. Also, if there is no 
discrete torsion between the generating elements $R_1$ and $R_2$ then
the Hodge numbers of this three-fold are given by $(h^{1,1},h^{2,1})=(51,3)$.) 
The orientifold group is given by 
${\cal O}=\{1,R_1,R_2,R_3,\Omega,\Omega R_1, \Omega R_2,\Omega R_3\}$.
This model contains 32 D9-branes and three sets of D5-branes with 32 D5-branes in 
each set. The locations of D5$_i$-branes are given by points in the $z_i$ complex plane.

{}We can T-dualize this model so that instead of D9- and D5-branes we have D3- and 
D7-branes. Then we can map this orientifold model to F-theory via the map of Refs \cite{sen}.
Here we would like to identify the Calabi-Yau four-fold corresponding to the F-theory dual.
Following our discussion in sections \ref{F} and \ref{FA} it is not difficult to see that the
four-fold is an orbifold $(T^2\otimes T^2\otimes T^2\otimes T^2)/({\bf Z}_2\otimes {\bf Z}_2
\otimes {\bf Z}_2)$, where the first $T^2$ is the fibre $T^2$, the other three $T^2$'s are those
of the original Calabi-Yau three-fold, the first two ${\bf Z}_2$'s act as above, and the third 
${\bf Z}_2$ (whose generator will be denoted by $S$) acts as follows: $Sz_0=-z_0$, 
$Sz_1=-z_1$, $Sz_{2,3}=z_{2,3}$. Here $z_0$ is the complex coordinate parametrizing the 
first $T^2$, and we have chosen $S$ to act non-trivially on $z_1$ without loss of generality.

{}The question that we need to address here is whether there is any discrete torsion between 
the generators $S$ and $R_{1,2}$. This is a non-trivial issue since in the six dimensional 
${\bf Z}_2$ model of Refs \cite{PS,GP} the choice of discrete torsion in mapping to F-theory
was crucial (see subsection B of this section for details). 
Here our discussion will be brief as the details 
are not difficult to reconstruct. Before giving the answer to the above question, we will 
discuss a class of Calabi-Yau four-folds (to which the four-fold under consideration belongs)
known as the Borcea four-folds \cite{Borcea}.

{}Consider $($K3$\otimes$K3$)/{\bf Z}_2$ where ${\bf Z}_2$ acts as an involution
labelled by $(r_1,a_1,\delta_1)$ on the first K3, and as an involution labelled by $(r_2,a_2,\delta_2)$ on the second K3. This quotient is a (singular) Calabi-Yau four-fold
with $SU(4)$ holonomy. Its Euler number is given by \cite{Borcea}
\begin{equation}
 {1\over 24}\chi=12+{1\over 4}(r_1-10)(r_2-10)~. 
\end{equation} 
Now consider F-theory compactified on such a four-fold. The space-time anomaly
can be cancelled via introducing three-branes if and only if $\chi/24$ 
is a non-negative integer (or else supersymmetry appears to be broken \cite{SVW}).

{}Let us return to the orbifold 
$(T^2\otimes T^2\otimes T^2\otimes T^2)/({\bf Z}_2\otimes {\bf Z}_2
\otimes {\bf Z}_2)$. It is not difficult to show that if there is no discrete torsion between 
any of the generating elements $S,R_1,R_2$, then this orbifold is a Borcea four-fold
with $r_1=r_2=18$ and $a_1=a_2=4$. The Euler number in this case is given by
$\chi/24=28$, and we need to introduce 28 three-branes to cancel the space-time anomaly.
This compactification for a specific distribution of three-branes corresponds to the T-dual
of the orientifold model of Ref \cite{BL} discussed above. In this T-dual 
model we have 32 D3-branes. These correspond to 4 dynamical three-branes. 
Each of these is made of 8 D3-branes. Here three pairings take place: one due to the 
$\Omega$ projection, and the other two due to the $R_1$ and $R_2$ orbifold projections.
The rest of the three-branes, namely, 24 three-branes, are ``dissolved'' into the seven-branes.
There are three kinds of seven-branes (different kinds of seven-branes are intersecting
at right angles). 8 three-branes are ``dissolved'' into each kind of seven-branes, which
corresponds to embedding a certain gauge bundle into the seven-brane gauge group.
In fact, the embedding here is the same as in the six dimensional ${\bf Z}_2$ case discussed
in subsection C of section \ref{FA}. Namely, from the six dimensional viewpoint (which is
applicable here as all the twisted sectors look six dimensional subject to additional
orbifold projections) we are embedding 16 instantons into the gauge group for each kind of
seven-branes. From the four dimensional viewpoint these correspond to 8 three-branes
``dissolved'' into each kind of seven-branes. The pairing here is due to the additional orbifold
projection in the four dimensional case compared with the six dimensional case.

{}It is not difficult to show that the  
$(T^2\otimes T^2\otimes T^2\otimes T^2)/({\bf Z}_2\otimes {\bf Z}_2
\otimes {\bf Z}_2)$ orbifold with non-trivial discrete torsion between any of the
generating elements $S,R_1,R_2$ is equivalent to the Borcea four-fold with
$r_1=18$, $r_2=2$ and $a_1=a_2=4$. The Euler number in this case is given by
$\chi/24=-4$. This implies that the space-time anomaly cannot be cancelled in this case
via introducing three-branes. This, in particular, explains the ``puzzle'' found in $\Omega$
orientifold of Type IIB on ${\widetilde {\cal M}}_3=
T^6/({\widetilde {\bf Z}}_2\otimes {\widetilde{\bf Z}}_2)$
with discrete torsion between the generating elements $R_1$ and $R_2$ (in this case
${\widetilde {\cal M}}_3$ has the Hodge numbers $(h^{1,1},h^{2,1})=(3,51)$): it is impossible to cancel
all the tadpoles in the corresponding orientifold model \cite{AP}\footnote{We would like to
thank C. Angelantonj for communications on this point.}. Here F-theory provides a
simple geometric explanation of this fact\footnote{We should point out that our
conclusions here disagree with those in section 4 of Ref \cite{Iba}.}.

{}Note that the ${\widetilde {\bf Z}}_2\otimes {\widetilde{\bf Z}}_2$ four dimensional example
discussed above is the only one that satisfies the world-sheet consistency 
conditions (\ref{wsc}) and (\ref{wscA}). Next, we would like to discuss other cases. 
In particular, from the F-theory viewpoint 
we will give evidence for the assertion made in            
subsection C of section \ref{other}
that $\Omega J^\prime$ action is not well defined in sectors twisted by orbifold elements
${\widetilde g}_a={\mbox{diag}}(\rho_a,\rho^\prime_a,(\rho_a\rho^\prime_a)^{-1})$ with
$\rho_a,\rho^\prime_a,(\rho_a\rho^\prime_a)\not=1$. (Here we are considering orientifolds
of Type IIB on ${\widetilde M}_3=T^6/{\widetilde G}$ where ${\widetilde G}=
\{{\widetilde g}_a\vert a=1,\dots{\mbox{dim}}({\widetilde G})\}$,
and ${\widetilde {\cal M}}_3$ has $SU(3)$ holonomy. Recall that the action of $J^\prime$
was defined to map the ${\widetilde g}_a$ twisted sector to the ${\widetilde g}^{-1}_a$ 
twisted sector where ${\widetilde g}_a^2\not=1$.)

{}Instead of being most general here\footnote{Generalization to other cases
should be clear from the following discussion.}, for illustrative purposes we will consider a
special class of cases, namely, orientifolds of Type IIB on ${\widetilde {\cal M}}_3=
T^6/{\bf Z}_N$ where ${\widetilde {\cal M}}_3$ has $SU(3)$ holonomy. (Here $N$ can be
$3,7,4,6,8,12$. See subsection B of section \ref{class} for details.) Let ${\widetilde g}$ be
the generator of the orbifold group ${\widetilde G}=\{{\widetilde g}^k\vert k=0,\dots,N-1\}$. 
The action of 
${\widetilde g}$ on the complex coordinates $z_i$ ($i=1,2,3$) parametrizing 
${\widetilde {\cal M}}_3$ is given by ${\widetilde g}z_1=\omega z_1$, 
${\widetilde g}z_2=\omega^p z_2$,
${\widetilde g}z_3=\omega^{-p-1} z_3$ where $\omega=\exp(2\pi i/N)$, 
and $p\in\{1,\dots,N-2\}$.
Suppose we intend to orientifold Type IIB on such ${\widetilde {\cal M}}_3$ so that
the orientifold projection is given by $\Omega {J} J^\prime (-1)^{F_L}$
where $J$ reverses the sign of one of the complex coordinates $z_i$,
and leaves the other two unaffected\footnote{This action is assumed to be compatible
with the symmetries of $T^6$.}. 
We also need to specify the action of $J^\prime$. It acts as identity in the untwisted
and ${\bf Z}_2$ twisted sectors\footnote{We can absorb possible discrete torsion in the
${\bf Z}_2$ twisted sector into the definition of ${J}$.},
and in other twisted sectors it acts only on ground states
by mapping the ${\widetilde g}^k$ twisted ground state to the inverse ${\widetilde g}^{-k}$
twisted ground state (just as in subsection A of  section \ref{other}). In the following we are
going to argue that such an action is not well defined if the ${\widetilde g}^k$ 
twist has fixed points in
$T^6$. 

{}To see this, let us assume that $J^\prime$ acts non-trivially in the ${\widetilde g}$ 
and ${\widetilde g}^{-1}$ twisted
sectors. By construction the ${\widetilde g}$ 
twist has fixed points in $T^6$ but no
fixed 2-tori. We can use the map of Refs \cite{sen} to map this orientifold to F-theory. Here
F-theory is compactified on a Calabi-Yau four-fold defined as
\begin{equation}
 {\widetilde {\cal X}}_4=(T^2\otimes{\widetilde {\cal M}}_3)/X~,
\end{equation}         
where $X=\{1,S\}\approx {\bf Z}_2$, and $S$ acts as $Sz_0=-z_0$ on $T^2$ ($z_0$ is 
a complex coordinate on $T^2$), and as ${J} J^\prime$ on 
${\widetilde {\cal M}}_3$. Let us see what the contribution of ${\widetilde g}$ 
and ${\widetilde g}^{-1}$ twisted
sectors into the Hodge numbers $h^{1,1}$ and $h^{2,1}$ of ${\widetilde {\cal X}}_4$
would look like for such an action of $X$. (Note that ${\widetilde g}$ and 
${\widetilde g}^{-1}$ twisted sectors
in the Calabi-Yau three-fold ${\widetilde {\cal M}}_3$ contribute only to $h^{1,1}$ but not
to $h^{2,1}$. The corresponding combined contribution of {\em both} 
${\widetilde g}$ and ${\widetilde g}^{-1}$
twisted sectors into $h^{1,1}$ of ${\widetilde {\cal M}}_3$
is simply given by the number of fixed points for the
twist ${\widetilde g}$. 
This number is given by $(4\sin^2(\pi/N))(4\sin^2(\pi p/N))(4\sin^2(\pi (p+1)/N))$.
Since there is no contribution to $h^{2,1}$ in ${\widetilde {\cal M}}_3$, the corresponding
contribution in ${\widetilde {\cal X}}_4$ can only be present for $h^{1,1}$ and $h^{2,1}$.)
It is not difficult to see that each fixed point (in ${\widetilde {\cal M}}_3$) of the twist 
${\widetilde g}$ 
would contribute {\em one half} into either $h^{1,1}$ or $h^{2,1}$ of 
${\widetilde {\cal X}}_4$ provided that $J^\prime$
acts non-trivially as described above. This is clearly inconsistent, so we conclude that
the action of $J^\prime$ must be trivial in twisted sectors where the corresponding twists
have fixed points.

{}The above discussion clearly implies that $\Omega J^\prime$ action is not well defined in sectors twisted by orbifold elements
${\widetilde g}_a={\mbox{diag}}(\rho_a,\rho^\prime_a,(\rho_a\rho^\prime_a)^{-1})$ with
$\rho_a,\rho^\prime_a,(\rho_a\rho^\prime_a)\not=1$ (for Calabi-Yau three-folds). That is,
in such sectors we are forced to consider $\Omega$ projection which in turn (as it should 
be clear from our previous discussions) is well defined only after we {\em blow up} the orbifold
singularities (except in the ${\bf Z}_2$ twisted sectors).
As a result of the
above discussion we, at least naively, expect non-perturbative (form the orientifold
viewpoint) states arising in the $\Omega {\widetilde g}_a$ ``twisted'' sectors for 
if ${\widetilde g}_a^2\not=1$.

{}Here we can ask whether such non-perturbative contributions can be absent in a given 
orientifold model so that the ``naive'' perturbative approach to the orientifold gives the
correct massless spectrum. Here we observe that we are forced to blow up the
orbifold singularities. In this process it is conceivable that all the non-perturbative
states become heavy due to existence of an appropriate superpotential. 
We will explore this possibility in the next section.

\subsection{An Explicit Map}

{}In this subsection we discuss a map between orientifolds of Type IIB on 
${\widetilde {\cal M}}_3$ and F-theory. For the $\Omega {J} (-1)^{F_L}$
orientifolds where in the diagonal basis ${J}={\mbox{diag}}(-1,+1,+1)$ this map
is straightforward. Suppose, however, we would like to find the map for the
$\Omega $ orientifolds\footnote{We will concentrate on these cases here. Other
cases can be treated analogously.}. These orientifolds
contain either only D9- or both D9- and D5-branes. Thus, we have to ``T-dualize'' to
obtain a setup with D7- and D3-branes.

{}Just as in the case of K3 discussed in subsection C of this section, ``T-dualizing'' is subtle.
In particular, starting with a toroidal orbifold ${\widetilde {\cal M}}_3$ which is a $T^2$
fibration over a base ${\cal B}_2$ we can attempt to T-dualize the fibre $T^2$ but the
resulting space need not be a toroidal orbifold. In particular, this is the case if the
orbifold group ${\widetilde G}$ contains elements of odd order only ({\em i.e.}, $\not\exists~
{\bf Z}_2\in {\widetilde G}$). There are three cases like this: the ${\bf Z}_3$, ${\bf Z}_7$ and
${\widetilde {\bf Z}}_3
\otimes{\widetilde {\bf Z}}_3$ orbifolds. Fortunately, these are precisely the cases (for they
do not contain any D5-branes) which have {\em perturbative} heterotic duals \cite{ZK,KS1,KS2}.
Type I-heterotic duality (which we discuss in section \ref{het}) suffices to understand these
orientifolds quite fully, so the map to F-theory (which does not appear to be so simple)
is not necessary in these cases. 
     
{}Let us therefore consider cases where ${\widetilde G}$ contains at least one ${\bf Z}_2$
subgroup\footnote{These are the cases whose heterotic duals are non-perturbative, so 
Type I-heterotic duality is not helpful in understanding them. Thus, the F-theory picture
is quite desirable as it provides certain independent checks.}. It turns out that the map to
F-theory in these cases is quite simple.
The approach that we would like to pursue here is that instead of T-dualizing in the fibre $T^2$
of ${\widetilde {\cal M}}_3$ we can T-dualize all six coordinates of ${\widetilde {\cal M}}_3$.
This operation is well defined and should not involve any subtleties. The D9-branes T-dualize
into D3-branes, and D5-branes (which are present since $\exists~
{\bf Z}_2\in {\widetilde G}$) T-dualize into D7-branes. This setup is now straightforward
to map to F-theory via the map of Refs \cite{sen}\footnote{Note that in the cases where
$\not\exists~{\bf Z}_2\in {\widetilde G}$ we only have D9-branes which T-dualize to 
D3-branes, but there are no D5-branes to T-dualize to D7-branes, so the map of Refs 
\cite{sen} is not applicable in these cases.}.

\section{${\cal N}=1$ $D=4$ Type I - Heterotic Duality}\label{het}

{}As we already noted, there are three cases, namely, the ${\bf Z}_3$, ${\bf Z}_7$ and
${\bf Z}_3\otimes{\bf Z}_3$ orbifold cases, where the $\Omega$ orientifold does not contain
D5-branes. Under Type I-heterotic duality, D5-branes map to heterotic NS 5-branes
which are non-perturbative objects. Absence of D5-branes, therefore, indicates that
the dual heterotic vacuum should be perturbative. Thus, we can use this observation to
learn about the expected non-perturbative states (coming from $\Omega {\widetilde g}_a$
sectors in Type I) by identifying them with presumably perturbative states on the heterotic
side.

{}This approach was originally taken in Ref \cite{ZK} where the Type I-heterotic 
duality matching was studied for the ${\bf Z}_3$ case of Ref \cite{Sagnotti}. It was 
subsequently extended to the ${\bf Z}_7$ and ${\bf Z}_3\otimes{\bf Z}_3$ cases
in Refs \cite{KS1,KS2}. Here we will briefly review the duality matching for the
${\bf Z}_3$ case as it will be important for understanding the subtleties pointed out
in the previous sections as well as for constructing consistent orientifold models
discussed in the next section. (Here we concentrate on the ${\bf Z}_3$ example 
as it is the simplest out of the three cases. The ${\bf Z}_7$ and ${\widetilde {\bf Z}}_3
\otimes {\widetilde {\bf Z}}_3$
cases work out similarly. All the details can be found in Refs \cite{KS1,KS2}.)

{}Let us start with the Type I ${\bf Z}_3$ orbifold model. There are 32 D9-branes in this
model, and the action of the orbifold group on the D9-brane Chan-Paton charges is
described by $16\times 16$ Chan-Paton matrices $\gamma_k$ (corresponding to
${\widetilde g}^k$ ($k=0,1,2$) elements of the orbifold group), where we have chosen
to work with $16\times 16$ matrices for we are not counting the orientifold images of 
D9-branes. The tadpole cancellation conditions \cite{Sagnotti,KS1,KS2} uniquely fix the
Chan-Paton matrices (up to equivalent representations):
\begin{equation}
 \gamma_1={\mbox{diag}}(\exp(2\pi i/3) ~(6 ~{\mbox{times}}),~
 \exp(-2\pi i/3)~ (6 ~{\mbox{times}}), ~1~(4 ~{\mbox{times}}))~.
\end{equation}
The gauge group is $U(12)\otimes SO(8)$, and the massless spectrum of this model
is given in Table \ref{Z3I}.

{}Next, let us consider the heterotic dual of this Type I model. We start from
${\mbox{Spin}}(32)/{\bf Z}_2$ heterotic string and compactify on $T^6/{\bf Z}_3$.
The choice of the gauge bundle is the same as in the Type I case, {\em i.e.}, the
${\bf Z}_3$ twists are accompanied by shifts in the ${\mbox{Spin}}(32)/{\bf Z}_2$
lattice with the corresponding Wilson lines given by the same $16\times 16$
matrices (in the $SO(32)$ basis) as the Chan-Paton matrices $\gamma_k$.
The gauge group of this model is also $U(12)\otimes SO(8)$, and its massless spectrum
is given in Table \ref{Z3h}.

{}The matching between the massless spectra of these two models is almost precise:
the only\footnote{There is another discrepancy which is the following. The orbifold blow-up
modes $S_{\alpha\beta\gamma}$ on the Type I side are neutral with respect to the
Chan-Paton gauge group whereas their heterotic counterparts are charged under
the $U(1)$ subgroup of the gauge group. This $U(1)$ can be seen to be anomalous
in both Type I and heterotic models, and on the Type I side the blow-up modes transform 
non-trivially under the $U(1)$ gauge transformations \cite{Sagnotti,ZK}. That is, they 
participate in breaking the anomalous $U(1)$ just as their heterotic counterparts.} 
discrepancy is that in the heterotic model we have extra twisted states charged under
the non-Abelian gauge group. These are the 27 spinors $T_{\alpha\beta\gamma}$
of $SO(8)$. These states are clearly non-perturbative from the Type I viewpoint (as 
perturbatively it is not possible to obtain spinorial representations from D-branes).
We identify these states with the expected $\Omega {\widetilde g}^k$ states which are non-perturbative from the orientifold viewpoint. Fortunately, however, these states do not
play any role at low energies as they {\em decouple} from the massless spectrum due to
the following effect.

{}The point here is that there are perturbative superpotentials on both Type I and heterotic 
sides \cite{ZK} (here we are interested in the general structure of the lowest order 
non-vanishing terms):
\begin{eqnarray}\label{sup1}
 {\cal W}_I =&&\lambda \epsilon_{abc} {\mbox{Tr}} (Q_a Q_b \Phi_c)+...~,\\
 \label{sup2}
  {\cal W}_H =&&\lambda^\prime  \epsilon_{abc} {\mbox{Tr}}(Q_a Q_b \Phi_c) +\nonumber\\
      && \Lambda_{(\alpha \alpha^\prime \alpha^{\prime\prime})
                   (\beta \beta^\prime \beta^{\prime\prime})
                   (\gamma \gamma^\prime \gamma^{\prime\prime})}
           {\mbox{Tr}} (S_{\alpha \beta \gamma} 
                       T_{\alpha^\prime \beta^\prime \gamma^\prime}
   T_{\alpha^{\prime\prime} \beta^{\prime\prime} \gamma^{\prime\prime}})+...~.
\end{eqnarray}
(The notation can be found in Tables \ref{Z3I} and \ref{Z3h}.) 
Note that the coupling $\Lambda_{(\alpha \alpha^\prime \alpha^{\prime\prime})
(\beta \beta^\prime \beta^{\prime\prime})
(\gamma \gamma^\prime \gamma^{\prime\prime})}
\not=0$ if and only if $\alpha=\alpha^\prime=\alpha^{\prime\prime}$ or $\alpha\not=\alpha^\prime \not=\alpha^{\prime\prime}\not=\alpha$, and similarly for the $\beta$- and $\gamma$-indices. This follows from the orbifold {\em {space group}} selection rules. Here we note that the couplings $\Lambda_{(\alpha \alpha^\prime \alpha^{\prime\prime})
(\beta \beta^\prime \beta^{\prime\prime})
(\gamma \gamma^\prime \gamma^{\prime\prime})}$ with $\alpha\not=\alpha^\prime \not=\alpha^{\prime\prime}\not=\alpha$, and similarly for the $\beta$- and $\gamma$-indices, are exponentially suppressed in the limit of large volume of the compactification manifold, whereas the couplings $\Lambda_{(\alpha \alpha\alpha)(\beta \beta\beta)(\gamma\gamma\gamma)}$ are not. This is because the corresponding $S_{\alpha \beta \gamma}$ and $T_{\alpha \beta \gamma}$ fields are coming from the same fixed point in the latter case, whereas in the former case they are sitting at different fixed points so that upon taking them apart (in the limit of large volume of the orbifold) their coupling becomes weak.

{}Here we immediately observe that upon the singlets $S_{\alpha \beta \gamma}$ (which are the 27 blow-up modes of the ${\bf Z}_3$ orbifold with non-standard embedding) acquiring vevs
(to cancel the Fayet-Iliopoulos D-term generated by the anomalous $U(1)$), the states $T_{\alpha \beta \gamma}$, that transform in the irrep $({\bf 1}, {\bf 8}_s)(+2)$ of $U(12)\otimes SO(8)$, become heavy and decouple from the massless spectrum. Thus, after blowing up the orbifold singularities on the heterotic side we can match the massless spectra 
of these two models.

{}We see that the original trouble with not having perturbative (from the orientifold
viewpoint) control over the expected extra $\Omega {\widetilde g}^k$ states in the
Type I model has evaporated and we can trust the ``naive'' orientifold answer. The
crucial check here is the Type I-heterotic duality which can be readily utilized since
the heterotic model is perturbative. In fact, the above ``perturbative'' matching is very 
natural from the following point of view. Thus, the tree-level relation between Type I 
and heterotic dilatons in $D$ space-time dimensions \cite{Sagnotti} 
(which follows from the conjectured Type I-heterotic duality in ten dimensions
\cite{PW}) reads:
\begin{equation}
 \phi_H={{6-D}\over 4} \phi_I -{{D-2}\over 16}\log(\det(g_I))~.
\end{equation} 
Here $g_I$ is the internal metric of the Type I compactification space, 
whereas $\phi_I$ and $\phi_H$ are the Type I and heterotic dilatons, 
respectively. From this one can see that (in four dimensions) 
there always exists a region in the moduli space where both Type I and 
heterotic string theories are weakly coupled, and there we can rely on perturbation theory.

{}As we will see in the next section, observations concerning (weak-weak)
Type I-heterotic duality in four dimensions \cite{ZK,KS1,KS2} which we reviewed in this 
section, will be crucial for consistency checks of other four dimensional ${\cal N}=1$
Type I models which are {\em non-perturbative} from the heterotic viewpoint.  

\section{${\cal N}=1$ $D=4$ Non-Perturbative Heterotic Vacua}\label{non-pert}

{}Having established that the non-perturbative states are ``harmless'' in the orientifolds
of Type IIB on four dimensional ${\bf Z}_3$, ${\bf Z}_7$ and ${\widetilde {\bf Z}}_3\otimes
{\widetilde {\bf Z}}_3$
orbifolds, it is natural to consider possible generalizations to cases with D5-branes
by combining these orbifolds with other twists which are also well defined perturbatively.
For example, we know that the six dimensional ${\bf Z}_2$ model of Refs \cite{PS,GP}
is perturbatively well defined. So, perhaps, by combining this ${\bf Z}_2$ twist with one of the
above twists we can obtain an orientifold model where all the naively expected 
non-perturbative states actually decouple along the lines of the previous subsection.
If so, the ``naive'' orientifold rules would produce a well defined vacuum. Such a vacuum
would be non-perturbative from the heterotic viewpoint (since it contains D5-branes) and 
would provide insight into non-perturbative dynamics of heterotic NS 5-branes which are
otherwise very difficult to deal with.

{}In moving along these lines some care is required. Let us first note that the ${\bf Z}_7$ twist
cannot be combined with any other twist to yield an ${\cal N}=1$ model. So we are left
with ${\bf Z}_3$ and ${\widetilde {\bf Z}}_3\otimes 
{\widetilde {\bf Z}}_3$ orbifolds. Here we will consider the ${\bf Z}_3$
orbifold in combination with other twists. (We will discuss the cases with the
${\bf Z}_3$ and ${\widetilde {\bf Z}}_3\otimes 
{\widetilde {\bf Z}}_3$ subgroup in the next section.) From our discussion in 
subsection C of section 
\ref{other} it is clear that we should confine our attention to {\em Abelian} orbifolds. There are
only three Abelian orbifolds (other than ${\bf Z}_3$ itself) that contain ${\bf Z}_3$ as a subgroup:
${\bf Z}_6(\approx{\widetilde {\bf Z}}_2\otimes{\bf Z}_3)$, ${\widetilde
{\bf Z}}_2\otimes{\widetilde {\bf Z}}_6^\prime 
(\approx{\widetilde {\bf Z}}_2\otimes{\widetilde {\bf Z}}_2\otimes{\bf Z}_3)$ and ${\bf Z}_{12}(\approx{\widetilde
{\bf Z}}_4\otimes{\bf Z}_3)$ (see subsection
B of section \ref{class} for details).
 
{}Let us first consider the ${\bf Z}_6$ case. Let ${\widetilde g}$ be the generator of 
${\bf Z}_6$. Consider the ${\widetilde g}^2$ and ${\widetilde g}^4$ 
(that is, the ${\bf Z}_3$ twisted sectors).
These are the same as in the ${\bf Z}_3$ model discussed in the previous section except
that we have to project onto ${\widetilde {\bf Z}}_2$ 
invariant states. It is not difficult to check that
upon performing this projection, the superpotential ${\cal W}_H$ in (\ref{sup2}) reduces in
such a way that all the ${\widetilde {\bf Z}}_2$ invariant twisted sector states
$T_{\alpha\beta\gamma}$ still decouple upon the ${\widetilde {\bf Z}}_2$ invariant
blow-up modes $S_{\alpha\beta\gamma}$ (there are 15 of such modes) acquiring vevs. 
Next, consider the ${\widetilde g}$ and ${\widetilde g}^5$ 
(that is, the ${\bf Z}_6$ twisted sectors). It is not difficult to see that the three fixed points in 
these sectors are the same three of the 15 fixed points in the ${\bf Z}_3$ twisted sectors.
Their blow up modes are therefore also identical. This implies that once the ${\bf Z}_3$
singularities are blown up all the non-perturbative states in the $\Omega {\widetilde g}$ 
and $\Omega {\widetilde g}^5$ sectors should decouple just as is the 
case for the non-perturbative states in the
$\Omega {\widetilde g}^2$ and $\Omega {\widetilde g}^4$ sectors. Finally, the ${\widetilde 
g}^3$ twisted sector is a ${\widetilde {\bf Z}}_2$ twisted sector so that all the states in the
$\Omega {\widetilde g}^3$ sector have a perturbative description. 

{}Thus, we conclude that
upon blowing up the orbifold singularities (except for the ${\widetilde {\bf Z}}_2$
singularities which are ``harmless''), all the non-perturbative (from the orientifold
viewpoint) states should decouple in this model. We can therefore use the
``naive'' tadpole cancellation conditions to compute the spectrum of this model. The 
${\bf Z}_6$ orientifold was first constructed in  
Ref \cite{KS2}. Its massless spectrum 
is summarized in Table \ref{Z6}. Note that the non-Abelian
gauge anomaly cancels in this model. This cancellation is rather non-trivial as the model is
chiral. This model contains D5-branes so the corresponding heterotic dual is 
non-perturbative. This is the first known example of a {\em non-perturbative} chiral ${\cal N}=1$
heterotic vacuum in four dimensions. 

{}It is not difficult to see that the above discussion straightforwardly generalizes to the
${\widetilde{\bf Z}}_2\otimes{\widetilde {\bf Z}}_6^\prime$ case. Here we can also use 
the ``naive'' orientifold approach to construct the corresponding model\footnote{This 
model will be discussed in detail in \cite{zk}.}.

{}Finally, let us consider the ${\bf Z}_{12}$ case.
Naively, one might expect that the arguments in the
${\bf Z}_6$ case concerning blowing up orbifold singularities apply in this case as well,
and all the non-perturbative states must decouple. This is, however, not completely
clear. The point is that in this case we expect non-perturbative contributions
in the ${\widetilde {\bf Z}}_4$ 
twisted sector. Blowing up orbifold singularities in the ${\bf Z}_3$, ${\bf Z}_6$
and ${\bf Z}_{12}$ twisted sectors need not result in decoupling of non-perturbative
states in the ${\widetilde {\bf Z}}_4$
 twisted sector (the latter has fixed 2-tori instead of fixed points).
Here Type I-heterotic duality is not very helpful as the corresponding heterotic dual is
non-perturbative. However, in the next section we will perform another test
for all of the models discussed in this section and we will argue that in the ${\bf Z}_{12}$
model some non-perturbative states do {\em not} decouple from the massless spectrum
after blowing up the orbifold singularities. This model, therefore, is {\em non-perturbative} from
the orientifold viewpoint. On the other hand, the same test will confirm that the ${\bf Z}_6$ and
${\widetilde{\bf Z}}_2\otimes{\widetilde {\bf Z}}_6^\prime$ models are indeed perturbative.

\section{Other Models}\label{anom}

{}In this section we discuss the rest of {\em Abelian} orbifolds. We start with 
a resolution of the following (longstanding\footnote{This ``puzzle'' has been
known to various people for awhile, albeit it appeared in print only in 
\cite{Zw} for the ${\widetilde {\bf Z}}_2\otimes {\widetilde {\bf Z}}_4$
and ${\widetilde {\bf Z}}_4\otimes {\widetilde {\bf Z}}_4$ cases,
and recently in Ref \cite{Iba} for the ${\bf Z}_4$, ${\bf Z}_8$,
${\bf Z}^\prime_8$ and ${\bf Z}^\prime_{12}$ cases.}) ``puzzle''. Namely, in the 
orientifolds of Type IIB on the ${\widetilde {\bf Z}}_2\otimes {\widetilde {\bf Z}}_4$ and
${\widetilde {\bf Z}}_4\otimes {\widetilde {\bf Z}}_4$ \cite{Zw} and 
${\bf Z}_4$, ${\bf Z}_8$,
${\bf Z}^\prime_8$ and ${\bf Z}^\prime_{12}$ \cite{Iba} orbifolds
the tadpole cancellation conditions have no solution. The resolution of this ``puzzle''
is that in all of these orientifolds there are additional non-perturbative contributions
coming from the $\Omega {\widetilde g}_a$ twisted sectors as we explained in sections
\ref{other} and \ref{FA}. For illustrative purposes we will discuss the ${\bf Z}_4$ model
in detail, and only briefly discuss other models of this type.

\subsection{``Anomalous'' Models}

{}Consider ``asymmetric'' Type IIB orientifolds where the orientifold projection
is given by $\Omega$ (so that $J=1$),
and the orbifold ${\widetilde {\cal M}}_3=T^6/{\widetilde G}$
(where ${\widetilde G}=\{{\widetilde g}_a\vert a=1,\dots,{\mbox{dim}}({\widetilde G)}\}$
is Abelian)
contains twisted sectors of the form ${\widetilde g}_a={\mbox{diag}}(-1,\rho_a,-\rho^{-1}_a)$,
where $\rho_a\not=\pm1$. Let $z_i$ be the complex coordinates on ${\widetilde {\cal M}}_3$
in the diagonal basis of ${\widetilde g}_a$ so that ${\widetilde g}_a z_1=-z_1$,
${\widetilde g}_a z_2=\rho_a z_2$, ${\widetilde g}_a z_3=-\rho^{-1}_a z_3$. 
Consider now the tree-channel amplitude corresponding to a cylinder with two cross-caps
(which is obtained via the modular transformation $t\rightarrow 1/t$ from the Klein bottle
amplitude). This amplitude is given by Eq (\ref{crossA}). (More precisely, 
Eq (\ref{crossA}) gives the contribution corresponding to the
{\em untwisted} sector contribution to the Klein bottle amplitude). Note that in the cases under 
consideration the lattice ${\widetilde \Lambda} (R{\widetilde J}_a)=
{\widetilde \Lambda} (R{\widetilde g}_a)$ is non-trivial and consists of momenta
in the $z_1$ direction only. (On the other hand, the winding lattice
${\Lambda} ({\widetilde J}_a)={\Lambda} ({\widetilde g}_a)$ is trivial, {\em i.e.}, it consists
of the origin only.) This implies that we have ``momentum flow'' through the 
corresponding cross-caps in the $z_1$ direction. Thus, we must introduce D-branes
such that the corresponding open strings have Dirichlet boundary conditions in the 
$z_1$ direction. In the other two complex directions $z_2$ and $z_3$, however, these
open strings would have to have {\em twisted} ({\em i.e.}, mixed) boundary conditions
(see subsection A of section \ref{F} for details). Such branes are not perturbative from
the orientifold viewpoint as we discussed at length in section \ref{F}. In this case these
boundary states would correspond to D5-branes wrapping collapsed ${\bf P}^1$'s of the 
orbifold ({\em i.e.}, these are D5-branes with ${\bf C}/{\bf Z}_N$ singularities in their
world-volumes). We therefore arrive at the conclusion that ``asymmetric'' Type IIB
orientifolds do not have perturbative description if the orbifold group ${\widetilde G}$
(which here we assume to be Abelian) contains elements of the form
\begin{equation}\label{bad}
{\widetilde g}_a={\mbox{diag}}(-1,\rho_a,-\rho^{-1}_a)~,~~~\rho_a\not=\pm1~. 
\end{equation}

{}In fact, the above resolves the following ``puzzle''.
In the $\Omega$ orientifold of Type IIB on $T^6/{\bf Z}_4$ (where the generator of
the orbifold group is defined as ${\widetilde g}z_1=-z_1$, ${\widetilde g}z_2=iz_2$,
${\widetilde g}z_3=iz_3$) it is impossible to cancel all the tadpoles. The tadpole that
is impossible to cancel is precisely the one that contains (in the tree-channel) 
the sum over momenta in the $z_1$ direction as discussed above. Other tadpoles can be
cancelled by a proper choice of the orbifold action on the Chan-Paton charges.
The latter is described via $16\times 16$ (here we choose not to count the
orientifold images of D9- and D5-branes)
matrices $\gamma_{{\widetilde g}^k}$ and ${\widetilde \gamma}_{{\widetilde g}^k}$ 
($k=1,2,3$) corresponding to 
D9- and D5-branes, respectively. The following choice is consistent 
with the ${\bf Z}_2$ model
of Ref \cite{PS,GP} (note that ${\bf Z}_2\subset {\widetilde G}\approx{\bf Z}_4$, where 
${\bf Z}_2$ acts as in the six dimensional model of Ref \cite{PS,GP}): 
\begin{eqnarray}\label{gamma}
 \gamma_{\widetilde g}={\widetilde \gamma}_{\widetilde g}=
 &&{\mbox{diag}}(\exp(\pi i/4)~ (4 ~{\mbox{times}}),
 \exp(-\pi i/4)~ (4 ~{\mbox{times}}),\nonumber\\ 
 &&\exp(3\pi i/4) ~(4 ~{\mbox{times}}),\exp(-3\pi i/4) ~(4~{\mbox{times}}))~.
\end{eqnarray}
The perturbative (from the orientifold viewpoint)
massless spectrum of this model
is given in Table \ref{Z4}\footnote{Here we should point out that
the brane configuration corresponding to the massless spectrum of Table
\ref{Z4} is such that all the D5-branes are located at the same fixed point.}.
That is, we purposefully ignore the non-perturbative states expected to arise 
in the $\Omega {\widetilde g}$ and $\Omega {\widetilde g}^3$ sectors (which is related
to the fact that some of the tadpoles have not been cancelled).   

{}Here we encounter an inconsistency. The massless spectrum 
in Table \ref{Z4} has non-Abelian gauge anomaly: the 
99 and 55 sectors possess $[SU(8)\otimes SU(8)]_{99}$ and 
$[SU(8)\otimes SU(8)]_{55}$
non-Abelian gauge anomalies, respectively, 
whereas the 59 sector is anomaly free. 
(Recall that the $M(M-1)/2$ dimensional antisymmetric representation of $SU(M)$ 
contributes as much as $M-4$ fundamentals of $SU(M)$ into the non-Abelian 
gauge anomaly.) Thus, ignoring the non-perturbative
contributions from the sectors of the type (\ref{bad})
leads (in this particular model) to an apparent space-time inconsistency.     

{}Similar remarks apply to the ${\bf Z}_8$, ${\bf Z}^\prime_8$ and ${\bf Z}^\prime_{12}$
cases. Also, the ${\widetilde {\bf Z}}_2\otimes {\widetilde {\bf Z}}_4$ 
and ${\widetilde {\bf Z}}_4\otimes{\widetilde {\bf Z}}_4$ orbifolds contain ${\bf Z}_4$ as 
a subgroup, so the fact that in the corresponding orientifold models there always are
leftover tadpoles \cite{Zw} is not surprising: these models too lack perturbative orientifold
description as there are non-perturbative contributions from the corresponding
sectors.

\subsection{Other Non-Perturbative Cases}

{}In the previous subsection we have asserted that if an Abelian orbifold group ${\widetilde
G}$ contains elements of type (\ref{bad}) then the corresponding orientifold ought to
include non-perturbative (from the orientifold viewpoint) sectors. This is readily observed in the
${\widetilde {\bf Z}}_2\otimes {\widetilde {\bf Z}}_4$,
${\widetilde {\bf Z}}_4\otimes {\widetilde {\bf Z}}_4$,
${\bf Z}_4$, ${\bf Z}_8$, ${\bf Z}^\prime_8$ and ${\bf Z}^\prime_{12}$ cases where
perturbatively there remain some uncanceled tadpoles. However, 
there are other cases that contain such elements, yet all the tadpoles can be cancelled.
These are the cases with the orbifold groups ${\bf Z}_6^\prime$, ${\widetilde {\bf Z}}_2
\otimes {\widetilde {\bf Z}}_6$, ${\widetilde {\bf Z}}_3 \otimes {\widetilde {\bf Z}}_6$,
${\widetilde {\bf Z}}_6 \otimes {\widetilde {\bf Z}}_6$ \cite{Zw} and ${\bf Z}_{12}$ \cite{Iba}.
Also, in these models the massless (open string) spectra computed using the ``naive'' tadpole
cancellation conditions are free of non-Abelian gauge anomalies \cite{Zw,Iba}. Naively this
appears to be in contradiction with some of the conclusions of the previous subsection. 
However, the issue here seems to be more subtle. We will discuss these subtleties in the
${\bf Z}_6^\prime$ case. Generalization to other cases should be clear.

{}Let us consider the ${\bf Z}_6^\prime$ case in more detail. Let ${\widetilde g}$ be the 
generator of ${\bf Z}_6^\prime$. The perturbative (from the orientifold viewpoint) massless 
spectrum of this model is given in Table \ref{Z6p}. Note that this spectrum is free of
non-Abelian gauge anomalies. Nonetheless, in the following we will argue that this spectrum
is incomplete.

{}According to our discussion in subsection C of section 
\ref{other} we expect non-perturbative (from the orientifold viewpoint) states arising in the
$\Omega {\widetilde g}^k$ sectors with $k=1,5$ and $k=2,4$. In fact, we can deduce the
extra states in the $\Omega {\widetilde g}^2$ plus $\Omega {\widetilde g}^4$ sectors
from the fact that the latter are the same as in the Type I compactification on $(T^4/{\bf Z}_3)
\otimes T^2$ with the same gauge bundle (which is perturbative from the heterotic viewpoint)
as in subsection C of section \ref{FA}. More precisely, these states must be further projected 
to those invariant with respect to the ${\bf Z}_2$ twist. It is not difficult to work out the 
quantum numbers of these states. In particular, we expect the following states (arising in the
$\Omega {\widetilde g}^2$ plus $\Omega {\widetilde g}^4$ sectors) charged under the 99
gauge group (which is $U(4)\otimes U(4)\otimes U(8)$): $9({\bf 6},{\bf 1},{\bf 1})$,
 $9({\bf 1},{\bf 6},{\bf 1})$, $6({\bf 4},{\overline {\bf 4}},{\bf 1})$ and 
$3({\overline {\bf 4}},{\bf 4},{\bf 1})$. (For the sake of simplicity we have suppressed the
$U(1)$ charges.) The multiplicities of these states come from the fixed points 
in the ${\bf Z}_3$ twisted sectors (or, more precisely, their linear combinations 
with respect to the ${\bf Z}_2$ twist). Note that these states give non-zero contributions
into non-Abelian gauge anomalies for the $SU(4)\otimes SU(4)$ subgroups. This implies 
that the $\Omega {\widetilde g}$ plus $\Omega {\widetilde g}^5$ sectors (which are also
expected to give rise to additional non-perturbative states) also contribute to the non-Abelian
gauge anomalies so that the total anomaly cancels. Note that we cannot reliably compute\footnote{Nonetheless, it is possible to guess what these states should look like
from the anomaly cancellation point of view.} these
states as the corresponding heterotic string sectors are 
non-perturbative\footnote{In particular, the level matching constraint is not satisfied
in these sectors for the corresponding choice of the gauge bundle.} (from the 
heterotic viewpoint). An important observation here is that the 
$\Omega {\widetilde g}^2$ plus $\Omega {\widetilde g}^4$ sector states must be included
(as including only 
the $\Omega {\widetilde g}$ plus $\Omega {\widetilde g}^5$ sector states would
result in an anomalous model)\footnote{It is not difficult to see
that the blow-ups cannot result in decoupling of the
extra non-perturbative states since the required terms in the superpotential are absent due to
the discrete symmetries of the ${\bf Z}_6^\prime$ orbifold.}.
This confirms our assertion in subsection C of section \ref{other}
that the orientifold projection must be the same in all twisted sectors (which in this case
corresponds to the $\Omega$ projection which after the required blow-ups results in
Type I compactification on the corresponding Calabi-Yau three-fold).
In the next subsection we will perform an independent check for the conclusion of this subsection that the perturbative orientifold approach to the ${\bf Z}_6^\prime$ model misses
relevant non-perturbative states. It is not difficult to see that the same conclusions 
extend to the ${\widetilde {\bf Z}}_2
\otimes {\widetilde {\bf Z}}_6$, ${\widetilde {\bf Z}}_3 \otimes {\widetilde {\bf Z}}_6$,
${\widetilde {\bf Z}}_6 \otimes {\widetilde {\bf Z}}_6$ and ${\bf Z}_{12}$ cases.
Note that these models are examples of orientifolds where non-perturbative (from the
orientifold viewpoint) states come in such combinations so that they do not contribute
into non-Abelian gauge anomalies (and this is precisely the reason why all the
``naive'' tadpoles are cancelled).

\subsection{Another Check}

{}The above discussion implies that the ${\bf Z}_6^\prime$, ${\widetilde {\bf Z}}_2
\otimes {\widetilde {\bf Z}}_6$, ${\widetilde {\bf Z}}_3 \otimes {\widetilde {\bf Z}}_6$,
${\widetilde {\bf Z}}_6 \otimes {\widetilde {\bf Z}}_6$, 
${\widetilde {\bf Z}}_2\otimes {\widetilde {\bf Z}}_4$ 
and ${\widetilde {\bf Z}}_4\otimes{\widetilde {\bf Z}}_4$ cases
of Ref \cite{Zw}, as well as the ${\bf Z}_8$, ${\bf Z}^\prime_8$, ${\bf Z}^\prime_{12}$ and
${\bf Z}_{12}$ cases of Ref \cite{Iba} should be non-perturbative from the orientifold 
viewpoint. On the other hand, the only cases that can be treated perturbatively in the
orientifold framework should be the ${\widetilde {\bf Z}}_2\otimes {\widetilde {\bf Z}}_2$
\cite{BL}, ${\bf Z}_3$ \cite{Sagnotti}, ${\bf Z}_7$ \cite{KS1},
${\widetilde {\bf Z}}_3\otimes {\widetilde {\bf Z}}_3$ and ${\bf Z}_6$ \cite{KS2} and 
${\widetilde {\bf Z}}_2\otimes {\widetilde {\bf Z}}_6^\prime$ \cite{zk} cases. The arguments 
presented up till now all indicate that this must be the case. On the other hand, due to a
rather involved (and intertwined) nature of these arguments it would be desirable to
perform a simple yet independent check for perturbative consistency of these models.
Fortunately, such a check can be performed.

{}Here we observe that the question of whether an orientifold of Type IIB on a given orbifold
contains extra non-perturbative states is really a {\em local} question as far as the geometry 
is concerned. That is, we should be able to test this issue in a local framework where
the ``compactification'' space is non-compact. This is because the question of whether there
are non-perturbative states in a given orbifold model depends on local considerations 
of whether there are states coming from sectors corresponding to certain D-branes
wrapping various 
collapsed 2-cycles at orbifold singularities. This observation can be utilized in the
framework recently discussed in Ref \cite{zura}.

{}Thus, consider the $\Omega J$ orientifold of Type IIB on ${\widetilde {\cal W}}_3=
{\bf C}^3/{\widetilde G}$ where ${\widetilde G}$ is any of the above (Abelian) 
orbifold groups, and
the action of $J$ is given by $Jz_i=-z_i$ ($z_i$, $i=1,2,3$, are the complex coordinates 
parametrizing ${\bf C}^3$). This orientifold contains orientifold 3-planes and an arbitrary
number of D3-branes\footnote{The number of the D3-branes is unconstrained due to the
fact that the space transverse to the D3-branes is non-compact.}. If the orbifold group
contains a ${\bf Z}_2$ subgroup, then there also are present the corresponding 
orientifold 7-planes
which are accompanied by 8 of the corresponding D7-branes. Here we can ask whether
such an orientifold model is consistent, in particular, if all the tadpoles can be cancelled.
Here we will skip all the details as the corresponding calculations are completely analogous 
to those discussed in Ref \cite{zura}, and will simply state the answer. The details can be found 
in Ref \cite{zura1}.

{}It is not difficult to show that the ``naive'' tadpole cancellation conditions
have a solution (which is unique in each of the following cases) only for the 
${\widetilde {\bf Z}}_2\otimes {\widetilde {\bf Z}}_2$
${\bf Z}_3$, ${\bf Z}_7$,
${\widetilde {\bf Z}}_3\otimes {\widetilde {\bf Z}}_3$, ${\bf Z}_6$ and 
${\widetilde {\bf Z}}_2\otimes {\widetilde {\bf Z}}_6^\prime$ cases\footnote{These solutions
give rise to four dimensional ${\cal N}=1$ supersymmetric gauge theories which are
free of non-Abelian gauge anomalies for any value $N$ of the number of D3-branes.
This has been explicitly checked for the ${\widetilde {\bf Z}}_2\otimes {\widetilde {\bf Z}}_2$,
${\bf Z}_3$, ${\bf Z}_7$ cases in Ref \cite{zura}. The remaining three cases are not difficult
to work out along the lines of Ref \cite{zura} - see Ref \cite{zura1} for details.}. On the other hand, in all of the
${\bf Z}_6^\prime$, ${\widetilde {\bf Z}}_2
\otimes {\widetilde {\bf Z}}_6$, ${\widetilde {\bf Z}}_3 \otimes {\widetilde {\bf Z}}_6$,
${\widetilde {\bf Z}}_6 \otimes {\widetilde {\bf Z}}_6$, 
${\widetilde {\bf Z}}_2\otimes {\widetilde {\bf Z}}_4$, 
${\widetilde {\bf Z}}_4\otimes{\widetilde {\bf Z}}_4$,
${\bf Z}_8$, ${\bf Z}^\prime_8$, ${\bf Z}^\prime_{12}$ and
${\bf Z}_{12}$ cases there are left-over uncanceled tadpoles (that is, the tadpole cancellation
conditions do not have a solution). This is precisely due to the fact that in these models
there are extra non-perturbative states which are not captured by the ``naive'' perturbative
orientifold construction. This test is a very non-trivial piece of evidence for correctness of our
previous discussions.

\section{Summary and Remarks}\label{disc}

{}Let us summarize some of the main conclusions of the previous discussions.\\
$\bullet$ Orientifolds of Type IIB on non-geometric (``symmetric'') toroidal orbifolds
always contain non-perturbative (from the orientifold viewpoint) sectors. The appropriate
framework for considering such orientifolds is F-theory.\\
$\bullet$ In six dimensions there are two choices for the orientifold projection
in Type IIB on geometric (``asymmetric'') orbifolds. The first one (once the
appropriate blow-ups are performed) corresponds to Type I compactifications on K3
(which have only one tensor multiplet in the massless spectrum)
with certain choices of the gauge bundle. These models contain non-perturbative
(from the orientifold viewpoint) sectors except for the case of $T^4/{\bf Z}_2$. The second
choice of the orientifold projection leads to the models of Refs \cite{GJ,DP} with more than 
one tensor multiplets. These models can be checked to be consistent {\em away} from the
orbifold conformal field theory points from various points of view (including the map to 
F-theory).\\
$\bullet$ The story with ${\cal N}=1$ orientifolds of Type IIB on geometric (``symmetric'')
orbifolds $T^6/{\widetilde G}$ is more involved, however. First, (unlike the
six dimensional cases) there is only one consistent choice of the orientifold projection.
This choice corresponds to Type I compactifications on Calabi-Yau three-folds
obtained by appropriately blowing up the corresponding orbifolds $T^6/{\widetilde G}$.
Such compactifications generically contain non-perturbative 
(from the orientifold viewpoint) sectors. An obvious exception is the ${\widetilde {\bf Z}}_2
\otimes {\widetilde {\bf Z}}_2$ model of Ref \cite{BL} which has perturbative 
orientifold description. More non-trivial examples are ${\bf Z}_3$ \cite{Sagnotti}, 
${\bf Z}_7$ \cite{KS1},
${\widetilde {\bf Z}}_3\otimes {\widetilde {\bf Z}}_3$ and ${\bf Z}_6$ \cite{KS2} and 
${\widetilde {\bf Z}}_2\otimes {\widetilde {\bf Z}}_6^\prime$ \cite{zk} cases. In these
models the expected non-perturbative states decouple from the massless spectrum
after blow-ups which can be explicitly checked using Type I-heterotic duality
along the lines of Ref \cite{ZK} (and also Refs \cite{KS1,KS2}).\\
$\bullet$ The other four dimensional examples, namely, the 
${\bf Z}_6^\prime$, ${\widetilde {\bf Z}}_2
\otimes {\widetilde {\bf Z}}_6$, ${\widetilde {\bf Z}}_3 \otimes {\widetilde {\bf Z}}_6$,
${\widetilde {\bf Z}}_6 \otimes {\widetilde {\bf Z}}_6$, 
${\widetilde {\bf Z}}_2\otimes {\widetilde {\bf Z}}_4$ and  
${\widetilde {\bf Z}}_4\otimes{\widetilde {\bf Z}}_4$ cases discussed in Ref \cite{Zw},
as well as the
${\bf Z}_8$, ${\bf Z}^\prime_8$, ${\bf Z}^\prime_{12}$ and
${\bf Z}_{12}$ cases discussed in Ref \cite{Iba} appear to suffer from non-perturbative
(from the orientifold viewpoint) contributions to the massless spectrum. The ``naive''
orientifold approach used in Refs \cite{Zw,Iba} to study these cases is therefore
inadequate.\\
$\bullet$ The ${\bf Z}_6$ model of Ref \cite{KS2} is the first known example of a 
consistent {\em chiral} ${\cal N}=1$ supersymmetric four dimensional vacuum which is 
{\em non-perturbative} from the heterotic viewpoint. Another example of such a vacuum
is the ${\widetilde {\bf Z}}_2\otimes {\widetilde {\bf Z}}_6^\prime$ model of Ref 
\cite{zk}. An example of a consistent {\em non-chiral} ${\cal N}=1$ 
supersymmetric four dimensional vacuum is the ${\widetilde {\bf Z}}_2
\otimes {\widetilde {\bf Z}}_2$ model of Ref \cite{BL}. The ${\bf Z}_3$ model of Ref 
\cite{Sagnotti}, the ${\bf Z}_7$ model of Ref \cite{KS1}, as well as the
${\widetilde {\bf Z}}_3\otimes {\widetilde {\bf Z}}_3$ model of Ref \cite{KS2}
are chiral but correspond to {\em perturbative} heterotic compactifications.\\
$\bullet$ Orientifolds of Type IIB on non-Abelian orbifolds with $SU(3)$ holonomy
contain mutually non-local orientifold planes and D-branes and, therefore, are 
non-perturbative from the orientifold viewpoint. The appropriate
framework for considering such orientifolds is F-theory.

{}Next, we would like to outline some directions for future study.\\
$\bullet$ It is clear from our previous discussions that four dimensional orientifolds
should be viewed as Type I compactifications on smooth (except for possible ${\bf Z}_2$
orbifold singularities) Calabi-Yau three-folds with certain choices of the gauge bundle.
It is therefore conceivable that a more geometric approach to Type I compactifications
could be useful, in particular, in determining which choices of the gauge bundle 
correspond to perturbative orientifolds for a given Calabi-Yau three-fold.\\
$\bullet$ Given the consistent four dimensional perturbative orientifolds of Type IIB
on the ${\widetilde {\bf Z}}_2
\otimes {\widetilde {\bf Z}}_2$, ${\bf Z}_3$, 
${\bf Z}_7$,
${\widetilde {\bf Z}}_3\otimes {\widetilde {\bf Z}}_3$, ${\bf Z}_6$ and 
${\widetilde {\bf Z}}_2\otimes {\widetilde {\bf Z}}_6^\prime$ orbifolds, it would be
interesting to extend the recent results of Ref \cite{KST} in six dimensions to four
dimensional orientifolds with non-trivial NS-NS antisymmetric tensor backgrounds.
(Such compactifications in the ${\bf Z}_3$ case were briefly discussed in Ref 
\cite{Sagnotti}.)\\
$\bullet$ Finally, it would be interesting to write down all ${\cal N}=1$ gauge theories
from orientifolds in the context of the setup recently discussed in Ref \cite{zura} such that
the orientifolds are perturbatively well defined. This would provide a list of
additional four dimensional gauge theories that possess certain nice properties in the
large $N$ limit. Also, as suggested in Ref \cite{zura}, it would be interesting to understand
tadpole (and anomaly) free ${\cal N}=0$ orientifolds that would also possess such 
properties.

\acknowledgements

{}We would like to thank 
Carlo Angelantonj, Philip Argyres, Oren Bergman,
Michael Bershadsky, Loriano Bonora, Chong-Sun Chu, Gregory Gabadadze,
Edi Gava,
Eric Gimon, Brian Greene, Roberto Iengo,
Andrei Johansen, Clifford Johnson, Albion Lawrence, K.S. Narain, Pran Nath, 
Jaemo Park, Augusto Sagnotti,
Ashoke Sen, Savdeep Sethi, 
Tom Taylor, Edward Witten and Piljin Yi for discussions. 
We are especially grateful to Cumrun Vafa for enlightening discussions 
and valuable observations. The research of G.S. and S.-H.H.T. was partially 
supported by the 
National Science Foundation. 
G.S. would like to thank the theory groups at SISSA and ICTP 
for their kind hospitality during his stay at Trieste.
G.S. would also like to thank
Joyce M. Kuok Foundation for financial support.
The work of Z.K. was supported in part by the grant NSF PHY-96-02074, 
and the DOE 1994 OJI award. Z.K. would like to thank the School of Natural Sciences at 
the Institute for Advanced Study for their kind hospitality while parts of this work were completed.
Z.K. would also like to thank Albert and Ribena Yu for 
financial support.     

\appendix

\section{Chiral Bosons}\label{Orbifold}

{}Consider a single free left-moving complex boson with the monodromy
\begin{equation}\label{monobos}
 \partial \phi_v (z e^{2\pi i} )=e^{-2\pi i v}\partial \phi_v (z)~,
 ~~~0\leq v<1~.
\end{equation}
The field $\partial \phi_v (z)$ has the following mode expansion
\begin{eqnarray}
 i\partial \phi_v (z) =&&\delta_{v,0} p z^{-1} + (1-\delta_{v,0} ) \sqrt{v}
 \, b_v z^{-v-1} + \nonumber\\
 &&\sum_{n=1}^{\infty} \{ {\sqrt{n+v}}\,
 b_{n+v} z^{-n-v-1} +{\sqrt{n-v}}\, d^{\dagger}_{n-v} z^{n-v-1} \}~.
\end{eqnarray}
Here $b^{\dagger}_r$ and $d^{\dagger}_s$ are creation operators, and $b_r$ and
$d_s$ are annihilation operators. The quantization conditions read
\begin{equation}
 [ b_r ,b^{\dagger}_{r^\prime} ]=\delta_{r r^\prime} ,~~~
 [ d_s ,d^{\dagger}_{s^\prime} ]=\delta_{s s^\prime} ,~~~
 [x^{\dagger} ,p]=[x,p^{\dagger} ]=i,~~~
 \mbox{others vanish}.
\end{equation}
The Hamiltonian $H_v$ and angular momentum operator $M_v$ are given by 
\begin{eqnarray}
 H_v& = &\delta_{v,0} pp^{\dagger} + (1-\delta_{v,0})vb^{\dagger}_{v} b_{v} + 
 \sum_{n=1}^{\infty} \{ (n+v)
   b^{\dagger}_{n+v} b_{n+v} 
   +(n-v)d^{\dagger}_{n-v} d_{n-v} \} +\nonumber\\
 &&{v(1-v) \over{2}}-{1\over{12}} ~,\\
 M_v &=&\delta_{v,0}i(xp^{\dagger}-x^{\dagger} p)-
 (1-\delta_{v,0})b^{\dagger}_{v} b_{v} -
   \sum_{n=1}^{\infty} \{ b^{\dagger}_{n+v} b_{n+v}
   -d^{\dagger}_{n-v} d_{n-v} \} ~.
\end{eqnarray}
Note that the vacuum energy is ${1 \over{2}}v(1-v)-{1\over{12}}$.

{}The operator $M_v$
is the generator of $U(1)$ rotations.
The corresponding characters read ($v+u\not=0$):
\begin{eqnarray}\label{bosonX}
 X^v_u =&&\mbox{Tr}(q^{H_v} g(u))= \mbox{Tr}(q^{H_v} \exp 
 (2\pi i u M_v ))= \nonumber\\
 &&q^{{v(1-v) \over{2}}-{1\over{12}}} 
 (1-(1-\delta_{v,0} ) q^v e^{-2\pi i u} )^{-1} 
 \prod_{n=1}^{\infty} (1-q^{n+v}
 e^{-2\pi i u} )^{-1} (1-q^{n-v} e^{2\pi i u} )^{-1} ~.
\end{eqnarray}

{}Under the generators of modular transformations
the characters (\ref{bosonX}) transform as
\begin{eqnarray}\label{modularS}
 X^v_u &\stackrel{S}{\rightarrow}& 
 ( 2\sin(\pi u) \delta_{v,0} +[2\sin(\pi v)]^{-1}
  \delta_{u,0} + (1-\delta_{vu,0})
 e^{-2\pi i(v-1/2)(u-1/2)} ) X^u_{-v} ~,\\
 \label{modularT}
 X^v_u &\stackrel{T}{\rightarrow}&
 e^{2\pi i({{v(1-v)}\over 2} -{1\over{12}})} X^v_{u-v} ~. 
\end{eqnarray}

{}Next, consider a single free right-moving complex boson with the monodromy
\begin{equation}\label{monobos1}
 {\overline\partial} {\overline \phi}_v ({\overline z} e^{-2\pi i} )=
 e^{+2\pi i v}{\overline\partial} {\overline \phi}_v ({\overline z})~,
 ~~~0\leq v<1~.
\end{equation}
The field ${\overline\partial} {\overline \phi}_v ({\overline z})$ has the same mode
expansion as the field $\partial \phi_v (z)$ (after replacing all left-moving quantities 
by their right-moving counterparts). The corresponding characters read 
($v+u\not=0$):
\begin{eqnarray}\label{bosonX1}
 {\overline X}^v_u =&&\mbox{Tr}({\overline q}^{{\overline H}_v} 
 {\overline g}(u))= \mbox{Tr}({\overline q}^{{\overline H}_v} \exp 
 (-2\pi i u {\overline M}_v ))= \nonumber\\
 &&{\overline q}^{{v(1-v) \over{2}}-{1\over{12}}} 
 (1-(1-\delta_{v,0} ) {\overline q}^v e^{2\pi i u} )^{-1} 
 \prod_{n=1}^{\infty} (1-{\overline q}^{n+v}
 e^{2\pi i u} )^{-1} (1-{\overline q}^{n-v} e^{-2\pi i u} )^{-1} ~.
\end{eqnarray}
Note that ${\overline X}^v_u$ is complex conjugate of $X^v_u$. The modular 
transformations for the characters ${\overline X}^v_u$ are therefore given by Eqs
(\ref{modularS}) and (\ref{modularT}) with all the quantities (including the phases) 
replaced by their complex conjugates.

{}Now consider an orbifold model where we have the following ground state
in the twisted sector: $\sigma_v\vert 0\rangle_L\otimes 
{\overline \sigma}_{\overline v}\vert 0\rangle_R$. Following the discussion in 
section \ref{prelim}, we have two possibilities: ${\overline v}=v$ (``symmetric'' 
orbifolds), and ${\overline v}=1-v$ (``asymmetric'' orbifolds). One of the twisted sector 
characters that enter the partition function is (up to a constant) given by $X^v_0 
{\overline X}^{\overline v}_0$. Under $S$ modular transformation this 
(up to a constant) is mapped  to an untwisted sector character $X^0_{1-v} 
{\overline X}^0_{1-{\overline v}}$. From this it is not difficult to see that the twist 
operator $g(v,{\overline v})$ in the untwisted sector is given by
\begin{equation}
 g(v,{\overline v}) =g(v){\overline g} ({\overline v})=\exp\left(2\pi i (vM_v-
 {\overline v} M_{\overline v})\right)~.
\end{equation}
Thus, for ``symmetric'' orbifolds the left- and right-moving contributions enter 
with the Lorentzian signature, whereas for the ``asymmetric'' orbifolds the left- and 
right-moving contributions enter with the Euclidean signature, as we pointed out in 
section \ref{prelim}. 

\section{Boundary Conditions}\label{MIX}

{}Consider a single free complex world-sheet boson $\phi(\sigma,\tau)$
with the following boundary conditions:
\begin{eqnarray}
 &&\left.\left(\cos(\pi v_1) \partial_\sigma \phi -\sin(\pi v_1) \partial_\tau \phi
 \right)\right|_{\sigma=0}=0~,\\
 &&\left.\left(\cos(\pi v_2) \partial_\sigma \phi -\sin(\pi v_2) \partial_\tau \phi
 \right)\right|_{\sigma=\pi}=0~,  
\end{eqnarray} 
where $\sigma$ and $\tau$ are the space-like world-sheet coordinates, respectively.
Without loss of generality we can assume that $0\leq v_1,v_2,v<1$, where 
$v\equiv v_2-v_1$. Then the mode expansion for $\phi(\sigma,\tau)$ is given by:
\begin{eqnarray}
 \phi(\sigma,\tau)=&&x+2(p\tau+w\sigma)-i\sum_{n=1}^\infty
 \left\{{\sqrt{n+v-1}}\, b_{n+v-1} \cos\left[(n+v-1)\sigma+\pi v_1\right]
 {\mbox{e}}^{-i(n+v-1)\tau}
 \right.\nonumber\\
 &&\left. +{\sqrt{n-v}}\, d^\dagger_{n-v} \cos\left[(n-v)\sigma-\pi v_1\right]
 {\mbox{e}}^{i(n-v)\tau}\right\}~. 
\end{eqnarray}
Here $b_{n+v-1},d_{n-v}$ are the annihilation operators, while $b^\dagger_{n+v-1},
d^\dagger_{n-v}$ are the creation operators. The momenta $p$ and windings $w$
cannot be arbitrary but satisfy the following conditions: $w=0$ if $v_1=v_2=0$;
$p=0$ if $v_1=v_2=1/2$; and $p=0$, $w=0$ in all the other cases. The physical
interpretation of these conditions is the familiar concept of momenta and/or windings
not flowing through the boundaries in the tree-channel amplitude.

{}The D-brane picture arises for the Dirichlet boundary conditions. Thus, for instance,
if $v_1=v_2=1/2$ then we have DD boundary conditions, and each endpoint of
the string (at $\sigma=0$ and $\sigma=\pi$) is stuck at the same position at all times 
$\tau$. We therefore have D-brane interpretation: D-branes are space-time defects
on which open strings can start and end. If, however, we have $v_1=v_2\not=0,1/2$
then the end-points harmonically oscillate around some fixed points in the 
corresponding space-like direction. This implies that there is no D-brane interpretation 
for such boundary conditions. 

\section{Some Voisin-Borcea Orbifolds}\label{VoBo}

{}In this section we provide some detail concerning the $\Omega J J^\prime (-1)^{F_L}$
orientifolds discussed in subsection B of section \ref{FA}. 
Thus, consider the $\Omega JJ^\prime (-1)^{F_L}$ orientifold
of Type IIB on ${\widetilde {\cal M}}_2=(T^2\otimes T^2)/{\bf Z}_N$, $N=3,4,6$ 
where $J$ acts as $Jz_1=-z_1$, $Jz_2=z_2$,
and the action of $J^\prime$ was discussed in subsection A of section \ref{other}. 
The corresponding Voisin-Borcea orbifold is given by $(T^2\otimes T^2\otimes
T^2)/({\bf Z}_2\otimes{\bf Z}_N)$. The generator $S$ of the ${\bf Z}_2$ twist acts as
follows: $Sz_0=-z_0$, $Sz_1=-z_1$, $Sz_2=z_2$. The generator ${\widetilde g}$
of the ${\bf Z}_N$ twist has the following action: ${\widetilde g}z_0=z_0$, 
${\widetilde g}z_1=\omega z_1$, ${\widetilde g}z_2=\omega^{-1} z_2$, where 
$\omega=\exp(2\pi i/N)$. In the ${\widetilde g}^k$, $k=1,\dots,N-1$, $2k\not=N$, the
$S$ twist is accompanied by the action of $J^\prime$. This interchanges 
${\widetilde g}^k$ and ${\widetilde g}^{N-k}$ twisted sectors. That is, states
from these sectors combine together into linear combinations that are invariant
under the action of the orbifold. Note that there are not $Sg^k$ twisted sectors with 
$k=1,\dots,N-1$, $2k\not=N$.
In the ${\widetilde g}^{N/2}$ twisted sector (for even $N$) we can have 
discrete torsion. 

{}Let us consider each case in a bit more detail. We will give the contributions
from each sector into the Hodge numbers $(h^{1,1},h^{2,1})$.\\ 
$\bullet$ $N=3$:\\
Untwisted: (3,1);
${\widetilde g}\oplus{\widetilde g}^2$: (9,9);
$S$: (8,4);\\
Total: (20,14).\\
$\bullet$ $N=4$, without discrete torsion:\\
Untwisted: (3,1);
${\widetilde g}\oplus{\widetilde g}^3$: (4,4);
${\widetilde g}^2$: (10,0);
$S$: (12,0);
$S{\widetilde g}^2$: (12,0);\\
Total: (41,5).\\
$\bullet$ $N=4$, with discrete torsion:\\
Untwisted: (3,1);
${\widetilde g}\oplus{\widetilde g}^3$: (4,4);
${\widetilde g}^2$: (0,10);
$S$: (0,12);
$S{\widetilde g}^2$: (0,4);\\
Total: (7,31).\\
$\bullet$ $N=6$, without discrete torsion:\\
Untwisted: (3,1);
${\widetilde g}\oplus{\widetilde g}^5$: (1,1);
${\widetilde g}^2\oplus{\widetilde g}^4$: (5,5);
${\widetilde g}^3$: (6,0);
$S$: (8,0);
$S{\widetilde g}^3$: (8,0);\\
Total: (31,7).\\
$\bullet$ $N=6$, with discrete torsion:\\
Untwisted: (3,1);
${\widetilde g}\oplus{\widetilde g}^5$: (1,1);
${\widetilde g}^2\oplus{\widetilde g}^4$: (5,5);
${\widetilde g}^3$: (0,6);
$S$: (0,4);
$S{\widetilde g}^3$: (0,4);\\
Total: (9,21).

\section{F-theory Duals of 6D CHL Strings}\label{CHL}

{}CHL heterotic strings in six dimensions are heterotic vacua with ${\cal N}=2$ 
supersymmetry and the rank of the gauge group (coming from the right-moving
world-sheet degrees of freedom) which is $r_L=12$ or $8$. In contrast, 
the Narain (that is, toroidal) compactifications of heterotic string yield 
${\cal N}=2$ supersymmetric vacua with $r_L=20$. In the latter case the we have a dual Type
IIA compactification, namely, on K3. This in turn is dual to F-theory on K3$\otimes T^2$. 
The Hodge numbers $(h^{1,1},h^{2,1})$ for this Calabi-Yau three-fold are
$(h^{1,1},h^{2,1})=(21,21)$. Note that this manifold has $SU(2)$ holonomy.

{}We can ask what would be the F-theory duals of CHL strings with $r_L=12$ and $8$. 
It is not difficult to see that these must be F-theory compactifications on Calabi-Yau
three-folds with $SU(2)$ holonomy and the Hodge numbers 
$(h^{1,1},h^{2,1})=(r_L+1,r_L+1)=(13,13)$ and $(9,9)$, respectively.
In the following we present explicit construction of these three-folds.

{}$\bullet$ Consider the following quotient: 
${\cal W}=(T^2\otimes T^2\otimes T^2)/{\bf Z}_2$. Let the complex coordinates 
corresponding to the three $T^2$'s be $z_1,z_2,z_3$. Then the generator $R$ of 
${\bf Z}_2$ acts as follows: $Rz_1=-z_1$, $Rz_2=z_3$, $Rz_3=z_2$. It is not difficult 
to see that this Calabi-Yau three-fold has $SU(2)$ holonomy and the Hodge numbers
$(h^{1,1},h^{2,1})=(9,9)$.

{}$\bullet$ Consider the following quotient: 
${\cal W}=(T^2\otimes S^1\otimes S^1 \otimes S^1\otimes S^1)/{\bf Z}_2$. 
Then the generator $R$ of 
${\bf Z}_2$ acts as follows. 
It reverses the sign of the complex coordinate on $T^2$, permutes the first two circles,
reverses the sign of the real coordinate on the third circle, and leaves the fourth circle
unaffected. It is not difficult 
to see that this Calabi-Yau three-fold has $SU(2)$ holonomy and the Hodge numbers
$(h^{1,1},h^{2,1})=(13,13)$.

%%%%%%%%%%%%%%% FIGURE 1 %%%%%%%%%%%%%%%%%%%%%%%%
\newpage
\begin{figure}[t]
\hspace*{2.15 cm}
%\vspace*{}
\epsfxsize=10 cm
\epsfbox{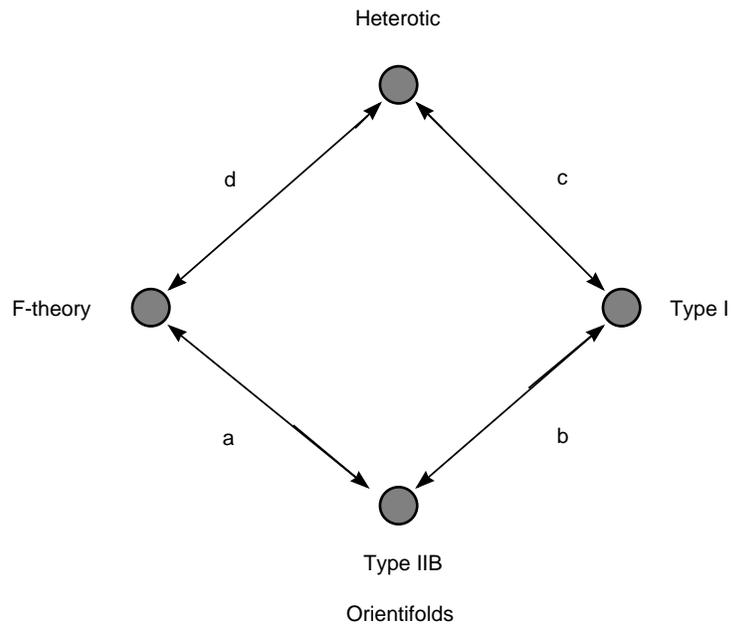}
\caption{The relations between Type IIB orientifolds, Type I, heterotic and F-theory.}
\end{figure}
%%%%%%%%%%%%%%%%%%%%%%%%%%%%%%%%%%%%%%%%%%%%%%%%

%%%%%%%%%%%%%%%% FIGURE 2 %%%%%%%%%%%%%%%
\newpage
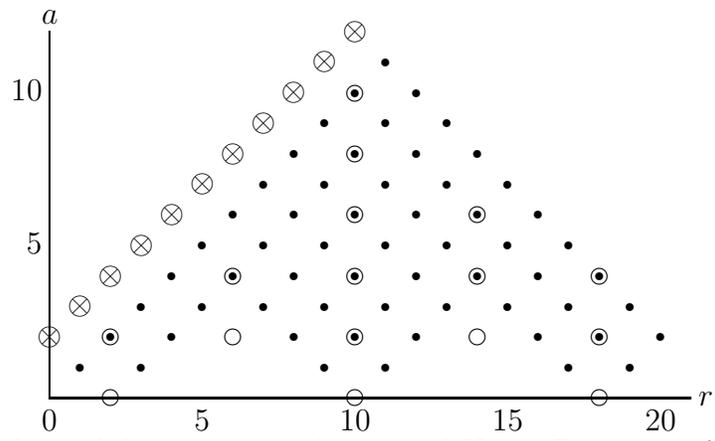
\begin{figure}
\setlength{\unitlength}{0.008in}%
$$\begin{picture}(445,266)(60,385)
\thinlines
\put(72,435){$\otimes$}
\put(92,455){$\otimes$}
\put(112,475){$\otimes$}
\put(132,495){$\otimes$}
\put(152,515){$\otimes$}
\put(172,535){$\otimes$}
\put(192,555){$\otimes$}
\put(212,575){$\otimes$}
\put(232,595){$\otimes$}
\put(252,615){$\otimes$}
\put(272,635){$\otimes$}

\put(100,420){\circle*{6}}

\put(140,420){\circle*{6}}
\put(260,420){\circle*{6}}
\put(300,420){\circle*{6}}
\put(420,420){\circle*{6}}
\put(460,420){\circle*{6}}
\put(120,440){\circle*{6}}
\put(160,440){\circle*{6}}
\put(240,440){\circle*{6}}
\put(280,440){\circle*{6}}
\put(320,440){\circle*{6}}
\put(400,440){\circle*{6}}
\put(440,440){\circle*{6}}
\put(140,460){\circle*{6}}
\put(180,460){\circle*{6}}
\put(220,460){\circle*{6}}
\put(260,460){\circle*{6}}
\put(300,460){\circle*{6}}
\put(340,460){\circle*{6}}
\put(380,460){\circle*{6}}
\put(420,460){\circle*{6}}
\put(160,480){\circle*{6}}
\put(200,480){\circle*{6}}
\put(240,480){\circle*{6}}
\put(280,480){\circle*{6}}
\put(360,480){\circle*{6}}
\put(400,480){\circle*{6}}
\put(180,500){\circle*{6}}
\put(220,500){\circle*{6}}
\put(260,500){\circle*{6}}
\put(300,500){\circle*{6}}
\put(340,500){\circle*{6}}
\put(380,500){\circle*{6}}
\put(200,520){\circle*{6}}
\put(240,520){\circle*{6}}
\put(280,520){\circle*{6}}
\put(320,520){\circle*{6}}
\put(360,520){\circle*{6}}
\put(220,540){\circle*{6}}
\put(260,540){\circle*{6}}
\put(300,540){\circle*{6}}
\put(340,540){\circle*{6}}
\put(240,560){\circle*{6}}
\put(280,560){\circle*{6}}
\put(320,560){\circle*{6}}
\put(260,580){\circle*{6}}
\put(300,580){\circle*{6}}
\put(280,600){\circle*{6}}
\put(320,480){\circle*{6}}
\put(120,400){\circle{10}}
\put(280,400){\circle{10}}
\put(440,400){\circle{10}}
\put(200,440){\circle{10}}
\put(200,480){\circle{10}}
\put(360,440){\circle{10}}
\put(360,480){\circle{10}}
\put(280,480){\circle{10}}
\put(280,520){\circle{10}}
\put(280,560){\circle{10}}
\put(280,600){\circle{10}}
\put(280,440){\circle{10}}
\put(120,440){\circle{10}}
\put(440,440){\circle{10}}
\put( 80,400){\line( 1, 0){420}}
\put( 80,400){\line( 0, 1){240}}
\put(300,620){\circle*{6}}
\put(320,600){\circle*{6}}
\put(340,580){\circle*{6}}
\put(380,540){\circle*{6}}
\put(400,520){\circle*{6}}
\put(420,500){\circle*{6}}
\put(440,480){\circle*{6}}
\put(440,480){\circle{10}}
\put(460,460){\circle*{6}}
\put(480,440){\circle*{6}}
\put(360,520){\circle{10}}
\put(360,560){\circle*{6}}
\put( 75,379){\makebox(0,0)[lb]{\raisebox{0pt}[0pt][0pt]{0}}}
\put(175,379){\makebox(0,0)[lb]{\raisebox{0pt}[0pt][0pt]{5}}}
\put(270,379){\makebox(0,0)[lb]{\raisebox{0pt}[0pt][0pt]{10}}}
\put(370,379){\makebox(0,0)[lb]{\raisebox{0pt}[0pt][0pt]{15}}}
\put(470,379){\makebox(0,0)[lb]{\raisebox{0pt}[0pt][0pt]{20}}}
\put( 65,495){\makebox(0,0)[lb]{\raisebox{0pt}[0pt][0pt]{5}}}
\put( 55,595){\makebox(0,0)[lb]{\raisebox{0pt}[0pt][0pt]{10}}}
\put(505,395){\makebox(0,0)[lb]{\raisebox{0pt}[0pt][0pt]{$r$}}}
\put( 75,645){\makebox(0,0)[lb]{\raisebox{0pt}[0pt][0pt]{$a$}}}
\end{picture}$$
	\caption{Open circles and dots represent the original
                         Voisin--Borcea orbifolds.
                         The line of $\otimes$'s corresponds to the extension
                         discussed in section VIII.}
	\label{figVB}
\end{figure}

%%%%%%%%%%%%%Table I %%%%%%%%
%%%%%%%%%%%%%%%%%%%%%%%%%%%%%%%%%%%%%%%%%%%%%%%%%%%%%%%%%%%%%%%%%%%%%%%%%%%%%%%
\begin{table}[t]
\begin{tabular}{|c|c|l|l|}
%%%%%%%%%%%%%%%%%%%%%%%%%%%%%%%%%%%%%%%%%%%%%%%%%%%%%%%%%%%%%%%%%%%%%%%%%%%%
Sector & Field & $SU(12)\otimes SO(8) \otimes U(1)$ & Comments\\
\hline
%%%%%%%%%%%%%%%%%%%%%%%%%%%%%%%%%%%%%%%%%%%%%%%%%%%%%%%%%%%%%%%%%%%%%%%%%%%%
Closed & & &\\
Untwisted & $\phi_{ab}$ & $9({\bf 1}, {\bf 1})(0)_L$ & $a,b=1,2,3$\\
\hline
%%%%%%%%%%%%%%%%%%%%%%%%%%%%%%%%%%%%%%%%%%%%%%%%%%%%%%%%%%%%%%%%%%%%%%%%%%%%
Closed & & &\\
Twisted & $S_{\alpha\beta\gamma}$ & $27({\bf 1}, {\bf 1})(0)_L$ & $\alpha,\beta,\gamma=1,2,3$ \\
\hline
%%%%%%%%%%%%%%%%%%%%%%%%%%%%%%%%%%%%%%%%%%%%%%%%%%%%%%%%%%%%%%%%%%%%%%%%%%%%
Open & $Q_a$ & $3({\bf 12},{\bf 8}_v)(-1)_L$ & \\
     & $\Phi_a$ & $3({\overline {\bf 66}}, {\bf 1})(+2)_L$ & $a=1,2,3$\\
%%%%%%%%%%%%%%%%%%%%%%%%%%%%%%%%%%%%%%%%%%%%%%%%%%%%%%%%%%%%%%%%%%%%%%%%%%%
\end{tabular}
%%%%%%%%%%%%%%%%%%%%%%%%%%%%%%%%%%%%%%%%%%%%%%%%%%%%%%%%%%%%%%%%%%%%%%%%%%%
\caption{The massless spectrum of the Type I ${\bf Z}_3$ orbifold
model with $N=1$ space-time supersymmetry 
and gauge group $SU(12)\otimes SO(8) \otimes U(1)$ discussed in section IX. 
The gravity, dilaton and gauge supermultiplets are not shown.}  
\label{Z3I}
\end{table}
%%%%%%%%%%%%%%%%%%%%%%%%%%%%%%%%%%%%%%%%%%%%%%%%%%%%%%%%%%%%%%%%%%%%%%%%%%%%%%%

%%%%%%%%%%%%%Table II %%%%%%%%
%%%%%%%%%%%%%%%%%%%%%%%%%%%%%%%%%%%%%%%%%%%%%%%%%%%%%%%%%%%%%%%%%%%%%%%%%%%%%%%
\begin{table}[t]
\begin{tabular}{|c|c|l|l|}
%%%%%%%%%%%%%%%%%%%%%%%%%%%%%%%%%%%%%%%%%%%%%%%%%%%%%%%%%%%%%%%%%%%%%%%%%%%%
Sector & Field & $SU(12)\otimes SO(8) \otimes U(1)$ & Comments\\
\hline
%%%%%%%%%%%%%%%%%%%%%%%%%%%%%%%%%%%%%%%%%%%%%%%%%%%%%%%%%%%%%%%%%%%%%%%%%%%%
  & $\phi_{ab}$ & $9({\bf 1}, {\bf 1})(0)_L$ & $a,b=1,2,3$\\
Untwisted & $Q_a$ & $3({\bf 12},{\bf 8}_v)(-1)_L$ & \\
     & $\Phi_a$ & $3({\overline {\bf 66}}, {\bf 1})(+2)_L$ & \\
\hline
%%%%%%%%%%%%%%%%%%%%%%%%%%%%%%%%%%%%%%%%%%%%%%%%%%%%%%%%%%%%%%%%%%%%%%%%%%%%
Twisted & $S_{\alpha\beta\gamma}$ & $27({\bf 1}, {\bf 1})(-4)_L$ & $\alpha,\beta,\gamma=1,2,3$ \\
   & $T_{\alpha\beta\gamma}$ & $27({\bf 1}, {\bf 8}_s)(+2)_L$ & \\
%%%%%%%%%%%%%%%%%%%%%%%%%%%%%%%%%%%%%%%%%%%%%%%%%%%%%%%%%%%%%%%%%%%%%%%%%%%
\end{tabular}
%%%%%%%%%%%%%%%%%%%%%%%%%%%%%%%%%%%%%%%%%%%%%%%%%%%%%%%%%%%%%%%%%%%%%%%%%%%
\caption{The massless spectrum of the heterotic ${\bf Z}_3$ orbifold
model with $N=1$ space-time 
supersymmetry and gauge group $SU(12)\otimes SO(8) \otimes U(1)$ discussed in 
section IX. The gravity, dilaton and gauge supermultiplets are not shown.}  
\label{Z3h}
\end{table}
%%%%%%%%%%%%%%%%%%%%%%%%%%%%%%%%%%%%%%%%%%%%%%%%%%%%%%%%%%%%%%%%%%%%%%%%%%%%%%%

%%%%%%%%%%%%%Table III %%%%%%%%
%%%%%%%%%%%%%%%%%%%%%%%%%%%%%%%%%%%%%%%%%%%%%%%%%%%%%%%%%%%%%%%%%%%%%%%%%%%%%%%
\begin{table}[t]
\begin{tabular}{|c|l|l|l|}
%%%%%%%%%%%%%%%%%%%%%%%%%%%%%%%%%%%%%%%%%%%%%%%%%%%%%%%%%%%%%%%%%%%%%%%%%%%%
 Sector & $[SU(6)\otimes SU(6)\otimes SU(4)\otimes U(1)^3]^2$
        & $(H_1,H_2,H_3)_{-1}$ & $(H_1,H_2,H_3)_{-1/2}$ \\
\hline
%%%%%%%%%%%%%%%%%%%%%%%%%%%%%%%%%%%%%%%%%%%%%%%%%%%%%%%%%%%%%%%%%%%%%%%%%%%%
Closed & & &\\
Untwisted & $5({\bf 1}, {\bf 1}, {\bf 1}; {\bf 1}, {\bf 1}, {\bf 1})
(0,0,0;0,0,0)_L$  & & \\
\hline
%%%%%%%%%%%%%%%%%%%%%%%%%%%%%%%%%%%%%%%%%%%%%%%%%%%%%%%%%%%%%%%%%%%%%%%%%%%%
Closed &  & & \\
${\bf Z}_3$ Twisted  & $15({\bf 1}, {\bf 1}, {\bf 1}; {\bf 1}, {\bf 1}, 
{\bf 1})
(0,0,0;0,0,0)_L$ & & \\
\hline
Closed & & &  \\
${\bf Z}_6$ Twisted  & $3({\bf 1}, {\bf 1}, {\bf 1}; {\bf 1}, {\bf 1}, {\bf 1})
(0,0,0;0,0,0)_L$ & & \\
\hline
Closed & & &  \\
${\bf Z}_2$ Twisted  & $11({\bf 1}, {\bf 1}, {\bf 1}; {\bf 1}, {\bf 1}, 
{\bf 1})
(0,0,0;0,0,0)_L$ & & \\
\hline
%%%%%%%%%%%%%%%%%%%%%%%%%%%%%%%%%%%%%%%%%%%%%%%%%%%%%%%%%%%%%%%%%%%%%%%%%%%%
           & $({\bf 15},{\bf 1},{\bf 1};{\bf 1},{\bf 1},{\bf 1})
(+2,0,0;0,0,0)_L$ & $(+1,0,0)$ & $(+{1\over 2},-{1\over 2},-{1\over 2})$ \\
           &  $({\bf 15},{\bf 1},{\bf 1};{\bf 1},{\bf 1},{\bf 1})
(+2,0,0;0,0,0)_L$    & $(0,+1,0)$ & $(-{1\over 2},+{1\over 2},-{1\over 2})$ \\   
           & $({\bf 1},\overline{\bf 15}, {\bf 1}; {\bf 1},{\bf 1},{\bf 1})
(0,-2,0;0,0,0)_L$ & $(+1,0,0)$ & $(+{1\over 2},-{1\over 2},-{1\over 2})$ \\
           & $({\bf 1},\overline{\bf 15}, {\bf 1}; {\bf 1},{\bf 1},{\bf 1})
(0,-2,0;0,0,0)_L$     & $(0,+1,0)$ & $(-{1\over 2},+{1\over 2},-{1\over 2})$ \\ 
           & $(\overline{\bf 6},{\bf 1},\overline{\bf 4};{\bf 1},{\bf 1},
{\bf 1})(-1,0,-1;0,0,0)_L$ & $(+1,0,0)$ 
& $(+{1\over 2},-{1\over 2},-{1\over 2})$ \\
           &   $(\overline{\bf 6},{\bf 1},\overline{\bf 4};{\bf 1},{\bf 1},
{\bf 1})(-1,0,-1;0,0,0)_L$   & $(0,+1,0)$ & $(-{1\over 2},+{1\over 2},-{1\over 2})$ \\ 
Open $99$ & $({\bf 1},{\bf 6},{\bf 4};{\bf 1},{\bf 1},{\bf 1})
(0,+1,+1;0,0,0)_L$ & $(+1,0,0)$ & $(+{1\over 2},-{1\over 2},-{1\over 2})$ \\
           &   $({\bf 1},{\bf 6},{\bf 4};{\bf 1},{\bf 1},{\bf 1})
(0,+1,+1;0,0,0)_L$    & $(0,+1,0)$ & $(-{1\over 2},+{1\over 2},-{1\over 2})$ \\ 
           & $({\bf 6},\overline{\bf 6},{\bf 1};{\bf 1},{\bf 1},{\bf 1})
(+1,-1,0;0,0,0)_L$ & $(0,0,+1)$ & $(-{1\over 2},-{1\over 2},+{1\over 2})$ \\
           & $(\overline{\bf 6},{\bf 1},{\bf 4};{\bf 1},{\bf 1},{\bf 1})
(-1,0,+1;0,0,0)_L$ & $(0,0,+1)$ & $(-{1\over 2},-{1\over 2},+{1\over 2})$ \\
           & $({\bf 1},{\bf 6},\overline{\bf 4};{\bf 1},{\bf 1},{\bf 1})
(0,+1,-1;0,0,0)_L$ & $(0,0,+1)$ & $(-{1\over 2},-{1\over 2},+{1\over 2})$ \\ 
\hline
%%%%%%%%%%%%%%%%%%%%%%%%%%%%%%%%%%%%%%%%%%%%%%%%%%%%%%%%%%%%%%%%%%%%%%%%%%%
           & $({\bf 1},{\bf 1},{\bf 1};{\bf 15},{\bf 1},{\bf 1})
(0,0,0;+2,0,0)_L$ & $(+1,0,0)$ & $(+{1\over 2},-{1\over 2},-{1\over 2})$ \\
           &   $({\bf 1},{\bf 1},{\bf 1};{\bf 15},{\bf 1},{\bf 1})
(0,0,0;+2,0,0)_L$ & $(+1,0,0)$ & $(+{1\over 2},-{1\over 2},-{1\over 2})$\\ 
           & $({\bf 1},{\bf 1},{\bf 1};{\bf 1},\overline{\bf 15},{\bf 1})
(0,0,0;0,-2,0)_L$ & $(+1,0,0)$ & $(+{1\over 2},-{1\over 2},-{1\over 2})$ \\
           &  $({\bf 1},{\bf 1},{\bf 1};{\bf 1},\overline{\bf 15},{\bf 1})
(0,0,0;0,-2,0)_L$    & $(0,+1,0)$ & $(-{1\over 2},+{1\over 2},-{1\over 2})$ \\ 
          & $({\bf 1},{\bf 1},{\bf 1};\overline{\bf 6},{\bf 1},
\overline{\bf 4})(0,0,0;-1,0,-1)_L$ & $(+1,0,0)$ 
                  & $(+{1\over 2},-{1\over 2},-{1\over 2})$ \\
           &  $({\bf 1},{\bf 1},{\bf 1};\overline{\bf 6},{\bf 1},
\overline{\bf 4})(0,0,0;-1,0,-1)_L$    & $(0,+1,0)$ & $(-{1\over 2},+{1\over 2},-{1\over 2})$ \\
Open  $55$ & $({\bf 1},{\bf 1},{\bf 1};{\bf 1},{\bf 6},{\bf 4})
(0,0,0;0,+1,+1)_L$ & $(+1,0,0)$ & $(+{1\over 2},-{1\over 2},-{1\over 2})$ \\
           &   $({\bf 1},{\bf 1},{\bf 1};{\bf 1},{\bf 6},{\bf 4})
(0,0,0;0,+1,+1)_L$   & $(0,+1,0)$ & $(-{1\over 2},+{1\over 2},-{1\over 2})$ \\ 
           & $({\bf 1},{\bf 1},{\bf 1};{\bf 6},\overline{\bf 6},{\bf 1})
(0,0,0;+1,-1,0)_L$ & $(0,0,+1)$ & $(-{1\over 2},-{1\over 2},+{1\over 2})$ \\
           & $({\bf 1},{\bf 1},{\bf 1};\overline{\bf 6},{\bf 1},{\bf 4})
(0,0,0;-1,0,+1)_L$ & $(0,0,+1)$ & $(-{1\over 2},-{1\over 2},+{1\over 2})$ \\
           & $({\bf 1},{\bf 1},{\bf 1};{\bf 1},{\bf 6},\overline{\bf 4})
(0,0,0;0,+1,-1)_L$ & $(0,0,+1)$ & $(-{1\over 2},-{1\over 2},+{1\over 2})$ \\
\hline
%%%%%%%%%%%%%%%%%%%%%%%%%%%%%%%%%%%%%%%%%%%%%%%%%%%%%%%%%%%%%%%%%%%%%%%%%%%
           & $({\bf 6},{\bf 1},{\bf 1};{\bf 6},{\bf 1},{\bf 1})
(+1,0,0;+1,0,0)_L$ & $(+{1\over 2},+{1\over 2},0)$ & $(0,0,-{1\over 2})$ \\
           & $({\bf 1},{\bf 6},{\bf 1};{\bf 1},{\bf 1},{\bf 4})
(0,+1,0;0,0,+1)_L$ & $(+{1\over 2},+{1\over 2},0)$ & $(0,0,-{1\over 2})$ \\
Open  $59$ & $({\bf 1},{\bf 1},{\bf 4};{\bf 1},{\bf 6},{\bf 1})
(0,0,+1;0,+1,0)_L$ & $(+{1\over 2},+{1\over 2},0)$ & $(0,0,-{1\over 2})$ \\
           & $(\overline{\bf 6},{\bf 1},{\bf 1};{\bf 1},{\bf 1},
\overline{\bf 4})(-1,0,0;0,0,-1)_L$ 
& $(+{1\over 2},+{1\over 2},0)$ & $(0,0,-{1\over 2})$ \\ 
          & $({\bf 1},\overline{\bf 6},{\bf 1};{\bf 1},\overline{\bf 6},
{\bf 1})(0,-1,0;0,-1,0)_L$ 
& $(+{1\over 2},+{1\over 2},0)$ & $(0,0,-{1\over 2})$ \\ 
          & $({\bf 1},{\bf 1},\overline{\bf 4};\overline{\bf 6},{\bf 1},
{\bf 1})(0,0,-1;-1,0,0)_L$
& $(+{1\over 2},+{1\over 2},0)$ & $(0,0,-{1\over 2})$ \\ 
%%%%%%%%%%%%%%%%%%%%%%%%%%%%%%%%%%%%%%%%%%%%%%%%%%%%%%%%%%%%%%%%%%%%%%%%%%%
\end{tabular}
%%%%%%%%%%%%%%%%%%%%%%%%%%%%%%%%%%%%%%%%%%%%%%%%%%%%%%%%%%%%%%%%%%%%%%%%%%%
\caption{The massless spectrum of the type I ${\bf Z}_6$ orbifold model 
with $N=1$ space-time 
supersymmetry and gauge group 
$[SU(6)\otimes SU(6) \otimes SU(4) \otimes U(1)^3]^2 $
discussed in 
section X. 
The $H$-charges in both the $-1$ picture and the $-1/2$ picture for states
in the open
string sector are also given. The gravity, dilaton and gauge supermultiplets 
are not shown.}  
\label{Z6}
\end{table}
%%%%%%%%%%%%%%%%%%%%%%%%%%%%%%%%%%%%%%%%%%%%%%%%%%%%%%%%%%%%%%%%%%%%%%%%%%%%%%%

%%%%%%%%%%%%%Table IV%%%%%%%%
%%%%%%%%%%%%%%%%%%%%%%%%%%%%%%%%%%%%%%%%%%%%%%%%%%%%%%%%%%%%%%%%%%%%%%%%%%%%%%%
\begin{table}[t]
\begin{tabular}{|c|l|l|l|}
%%%%%%%%%%%%%%%%%%%%%%%%%%%%%%%%%%%%%%%%%%%%%%%%%%%%%%%%%%%%%%%%%%%%%%%%%%%%
 Sector & $[SU(8)\otimes SU(8)\otimes U(1)^2]^2$
        & $(H_1,H_2,H_3)_{-1}$ & $(H_1,H_2,H_3)_{-1/2}$ \\
\hline
%%%%%%%%%%%%%%%%%%%%%%%%%%%%%%%%%%%%%%%%%%%%%%%%%%%%%%%%%%%%%%%%%%%%%%%%%%%%
%& & & \\
Closed & $6 ({\bf 1}, {\bf 1}, {\bf 1}; {\bf 1}, {\bf 1}, {\bf 1})
 (0,0,0;0,0,0)_L$& &\\
Untwisted & & & \\
%& & & \\
\hline
%%%%%%%%%%%%%%%%%%%%%%%%%%%%%%%%%%%%%%%%%%%%%%%%%%%%%%%%%%%%%%%%%%%%%%%%%%%%
%& & & \\
Closed &  $16 ({\bf 1}, {\bf 1}, {\bf 1}; {\bf 1}, {\bf 1}, {\bf 1})
(0,0,0;0,0,0)_L$& & \\
${\bf Z}_4$ Twisted  &  & & \\
%& & & \\
\hline
%& & & \\
Closed & $16 ({\bf 1}, {\bf 1}, {\bf 1}; {\bf 1}, {\bf 1}, {\bf 1})
(0,0,0;0,0,0)_L$ & &  \\
${\bf Z}_2$ Twisted  &  & & \\
%& & & \\
\hline
%%%%%%%%%%%%%%%%%%%%%%%%%%%%%%%%%%%%%%%%%%%%%%%%%%%%%%%%%%%%%%%%%%%%%%%%%%%%
 & & & \\
 & $({\bf 8},\overline{\bf 8};{\bf 1},{\bf 1})(+1,-1;0,0)_L$
 & $(+1,0,0)$ & $(+{1\over 2},-{1\over 2},-{1\over 2})$ \\
 & $(\overline{\bf 28},{\bf 1};{\bf 1},{\bf 1})(-2,0;0,0)_L$ 
 & $(+1,0,0)$ & $(+{1\over 2},-{1\over 2},-{1\over 2})$ \\
 & $({\bf 1},{\bf 28};{\bf 1},{\bf 1})(0,+2;0,0)_L$ 
 & $(+1,0,0)$ & $(+{1\over 2},-{1\over 2},-{1\over 2})$ \\
Open $99$
 & $({\bf 8},\overline{\bf 8};{\bf 1},{\bf 1})(+1,-1;0,0)_L$
 & $(0,+1,0)$ & $(-{1\over 2},+{1\over 2},-{1\over 2})$ \\
 & $(\overline{\bf 28},{\bf 1};{\bf 1},{\bf 1})(-2,0;0,0)_L$ 
 & $(0,+1,0)$ & $(-{1\over 2},+{1\over 2},-{1\over 2})$ \\
 & $({\bf 1},{\bf 28};{\bf 1},{\bf 1})(0,+2;0,0)_L$ 
 & $(0,+1,0)$ & $(-{1\over 2},+{1\over 2},-{1\over 2})$ \\
 & $({\bf 8},{\bf 8};{\bf 1},{\bf 1})(+1,+1;0,0)_L$
 & $(0,0,+1)$ & $(-{1\over 2},-{1\over 2},+{1\over 2})$ \\
 & $(\overline{\bf 8},\overline{\bf 8};{\bf 1},{\bf 1})(-1,-1;0,0)_L$
 & $(0,0,+1)$ & $(-{1\over 2},-{1\over 2},+{1\over 2})$ \\
 & & & \\
\hline
%%%%%%%%%%%%%%%%%%%%%%%%%%%%%%%%%%%%%%%%%%%%%%%%%%%%%%%%%%%%%%%%%%%%%%%%%%%%
 & & & \\
 & $({\bf 1},{\bf 1};{\bf 8},\overline{\bf 8})(0,0;+1,-1)_L$
 & $(+1,0,0)$ & $(+{1\over 2},-{1\over 2},-{1\over 2})$ \\
 & $({\bf 1},{\bf 1};\overline{\bf 28},{\bf 1})(0,0;-2,0)_L$ 
 & $(+1,0,0)$ & $(+{1\over 2},-{1\over 2},-{1\over 2})$ \\
 & $({\bf 1},{\bf 1};{\bf 1},{\bf 28})(0,0;0,+2)_L$ 
 & $(+1,0,0)$ & $(+{1\over 2},-{1\over 2},-{1\over 2})$ \\
Open $55$
 & $({\bf 1},{\bf 1};{\bf 8},\overline{\bf 8})(0,0;+1,-1)_L$
 & $(0,+1,0)$ & $(-{1\over 2},+{1\over 2},-{1\over 2})$ \\
 & $({\bf 1},{\bf 1};\overline{\bf 28},{\bf 1})(0,0;-2,0)_L$ 
 & $(0,+1,0)$ & $(-{1\over 2},+{1\over 2},-{1\over 2})$ \\
 & $({\bf 1},{\bf 1};{\bf 1},{\bf 28})(0,0;0,+2)_L$ 
 & $(0,+1,0)$ & $(-{1\over 2},+{1\over 2},-{1\over 2})$ \\
 & $({\bf 1},{\bf 1};{\bf 8},{\bf 8})(0,0;+1,+1)_L$
 & $(0,0,+1)$ & $(-{1\over 2},-{1\over 2},+{1\over 2})$ \\
 & $({\bf 1},{\bf 1};\overline{\bf 8},\overline{\bf 8})(0,0;-1,-1)_L$
 & $(0,0,+1)$ & $(-{1\over 2},-{1\over 2},+{1\over 2})$ \\
 & & & \\
\hline
%%%%%%%%%%%%%%%%%%%%%%%%%%%%%%%%%%%%%%%%%%%%%%%%%%%%%%%%%%%%%%%%%%%%%%%%%%%%
 & & & \\
 & $(\overline{\bf 8},{\bf 1};\overline{\bf 8},{\bf 1})(-1,0;-1,0)_L$
 & $(+{1\over 2},+{1\over 2},0)$ & $(0,0,-{1\over 2})$ \\
 Open $59$ & $({\bf 1},{\bf 8};{\bf 1},{\bf 8})(0,+1;0,+1)_L$ 
 & $(+{1\over 2},+{1\over 2},0)$ & $(0,0,-{1\over 2})$ \\
 & $({\bf 8},{\bf 1};{\bf 1},{\overline {\bf 8}})(+1,0;0,-1)_L$ 
 & $(+{1\over 2},+{1\over 2},0)$ & $(0,0,-{1\over 2})$ \\
  & $({\bf 1},{\overline {\bf 8}};{\bf 8},{\bf 1})(0,-1;+1,0)_L$ 
 & $(+{1\over 2},+{1\over 2},0)$ & $(0,0,-{1\over 2})$ \\
 & & & \\
\end{tabular}
%%%%%%%%%%%%%%%%%%%%%%%%%%%%%%%%%%%%%%%%%%%%%%%%%%%%%%%%%%%%%%%%%%%%%%%%%%%
\caption{The perturbative (from the orientifold viewpoint) 
massless spectrum of the four dimensional 
${\cal N}=1$ space-time supersymmetric orientifold of Type IIB on $T^6/{\bf Z}_4$ 
orbifold discussed in section XI. The gauge group is 
$[U(8)\otimes U(8)]_{99}\otimes [U(8)\otimes U(8)]_{55}$. 
The $H$-charges in both the $-1$ picture and the $-1/2$ picture for states
in the open
string sectors are also given. The gravity, dilaton and gauge supermultiplets 
are not shown.} 
\label{Z4} 
\end{table}
%%%%%%%%%%%%%%%%%%%%%%%%%%%%%%%%%%%%%%%%%%%%%%%%%%%%%%%%%%%%%%%%%%%%%%%%%%%%

%%%%%%%%%%%%%Table V %%%%%%%%
%%%%%%%%%%%%%%%%%%%%%%%%%%%%%%%%%%%%%%%%%%%%%%%%%%%%%%%%%%%%%%%%%%%%%%%%%%%%%%%
\begin{table}[t]
\begin{tabular}{|c|l|l|l|}
%%%%%%%%%%%%%%%%%%%%%%%%%%%%%%%%%%%%%%%%%%%%%%%%%%%%%%%%%%%%%%%%%%%%%%%%%%%%
 Sector & $[SU(4)\otimes SU(4)\otimes SU(8)\otimes U(1)^3]^2$
        & $(H_1,H_2,H_3)_{-1}$ & $(H_1,H_2,H_3)_{-1/2}$ \\
\hline
%%%%%%%%%%%%%%%%%%%%%%%%%%%%%%%%%%%%%%%%%%%%%%%%%%%%%%%%%%%%%%%%%%%%%%%%%%%%
Closed & & &\\
Untwisted & $4({\bf 1}, {\bf 1}, {\bf 1}; {\bf 1}, {\bf 1}, {\bf 1})
(0,0,0;0,0,0)_L$  & & \\
\hline
%%%%%%%%%%%%%%%%%%%%%%%%%%%%%%%%%%%%%%%%%%%%%%%%%%%%%%%%%%%%%%%%%%%%%%%%%%%%
Closed &  & & \\
${\bf Z}_3$ Twisted  & $18({\bf 1}, {\bf 1}, {\bf 1}; {\bf 1}, {\bf 1}, 
{\bf 1})
(0,0,0;0,0,0)_L$ & & \\
\hline
Closed & & &  \\
${\bf Z}_6$ Twisted  & $12({\bf 1}, {\bf 1}, {\bf 1}; {\bf 1}, {\bf 1}, 
{\bf 1})
(0,0,0;0,0,0)_L$ & & \\
\hline
Closed & & &  \\
${\bf Z}_2$ Twisted  & $12({\bf 1}, {\bf 1}, {\bf 1}; {\bf 1}, {\bf 1}, 
{\bf 1})
(0,0,0;0,0,0)_L$ & & \\
\hline
%%%%%%%%%%%%%%%%%%%%%%%%%%%%%%%%%%%%%%%%%%%%%%%%%%%%%%%%%%%%%%%%%%%%%%%%%%%%
           & $(\overline{\bf 4},{\bf 1},{\bf 8};{\bf 1},{\bf 1},{\bf 1})
(-1,0,+1;0,0,0)_L$ & $(0,0,+1)$ & $(-{1\over 2},-{1\over 2},+{1\over 2})$ \\
           & $({\bf 1},{\bf 4},\overline{\bf 8};{\bf 1},{\bf 1},{\bf 1})
(0,+1,-1;0,0,0)_L$ & $(0,0,+1)$ & $(-{1\over 2},-{1\over 2},+{1\over 2})$ \\
           & $({\bf 4},\overline{\bf 4},{\bf 1};{\bf 1},{\bf 1},{\bf 1})
(+1,-1,0;0,0,0)_L$ & $(0,0,+1)$ & $(-{1\over 2},-{1\over 2},+{1\over 2})$ \\
           & $({\bf 4},{\bf 1},{\bf 8};{\bf 1},{\bf 1},{\bf 1})
(+1,0,+1;0,0,0)_L$ & $(0,+1,0)$ & $(-{1\over 2},+{1\over 2},-{1\over 2})$ \\
           & $({\bf 1},\overline{\bf 4},\overline{\bf 8};{\bf 1},{\bf 1},
{\bf 1})
(0,-1,-1;0,0,0)_L$ & $(0,+1,0)$ & $(-{1\over 2},+{1\over 2},-{1\over 2})$ \\
Open $99$  & $({\bf 6},{\bf 1},{\bf 1};{\bf 1},{\bf 1},{\bf 1})
(-2,0,0;0,0,0)_L$ & $(0,+1,0)$ & $(-{1\over 2},+{1\over 2},-{1\over 2})$ \\
           & $({\bf 1},{\bf 6},{\bf 1};{\bf 1},{\bf 1},{\bf 1})
(0,+2,0;0,0,0)_L$ & $(0,+1,0)$ & $(-{1\over 2},+{1\over 2},-{1\over 2})$ \\ 
           & $({\bf 1},{\bf 1},{\bf 28};{\bf 1},{\bf 1},{\bf 1})
(0,0,+2;0,0,0)_L$ & $(+1,0,0)$ & $(+{1\over 2},-{1\over 2},-{1\over 2})$ \\
           & $({\bf 1},{\bf 1},\overline{\bf 28};{\bf 1},{\bf 1},{\bf 1})
(0,0,-2;0,0,0)_L$ & $(+1,0,0)$ & $(+{1\over 2},-{1\over 2},-{1\over 2})$ \\
           & $({\bf 4},{\bf 4},{\bf 1};{\bf 1},{\bf 1},{\bf 1})
(+1,+1,0;0,0,0)_L$ & $(+1,0,0)$ & $(+{1\over 2},-{1\over 2},-{1\over 2})$ \\
           & $(\overline{\bf 4},\overline{\bf 4},{\bf 1};{\bf 1},{\bf 1},
{\bf 1})
(-1,-1,0;0,0,0)_L$ & $(+1,0,0)$ & $(+{1\over 2},-{1\over 2},-{1\over 2})$ \\
\hline
%%%%%%%%%%%%%%%%%%%%%%%%%%%%%%%%%%%%%%%%%%%%%%%%%%%%%%%%%%%%%%%%%%%%%%%%%%%
           & $({\bf 1},{\bf 1},{\bf 1};\overline{\bf 4},{\bf 1},{\bf 8})
(0,0,0;-1,0,+1)_L$ & $(0,0,+1)$ & $(-{1\over 2},-{1\over 2},+{1\over 2})$ \\
           & $({\bf 1},{\bf 1},{\bf 1};{\bf 1},{\bf 4},\overline{\bf 8})
(0,0,0;0,+1,-1)_L$ & $(0,0,+1)$ & $(-{1\over 2},-{1\over 2},+{1\over 2})$ \\
           & $({\bf 1},{\bf 1},{\bf 1};{\bf 4},\overline{\bf 4},{\bf 1})
(0,0,0;+1,-1,0)_L$ & $(0,0,+1)$ & $(-{1\over 2},-{1\over 2},+{1\over 2})$ \\
           & $({\bf 1},{\bf 1},{\bf 1};{\bf 4},{\bf 1},{\bf 8})
(0,0,0;+1,0,+1)_L$ & $(0,+1,0)$ & $(-{1\over 2},+{1\over 2},-{1\over 2})$ \\
           & $({\bf 1},{\bf 1},{\bf 1};{\bf 1},\overline{\bf 4},
\overline{\bf 8})
(0,0,0;0,-1,-1)_L$ & $(0,+1,0)$ & $(-{1\over 2},+{1\over 2},-{1\over 2})$ \\
Open $55$  & $({\bf 1},{\bf 1},{\bf 1};{\bf 6},{\bf 1},{\bf 1})
(0,0,0;-2,0,0)_L$ & $(0,+1,0)$ & $(-{1\over 2},+{1\over 2},-{1\over 2})$ \\
           & $({\bf 1},{\bf 1},{\bf 1};{\bf 1},{\bf 6},{\bf 1})
(0,0,0;0,+2,0)_L$ & $(0,+1,0)$ & $(-{1\over 2},+{1\over 2},-{1\over 2})$ \\ 
           & $({\bf 1},{\bf 1},{\bf 1};{\bf 1},{\bf 1},{\bf 28})
(0,0,0;0,0,+2)_L$ & $(+1,0,0)$ & $(+{1\over 2},-{1\over 2},-{1\over 2})$ \\
           & $({\bf 1},{\bf 1},{\bf 1};{\bf 1},{\bf 1},\overline{\bf 28})
(0,0,0;0,0,-2)_L$ & $(+1,0,0)$ & $(+{1\over 2},-{1\over 2},-{1\over 2})$ \\
           & $({\bf 1},{\bf 1},{\bf 1};{\bf 4},{\bf 4},{\bf 1})
(0,0,0;+1,+1,0)_L$ & $(+1,0,0)$ & $(+{1\over 2},-{1\over 2},-{1\over 2})$ \\
           & $({\bf 1},{\bf 1},{\bf 1};\overline{\bf 4},\overline{\bf 4},
{\bf 1})(0,0,0;-1,-1,0)_L$ 
           & $(+1,0,0)$ & $(+{1\over 2},-{1\over 2},-{1\over 2})$ \\
\hline
%%%%%%%%%%%%%%%%%%%%%%%%%%%%%%%%%%%%%%%%%%%%%%%%%%%%%%%%%%%%%%%%%%%%%%%%%%%
           & $({\bf 4},{\bf 1},{\bf 1};{\bf 4},{\bf 1},{\bf 1})
(+1,0,0;+1,0,0)_L$ & $(+{1\over 2},+{1\over 2},0)$ & $(0,0,-{1\over 2})$ \\
           & $({\bf 1},{\bf 4},{\bf 1};{\bf 1},{\bf 1},{\bf 8})
(0,+1,0;0,0,+1)_L$ & $(+{1\over 2},+{1\over 2},0)$ & $(0,0,-{1\over 2})$ \\
Open  $59$ & $({\bf 1},{\bf 1},{\bf 8};{\bf 1},{\bf 4},{\bf 1})
(0,0,+1;0,+1,0)_L$ & $(+{1\over 2},+{1\over 2},0)$ & $(0,0,-{1\over 2})$ \\
           & $(\overline{\bf 4},{\bf 1},{\bf 1};{\bf 1},{\bf 1},
\overline{\bf 8})(-1,0,0;0,0,-1)_L$ 
& $(+{1\over 2},+{1\over 2},0)$ & $(0,0,-{1\over 2})$ \\ 
          & $({\bf 1},\overline{\bf 4},{\bf 1};{\bf 1},\overline{\bf 4},
{\bf 1})(0,-1,0;0,-1,0)_L$ 
& $(+{1\over 2},+{1\over 2},0)$ & $(0,0,-{1\over 2})$ \\ 
          & $({\bf 1},{\bf 1},\overline{\bf 8};\overline{\bf 4},{\bf 1},
{\bf 1})(0,0,-1;-1,0,0)_L$
& $(+{1\over 2},+{1\over 2},0)$ & $(0,0,-{1\over 2})$ \\ 
%%%%%%%%%%%%%%%%%%%%%%%%%%%%%%%%%%%%%%%%%%%%%%%%%%%%%%%%%%%%%%%%%%%%%%%%%%%
\end{tabular}
%%%%%%%%%%%%%%%%%%%%%%%%%%%%%%%%%%%%%%%%%%%%%%%%%%%%%%%%%%%%%%%%%%%%%%%%%%%
\caption{The perturbative (from the orientifold viewpoint) massless spectrum of the four dimensional ${\cal N}=1$ space-time supersymmetric orientifold of Type IIB on $T^6/{\bf Z}_6^\prime$ orbifold discussed 
in section XI. The gauge group is
$[U(4)\otimes U(4) \otimes U(8)]_{99}\otimes [U(4)\otimes U(4) \otimes U(8)]_{55}$.
The $H$-charges in both the $-1$ picture and the $-1/2$ picture for states
in the open
string sectors are also given. The gravity, dilaton and gauge supermultiplets 
are not shown.} 
\label{Z6p} 
\end{table}
%%%%%%%%%%%%%%%%%%%%%%%%%%%%%%%%%%%%%%%%%%%%%%%%%%%%%%%%%%%%%%%%%%%%%%%%%%%%%%%

%%%%%%%%%%%%%%%%%%%%%%%%%%%%%%%%%%%%%

\end{document}